\documentclass[twoside,tighten]{revtex4} 
\usepackage{graphicx}
\usepackage{hyperref}

\def\lesssim{\mathrel{\hbox{\rlap{\hbox{\lower4pt\hbox{$\sim$}}}\hbox{$<$}}}}
\def\gtrsim{\mathrel{\hbox{\rlap{\hbox{\lower4pt\hbox{$\sim$}}}\hbox{$>$}}}}

\begin{document}

\title{Gamma-ray and X-ray Properties of Pulsar Wind Nebulae and Unidentified Galactic TeV Sources\footnote{This is an abridged version of the review. The full review article is to be published in  "The Universe Evolution. Astrophysical and Nuclear Aspects",  2013,  Nova Science Publishers, Inc. Reproduced with  permission of the publisher.}}

\author{Oleg Kargaltsev,$^1$ Blagoy Rangelov,$^1$, and George G.\ Pavlov,$^{2}$ }
\affiliation{$^1$ George Washington University, 105 Corcoran Hall, Washington,
 DC 20052}
\affiliation{$^2$ Pennsylvania State University, 525 Davey Lab., University Park,
PA 16802}

\vspace{2in}


\newcommand{\edot}{\dot{E}}
\newcommand{\pdot}{\dot{P}}
\newcommand{\lpwn}{L_{\rm pwn}}
\newcommand{\lpsr}{L_{\rm psr}}
\newcommand{\etapwn}{\eta_{\rm pwn}}
\newcommand{\etapsr}{\eta_{\rm psr}}
\newcommand{\chan}{{\sl Chandra}\/}
\newcommand{\be}{\begin{equation}}
\newcommand{\ee}{\end{equation}}
\newcommand{\xmm}{{\sl XMM-Newton\/}}
\newcommand{\suz}{{\sl Suzaku\/}}
\newcommand{\fermi}{{\sl Fermi\/}}

\begin{abstract}

We review properties of Galactic VHE sources detected at TeV energies. The number of associations between the VHE  sources and pulsars has grown in recent years, making pulsar-wind nebulae the dominant population, although there is still a substantial number of VHE sources which remain to be identified. Among the latter  there are several  ``dark'' sources which do not have   plausible counterparts at any other wavelengths. In this review we compile and compare the TeV and X-ray properties of pulsar wind nebulae (PWNe),  PWN candidates, and unidentified TeV sources.

\end{abstract}

\maketitle

\section{Current census of Galactic VHE sources}

During the past decade observations with the H.E.S.S.,~VERITAS, Milagro, and MAGIC  TeV $\gamma$-ray observatories revealed a large number of very-high energy (VHE) sources in the Galactic plane. Pulsar-wind nebulae (PWNe), shell-type supernova remnants (SNRs), and microquasar-type high-mass X-ray binaries (HMXBs)   appear to be prominent sources of the leptonic cosmic rays in our Galaxy.  Firmly identified sources of these types account for  48\% of the total number ($\sim90$) of Galactic VHE sources, with 28 PWNe, 10 SNRs and 5 HMXBs. There is also a large number of extended TeV sources positionally coincident with young energetic pulsars; they can be considered as TeV PWN candidates and are listed in Table~1. Some of these associations are more secure than others (e.g., in those cases when X-ray PWNe have been detected). In addition, there remains a sizable fraction of unidentified VHE sources (20 are listed in Table~4). For some of these sources, multiwavelength observations suggest a possible counterpart (such as an SNR interacting with a molecular cloud, or a star-forming region), but most of these associations are still uncertain because at least some of these sources still could  be powered by offset pulsars whose PWNe are faint in X-rays (see below). Finally, there are ``dark'' VHE sources, for which neither radio nor X-ray images reveal any plausible counterparts (last section in Table~4).

\section{Pulsar-Wind Nebulae}
 
Tables~1, 2, and 3  provide basic properties of 91 known PWNe and PWN candidates, most of which were initially found in X-rays. Note that for some of these  PWNe no pulsars have  been detected yet (Table~3) and only 52 of them have VHE associations or possible VHE counterparts.  Although PWNe represent  the dominant population among identified Galactic TeV sources,  establishing the association between a VHE source and a PWN/pulsar is often a complex task.  

Thanks to the \emph{Chandra X-ray Observatory},  $\sim70$  PWNe have been detected in X-rays  (Tables~1 and 2). X-ray nebulae around pulsars are created by synchrotron radiation of relativistic pulsar winds shocked in the ambient medium (e.g., \cite{1984ApJ...283..694K,2006ARAnA..44...17G,2008AIPC..983..171K}). Many of them show symmetric torus-jet or bow shock morphologies \cite{2008AIPC..983..171K}, but some PWNe associated with extended TeV sources consist of a compact brighter core and a faint asymmetric component elongated toward the offset TeV source (e.g., Vela X and HESS~J1825--137). Such an asymmetry could be created by the reverse SNR shock that reached one side of the PWN sooner than the other side because of nonhomogeneity of the SNR interiors, crushed the PWN, and pushed the PWN away from the pulsar \cite{2001ApJ...563..806B}.  Generally, in this scenario a PWN would consist of two distinct parts -- an offset  ``relic'' PWN filled with aged pulsar wind particles and a compact PWN near the pulsar filled with ``fresh'' pulsar wind particles. In many cases X-ray emission from a  relic PWN can be very faint or absent because the pulsar wind electrons become too cold and their characteristic synchrotron frequencies move outside the X-ray band. On the other hand, these electrons still can produce TeV emission via inverse Compton scattering (ICS) of the ambient low-energy background photons (such as CMB, diffuse Galactic infrared background,  or  starlight).

Typical energies of the synchrotron or ICS photons ($E_{\rm syn}$ and $E_{\rm ICS}$, respectively) depend on the energy of emitting electrons and the magnetic field strength or the energy of seed photons (e.g., the photon energy $\epsilon=kT_{\rm CMB} \sim 6\times 10^{-4}$ eV for the CMB radiation, where $k$ is the Boltzmann constant, and $T\sim2.7$~K):
\begin{equation}
E_{\rm syn} \sim 4 (E_e/100\,{\rm TeV})^2 (B/10^{-5}\,{\rm G})\,\, {\rm keV},
\end{equation}
\begin{equation}
E_{\rm ICS} \sim 1 (E_e/20\,{\rm TeV})^2 (\epsilon/6\times 10^{-4}\,{\rm eV})\,\, {\rm TeV},
\end{equation}
\noindent where $E_e$ is the electron energy and $B = 10^{-5}B_{-5}$~G is the magnetic field. In Eq.~(2) the Thomson scattering regime [$E_e \ll 400 (\epsilon/6\times 10^{-4}\,{\rm eV})^{-1}$ TeV] is assumed. One can also use Eqs.(1) and (2) to show that $E_{\rm syn} \sim 0.18 (E_{\rm ICS}/1\,{\rm TeV})(B/10^{-5}\,{\rm G}) (\epsilon/6\times 10^{-4}\,{\rm eV})^{-1}\,\, {\rm keV}$.
 The synchrotron cooling time for a compact PWN,
\begin{equation}
\tau_{X}\approx1.2~B_{-5}^{-3/2}(E_{\rm syn}/1~{\rm keV})^{-1/2}~{\rm kyr},
\end{equation} 
\noindent can be much shorter than the ICS+synchrotron cooling time for TeV emitting electrons in an extended PWN with a lower magnetic field,
\begin{equation}
\tau_{\gamma}\approx100(1+0.144B^2_{-6})^{-1}(E_{ICS}/1{\rm TeV})^{-1/2}~{\rm kyr},
\end{equation} 

\noindent where $B = 10^{-6}B_{-6}$~G is the magnetic field. The latter equation includes the CMB contribution only and an approximate correction for the Klein-Nishina effect \cite{2009ASSL..357..451D}. 
 More realistically, the cooling times ($\tau_X$, $\tau_\gamma$), the luminosities ($L_X$, $L_\gamma$), and the efficiencies ($\eta_{X}=L_X/\dot{E}$, $\eta_{\gamma}=L_\gamma/\dot{E}$) of X-ray and TeV PWNe also depend on other factors, such as the presence of local sources of IR photons, the size of the region from which luminosities are measured, and  the validity of the Thomson regime approximation.

Due to the longer cooling time for the electrons responsible for the TeV emission, the TeV PWN properties reflect the cumulative history of the pulsar wind losses, and therefore the apparent efficiency of a relic PWN can be large because the pulsar's spin-down power $\dot{E}$ was larger earlier in its life: $\dot{E}(t)=\dot{E}_0(1+t/\tau_{sd})^{-2}$, where $\dot{E}_0$ is the initial rotational energy loss rate, and a dipolar magnetic field is assumed. This is reflected in the left panel of Figure~1, where younger pulsars have more luminous X-ray PWNe while the TeV PWN luminosities  do not show an obvious correlation with the age. The lack of TeV PWN detections for pulsars older than  few hundred kilo-years can be explained by the fact that the characteristic cooling lifetime of the relic PWN due to the ICS on CMB photons is limited by $\sim100$~kyr. (Note that the characteristic pulsar age, $\tau_{\rm sd}=P/2\dot{P}$, may differ from its true age by a factor of a few). Because of the aging effect, one would expect the peak of the ICS TeV spectrum to be shifting toward lower energies as a relic PWN becomes older. Such very old  PWNe still could be sources of   GeV and radio-optical radiation unless the diffusion and advection completely dissolve the relic PWN bubble on that timescale.

\begin{figure}
 \centering
\includegraphics[width=0.37\textwidth,angle=90]{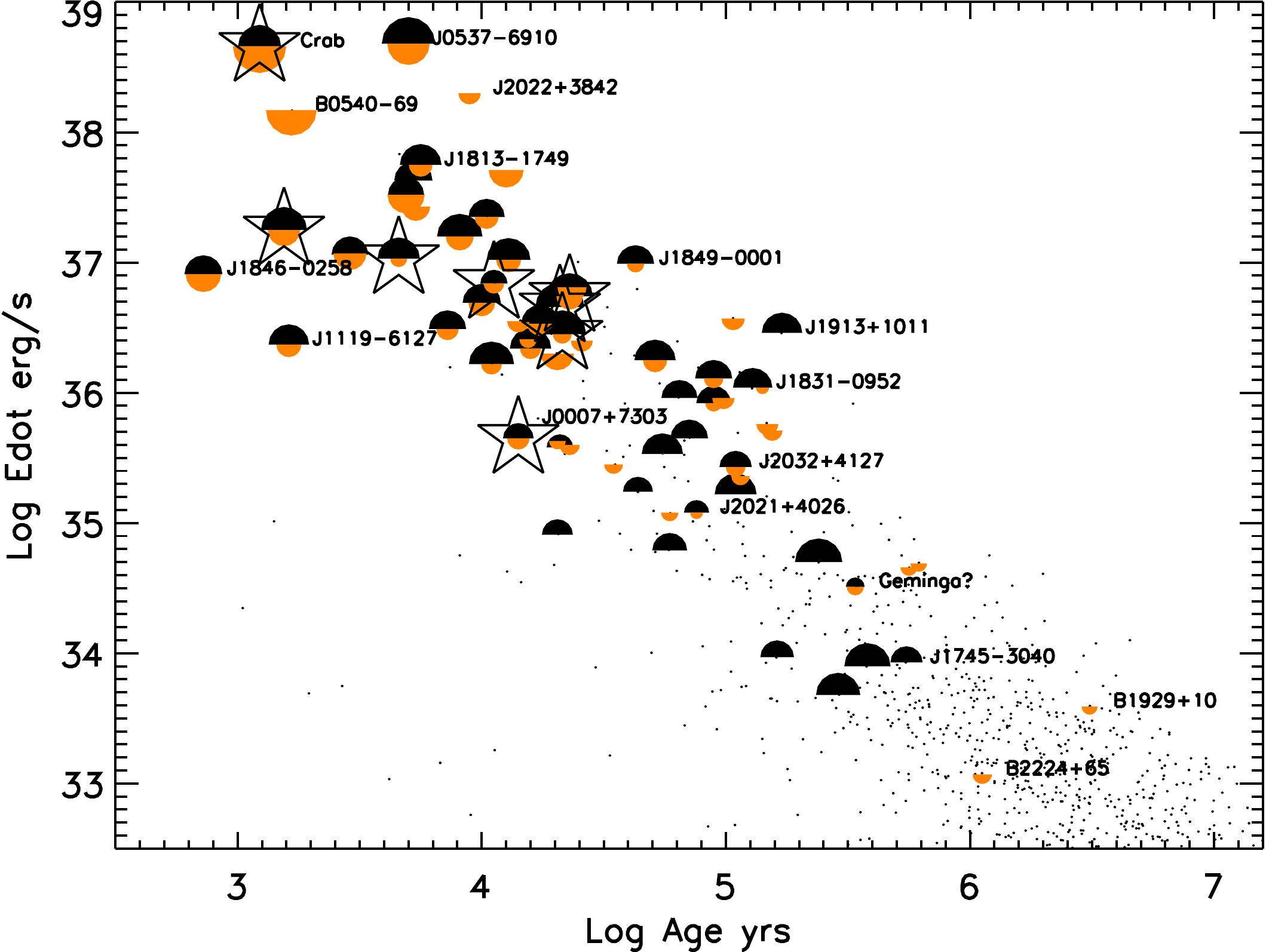}
\includegraphics[width=0.37\textwidth,angle=90]{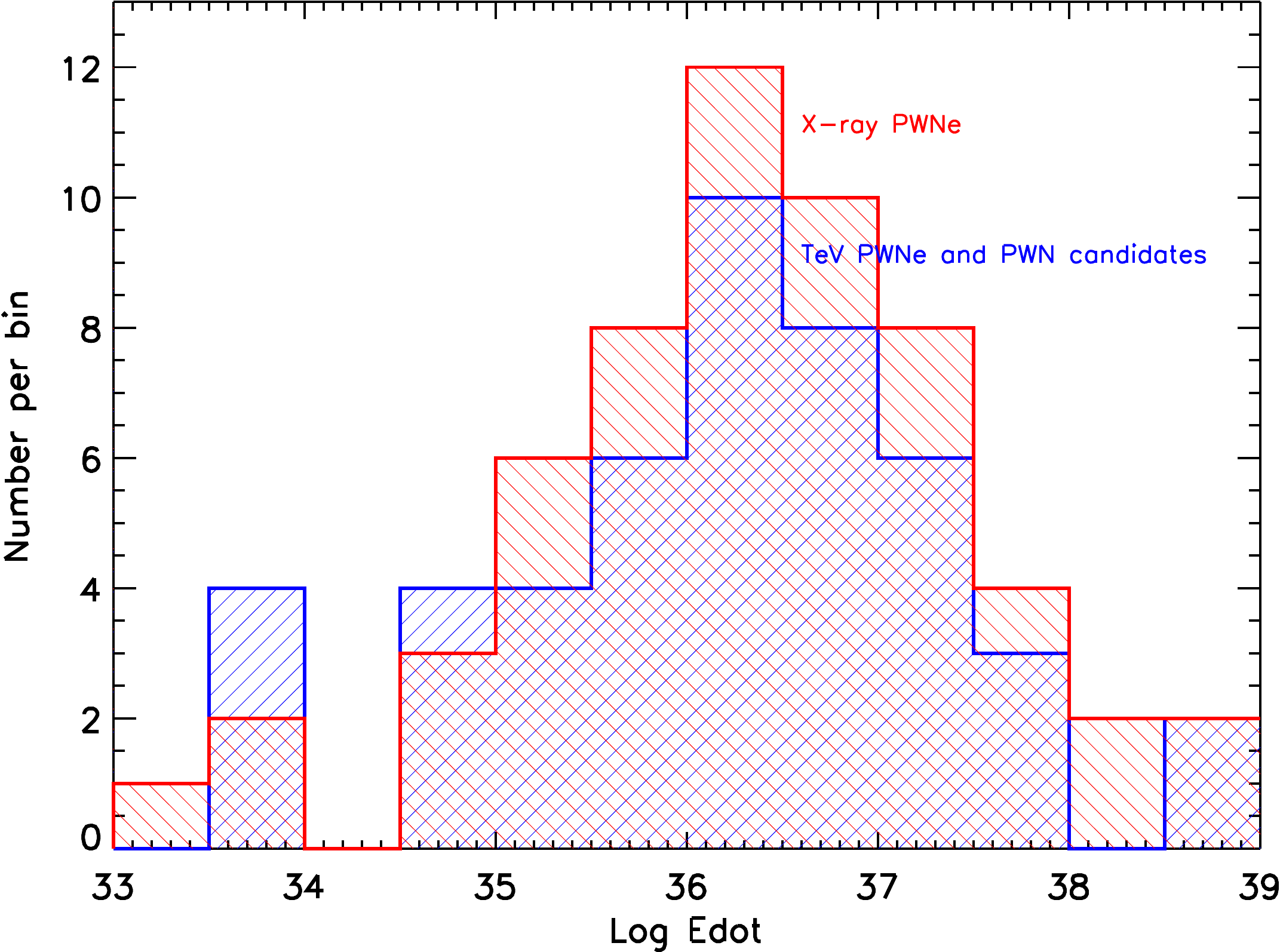}
\caption{{\em Left:} Pulsars with detected PWNe (or PWN candidates) in the $\tau_{\rm sd}$-$\edot$ diagram.
The semi-circles correspond to X-ray (orange) and TeV (black) PWNe, their sizes are proportional to logarithms of the corresponding PWN luminosities. The small black  dots denote the pulsars from the ATNF catalog \cite{2005AJ....129.1993M}. Pulsars with PWNe detected by {\sl Fermi} are marked by stars. {\em Right:}  Distribution of  X-ray (red) and TeV (blue) PWN and PWN candidates  over pulsar's $\dot{E}$. Two distributions closely follow each other.}
\end{figure}

\begin{figure}
\centering
\includegraphics[width=0.37\textwidth,angle=90]{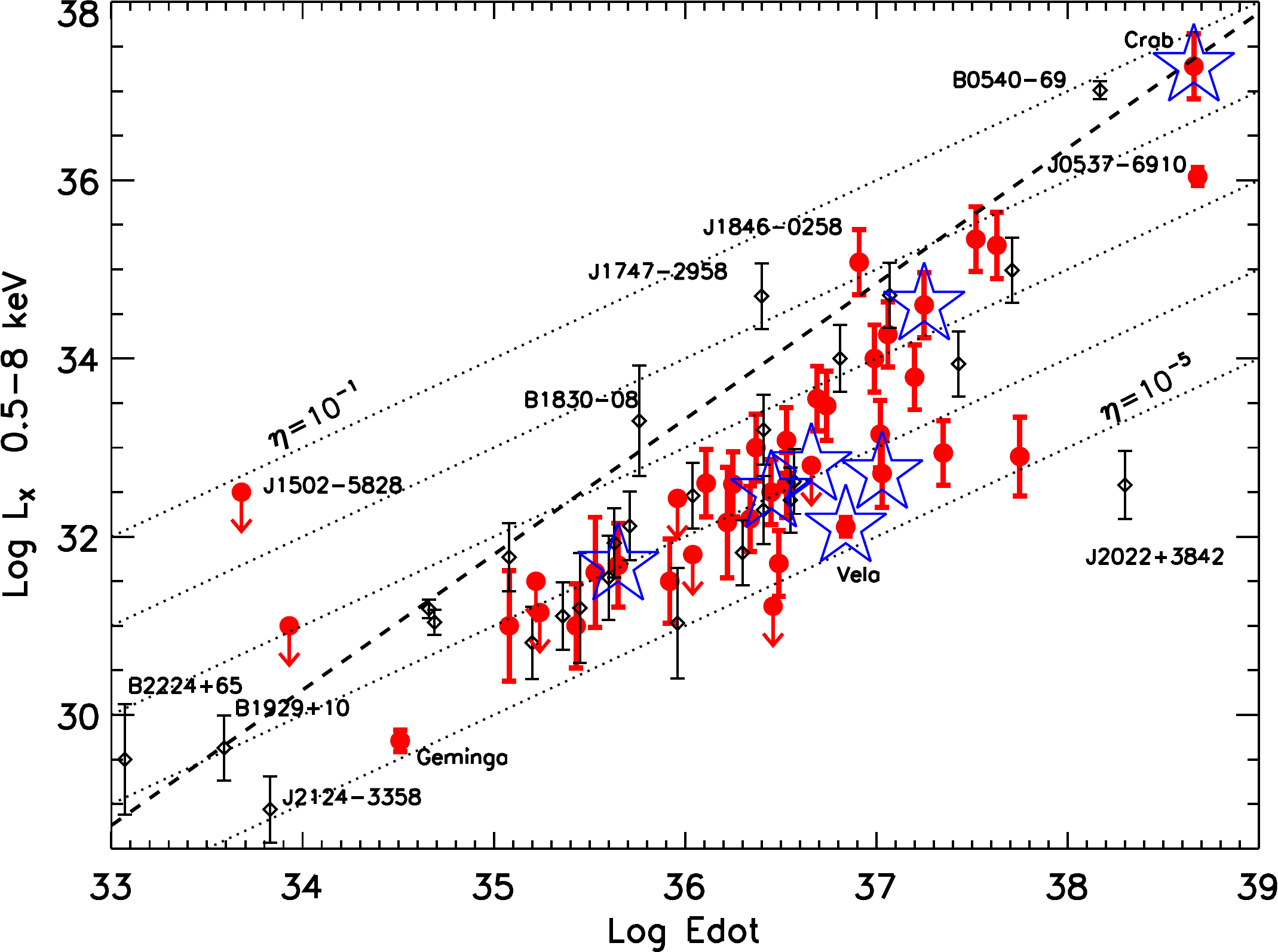}
\includegraphics[width=0.37\textwidth,angle=90]{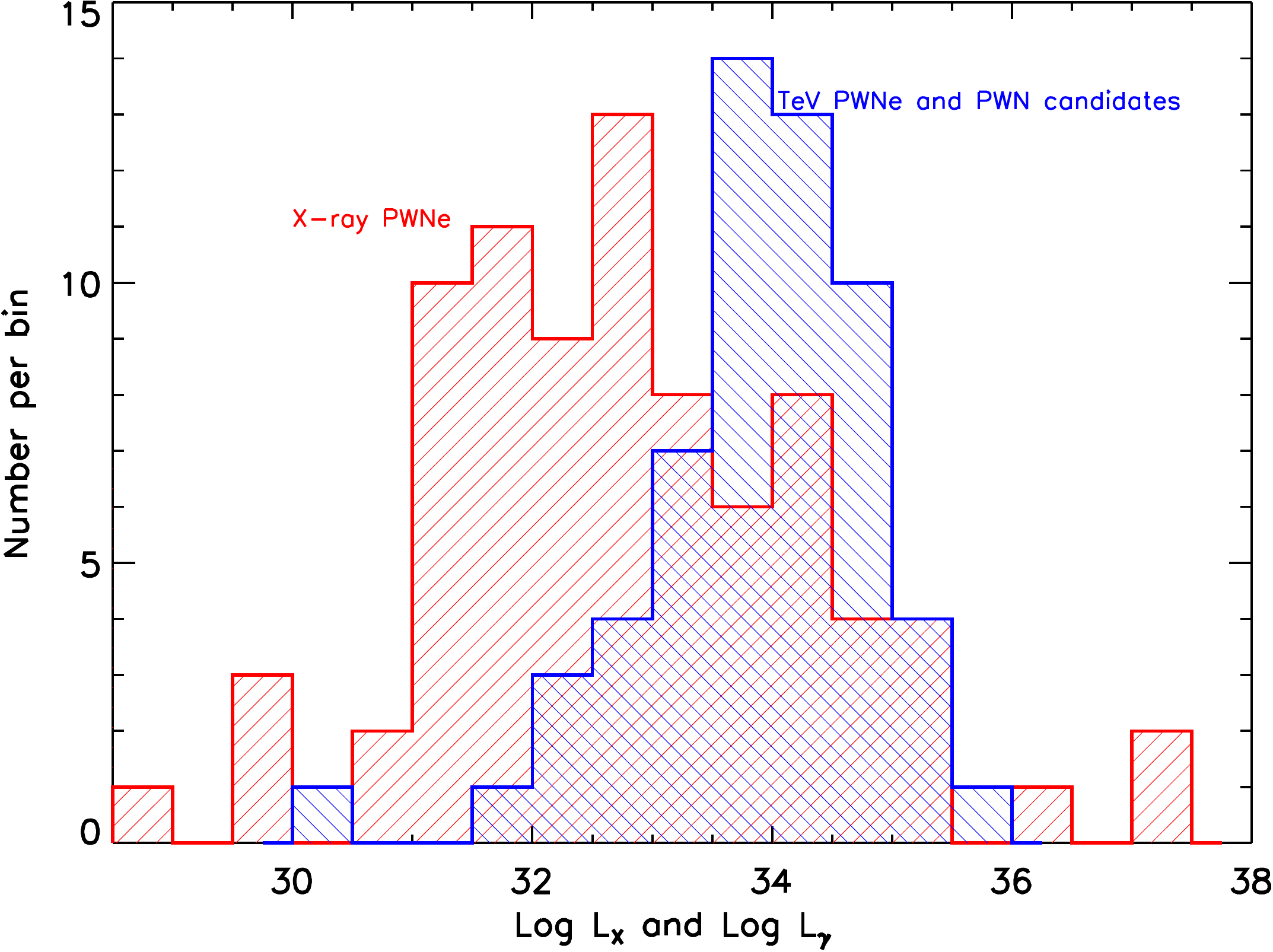}
 \caption{{\em Left:} X-ray luminosities of PWNe and PWN candidates vs.\  pulsar's $\dot{E}$. TeV PWNe and TeV PWN candidates  are shown in red. The dotted straight lines correspond to constant X-ray efficiencies; the upper bound, $\log L_X^{\rm cr} = 1.51\log\edot -21.4$ \cite{2012arXiv1202.3838K}, is shown by a dashed line. The PWNe detected in GeV by {\sl Fermi} are marked by stars. {\em Right}: Distributions of PWNe and PWN candidates over the X-ray and TeV luminosities.}
\end{figure}

As one can see from Figure~1 and Table~1,  several energetic pulsars with prominent X-ray PWNe  are not detected as TeV sources (e.g., G76.9+1.0, G310.6--1.6, 3C 58). On the other hand, the histogram in the right panel of Figure~1 shows that the X-ray PWNe and TeV PWNe (and PWN candidates) have very similar distributions over the pulsar's $\dot{E}$ (which is the ultimate source of energy powering PWN). This suggests that other factors (such as the local background photon density and confinement/compression by the reverse shock) are more important for TeV PWNe  than the current power output ($\dot{E}$) of the pulsar. X-ray luminosities of PWNe in Figure~2 show some correlation with $\dot{E}$ albeit with a very large spread for pulsars with $\dot{E}\gtrsim10^{36}$~erg~s$^{-1}$. The $L_{X}-\dot{E}$ distribution has a fairly well defined upper bound which likely describes the $L_{X}-\dot{E}$ correlation  when other parameters influencing $L_X$ are at their ``optimal'' values. On the contrary, the TeV luminosities do not show a significant correlation with $\dot{E}$, and the upper bound, $L_\gamma^{\rm cr} \sim 10^{35}$ erg s$^{-1}$, does not show a significant dependence on the pulsar's spin-down power (Figure~3).

Moreover, there is no obvious correlation between $L_{\gamma}$ and  $\dot{E}^{1/2}\tau_{sd}$ product  (Figure~3, right panel).  Such a correlation could be expected because the number of particles injected into a PWN  should be proportional to the Goldreich-Julian current \cite{1969ApJ...157..869G}, $\dot{N}\propto B_{\rm sd}/P^2\propto\dot{P}^{1/2}P^{-3/2}\propto\dot{E}^{1/2}$, where $B_{\rm sd}=3.2\times10^{19}(P\dot{P})^{1/2}$~G is the strength of the dipole magnetic field at the NS equator. Therefore, the number of particles accumulated in the PWN over a timescale comparable with the pulsar's age is $\propto \int_{0}^{\tau_{\rm sd}}\dot{E}(t)^{1/2}dt \propto \dot{E}^{1/2} \tau_{\rm sd}$. The poor correlation could be explained by the varying pair production multiplicity, deviations from the dipolar magnetic filed, different environmental conditions, and by the possible interaction with the SNR reverse shock.

Despite the lack of correlation between $L_{\gamma}$ and $\dot{E}$, the scatter in $L_{\gamma}$ is smaller compared to that in $L_X$ (cf.\ the left panels of Figures~2 and 3; see also the right panels of Figure~2  and the left panel of Figure~4). The left panel of Figure~4 also shows that for older PWNe most of the power is radiated at $\gamma$-ray energies. Some of the scatter in the $L_{X}-\dot{E}$ and $L_{\gamma}-\dot{E}$ diagrams must be due to distance errors that are very uncertain in some cases. This uncertainty can be eliminated by using a distance-independent $L_{\gamma}/L_X$ ratio which exhibits some correlation with age (Figure~4, right panel) and hence with $\dot{E}$ (as $\dot{E}$ and $\tau_{\rm sd}$ are correlated; see Figure~1, left panel). This correlation still shows a substantial scatter, suggesting that the origin of the scatter is not simply  due to the  distance errors.  On the other hand, the correlation is  similar in shape to the one suggested by \cite{2009arXiv0906.2644D,2009ApJ...694...12M} for simplistic one-zone PWN models, which, however, may not be directly relevant for  relic PWNe as they are significantly displaced due to the interaction with the SNR reverse shock.  
  
\begin{figure}
\centering
\includegraphics[width=0.37\textwidth,angle=90]{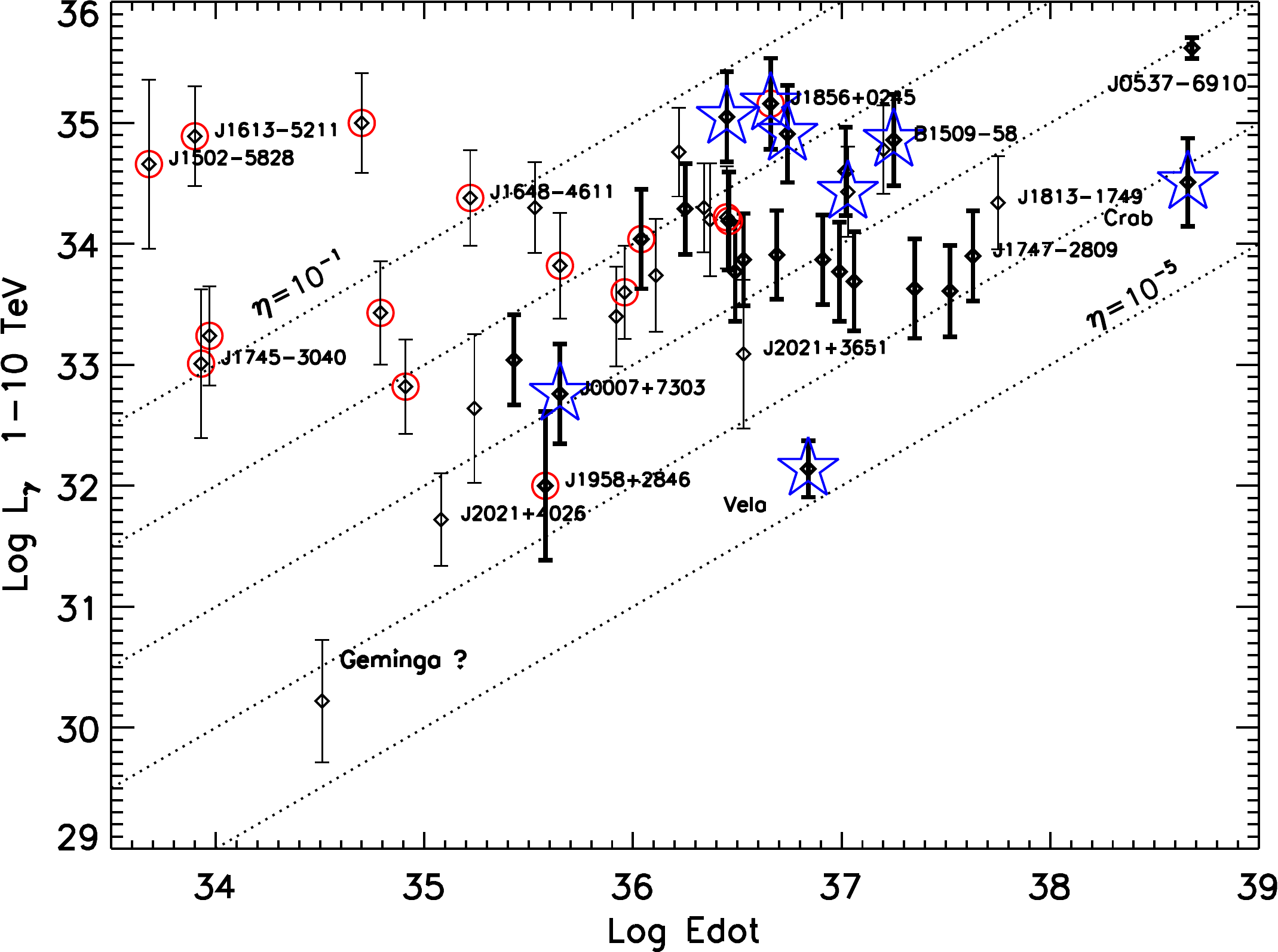}
\includegraphics[width=0.37\textwidth,angle=90,trim=4mm 0 0 5mm,clip]{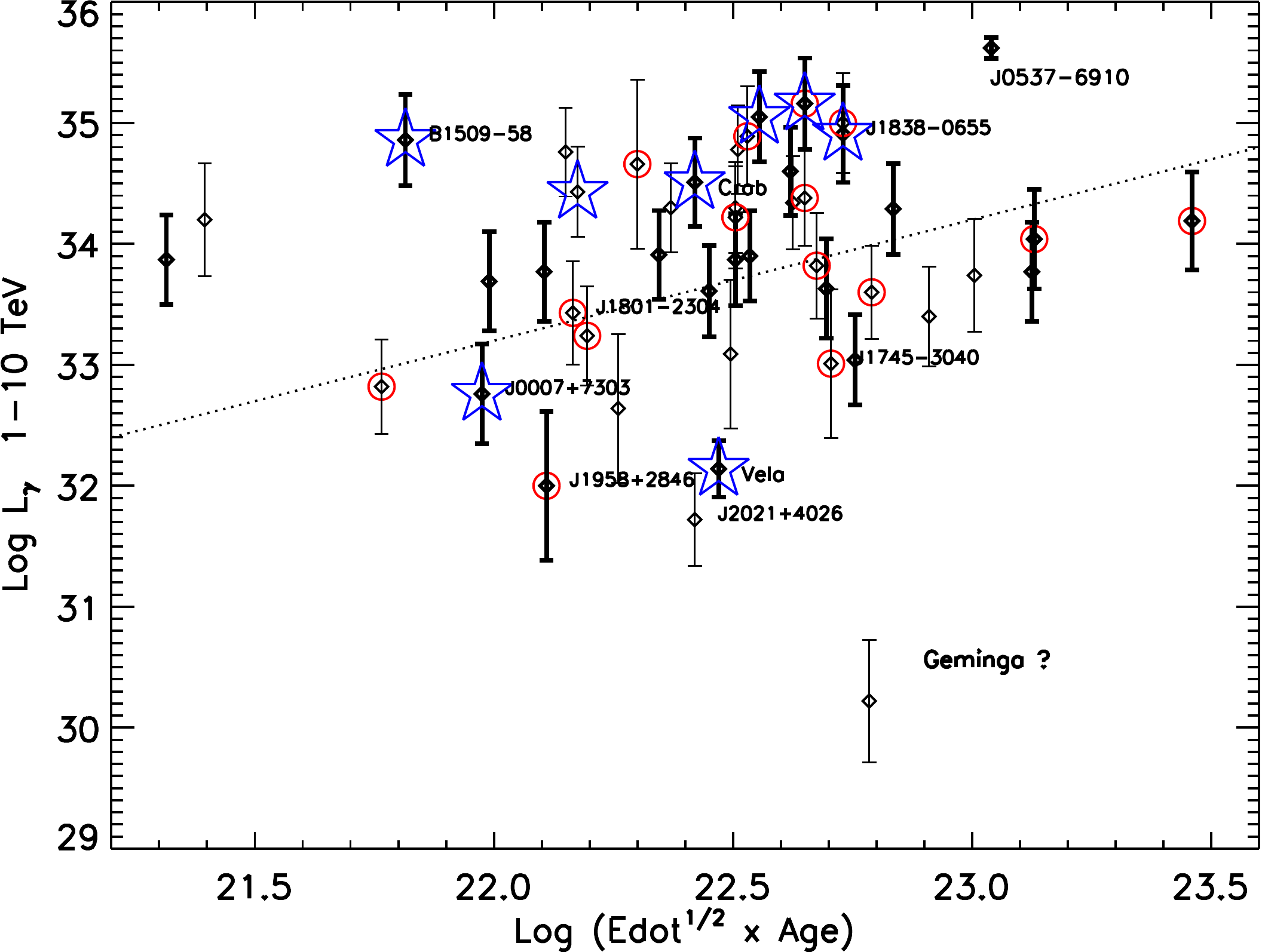}
\caption{TeV luminosities of PWNe and PWN candidates  vs.\  pulsar's $\dot{E}$  ({\em left}) and $\edot^{1/2}\tau_{\rm sd}$ ({\em right}).  Thin error bars mark questionable associations. The PWNe undetected in X-rays are marked by circles.  PWNe detected by {\sl Fermi} are marked by stars. The dotted lines in the left panel correspond to constant values of the TeV $\gamma$-ray efficiency $\eta_\gamma=L_\gamma/\edot$. The dotted line in the right panel corresponds to $\log L_{\gamma}=\log(\edot^{1/2}\tau_{\rm sd})+11.2$.}
\end{figure}
  
In most cases the TeV PWN spectra fit well a power-law (PL) model ($F_{\nu}\propto \nu^{-\Gamma+1}$, where $\Gamma$ is the photon index), but   a more complex spectral shape is suggested by the data in several cases (e.g., Vela X and HESS~J1718--385). Also,  spectral cut-offs at $\gtrsim10$ TeV are seen  in some cases. The X-ray spectra of  PWNe  are typically well characterized by an absorbed PL, which may exhibit softening with increasing distance from the pulsar due to the synchrotron cooling. There are, however, only few bright PWNe where the quality of the data allows one to detect such changes. In most cases $\Gamma_{\gamma}$ and $\Gamma_X$ are measured from regions of different sizes. It is interesting to note that there appears to be some correlation between $L_{\gamma}$ and $\Gamma_{\gamma}$, such that $\Gamma_{\gamma}$ increases with $L_{\gamma}$ up to $L_{\gamma}\sim$~a few $\times 10^{35}$ erg s$^{-1}$, but then the trend seems to be reversed (see Figure~5, left panel). More precise measurements of the luminosities and spectral slopes are needed to confirm these trends.

X-ray and TeV observations of PWNe provide a useful diagnostic of pulsar wind properties since the PWN emission is intrinsically a multiwavelength phenomenon. However, even the mere detection of a PWN in some energy band indicates the emission mechanism and the electron energies involved. For instance,  detecting a PWN in the X-ray band, where  the synchrotron emission dominates ICS, implies that the wind particles have been accelerated up to $\sim 100$ TeV (note that particles with such energies cannot leave the pulsar magnetosphere because of strong radiative losses) and that the same particles should produce ICS emission in the TeV energy range. For young PWNe unaffected by synchrotron burn-off ($\tau_{\rm sd}\lesssim\tau_{X}<\tau_{\gamma}$) one can crudely estimate the PWN magnetic field  from the equation $L_{\gamma}/L_{X}\sim u_{\rm rad}/u_{B}\approx 10(u_{\rm rad}/u_{\rm CMB} )B_{-6}^{-2}$, provided that $L_{\gamma}$ and  $L_{X}$ are measured from the same regions\cite{1997MNRAS.291..162A}, which is only possible for youngest compact PWNe such as the Crab (cf.\ Figure~4, right panel). (Here, $u_{\rm rad}$ is the energy density in the radiation field, $u_{\rm CMB}=0.26$ eV cm$^{-3}$, and $u_B=B^2/8\pi$ is the magnetic field energy density.)

More detailed modeling may allow one to constrain the ambient pressure, pulsar-wind electron density, boundary energies and the shape of the injected  electron SED, true PWN age, ion fraction in the wind, and pair production multiplicity \cite{2008AIPC.1085..199D,2009ApJ...703.2051G, 2011MNRAS.410..381B, 2013MNRAS.429.2945T}.  However, the current PWN emission models are often limited to the one-zone approximation, spherical symmetry, and either the advection or diffusion transport (but not both), although some attempts to go beyond these approximations have been recently made (e.g., \cite{2011ApJ...742...62V,2012MNRAS.427..415M,2013ApJ...765...30V}). In relic PWNe, the particle escape via diffusion can be more important than advection.  The Bohm diffusion with the diffusion coefficient $D = 8.5\times10^{26}E_{\gamma}^{1/2}B_{-6}^{-1}  {\rm\ cm}^{2} {\rm\ s}^{-1}$ implies that on the timescale of $\tau_{\gamma}$ the particles will diffuse out to distances  $\sim(6 D \tau_{\gamma})^{1/2}\sim40 B_{-6}^{-1/2} (1+0.144B_{-6}^2)^{-1/2}$~pc (which translates into the angular size of   $\sim70' d_{4}^{-1}$ for a typical distance $d=4$ kpc and  $B=1$~$\mu$G), comparable to the observed sizes of some extended VHE sources (Figure~6). Once the relic PWN models become more realistic, one might be able to determine whether any of the observed PWNe require an addition of a hadronic mechanism ($\pi^{0}$ decay in proton-proton interactions; see, e.g., \cite{2007ApnSS.309..189H}) in order to account for the observed multiwavelength emission and spectrum. In principle, the pulsar wind may contain a fraction of relativistic protons (or ions; \cite{1994ApJS...90..797A}), which would produce a detectable amount of VHE emission.  In this respect, particular attention should be paid to the sources in which powerful PWNe are located near SNR shells interacting with (or expanding into) dense molecular clouds \cite{2008MNRAS.385.1105B}. 
 
\begin{figure}
\centering
\includegraphics[width=0.37\textwidth,angle=90]{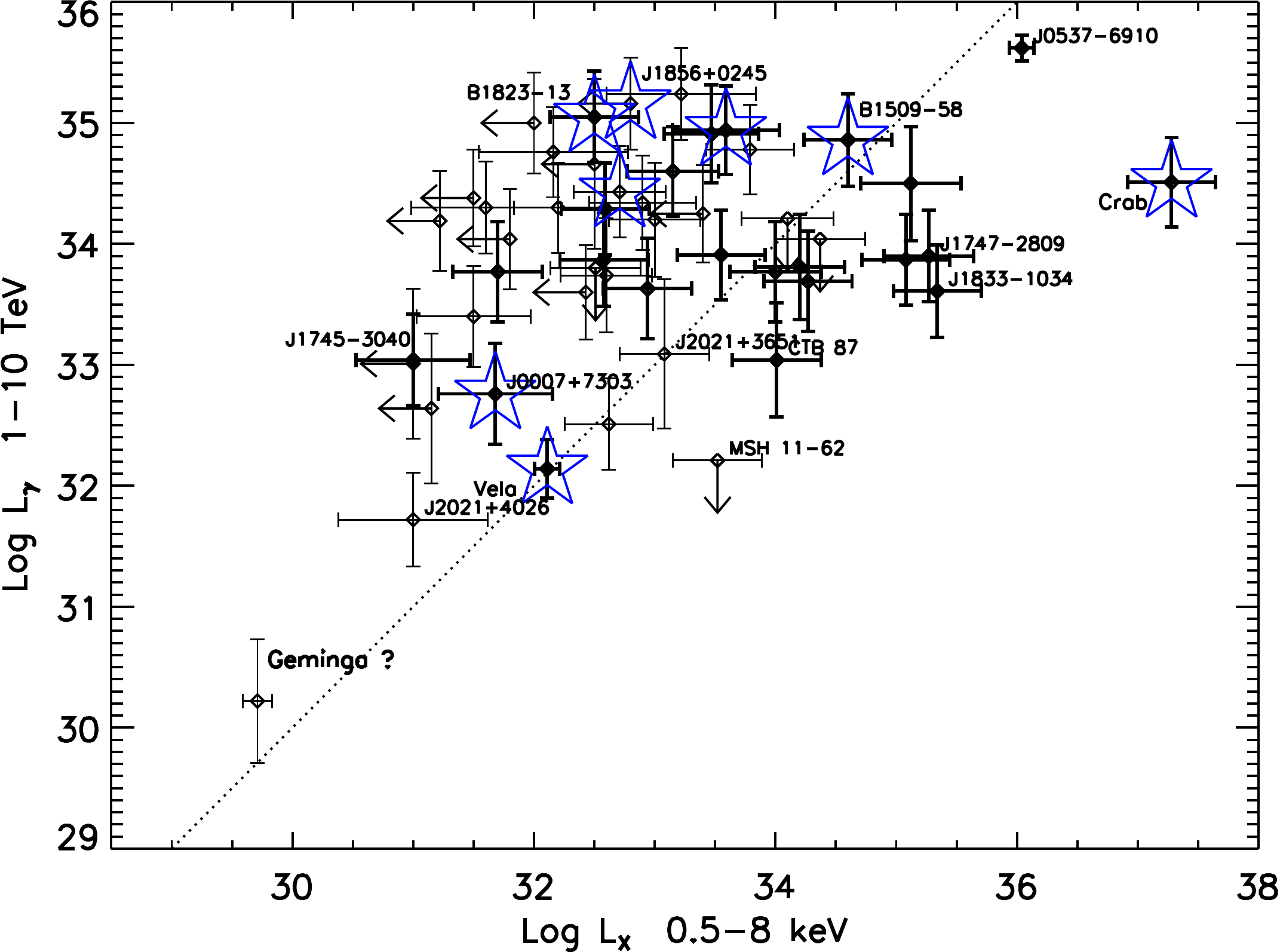}
\includegraphics[width=0.37\textwidth,angle=90]{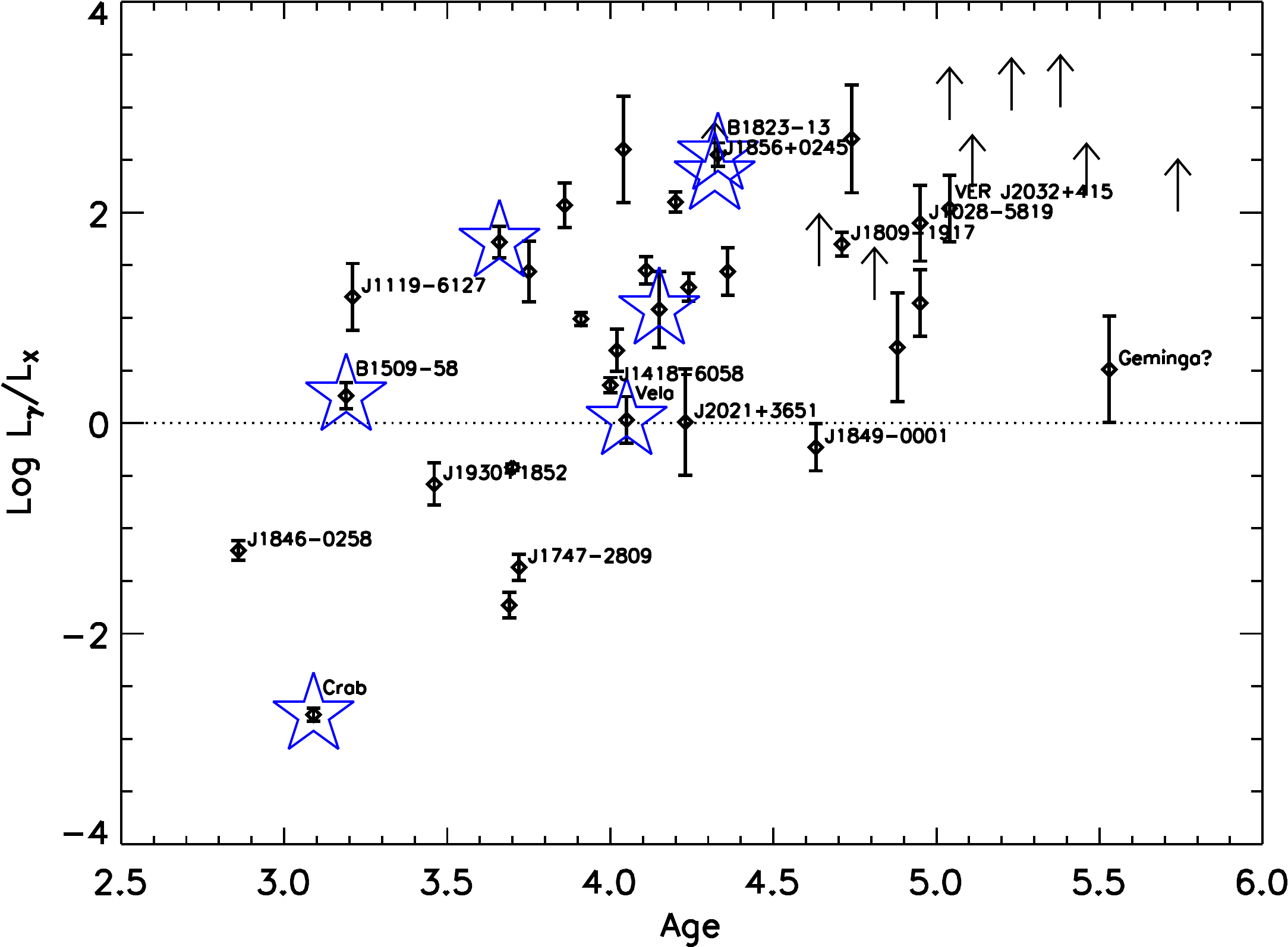}
\caption{TeV luminosity vs.\  X-ray luminosity ({\em left}) and TeV-to-X-ray luminosity ratio vs. pulsar's age  ({\em right}) for PWNe and PWN candidates. Limits are shown by arrows. The PWNe detected by {\sl Fermi} are marked by stars. Uncertain detections are shown by thin lines. The dotted lines corresponds to $L_\gamma = L_X$.}
\end{figure}

 

Another type of PWNe are those around fast moving pulsars, some of which may have escaped from their host SNRs and are now moving supersonically through the rarefied ISM. These PWNe tend to have cometary tail-like shapes in X-ray and radio images \cite{2008AIPC..983..171K}. Interestingly, the TeV emission from PWNe of this type  (marked by $^{\ast}$ next to the pulsar name in Table~1) has not been firmly detected yet although about a dozen of such pulsars/PWNe are known (e.g., PSRs J1747--2958, J1509--5850, B0355+54, J0633+1746, B1957+20). If a substantial fraction of  TeV emission in PWNe  is due to a hadronic component of the pulsar wind, the nondetections  of pulsar tails might be explained by the lower ambient density. In the case of purely leptonic pulsar wind, the bow-shock TeV PWNe, which might be created by freshly shocked electrons in the pulsar vicinity, are perhaps too faint because of lower spin-down powers of these relatively old pulsars. A possible explanation of nondetections of the long, up to $\sim20$~pc, tails filled by older wind particles  is a lower sensitivity of the existing TeV imaging techniques to the extended linear structures. It is, however, possible that the wind particles channeled into the tails behind the pulsars accumulate in lobes, which are not seen in X-rays (due to the relatively short synchrotron cooling time) but could be sources of ICS emission in TeV/GeV and synchrotron emission in the radio (or IR/mm bands), strongly offset from the pulsar. As the surface brightness of such lobes may  be relatively low, deep TeV (and radio) observations are required for their detection.

With the launch of {\sl Fermi Gamma-ray Observatory} it has become apparent that the dominant population of GeV sources in the Galactic plane are young and energetic pulsars. However, pulsar emission usually dominates PWN emission at GeV energies, and therefore it is often challenging to isolate the PWN component. So far, there are about 8 cases (see  the last column of Table~1) where it has been more or less convincingly demonstrated that, in addition to the pulsed GeV emission, there is a significant unpulsed (and sometimes spatially extended) component due to a PWN.

\begin{figure}
\centering
\includegraphics[width=0.37\textwidth,angle=90]{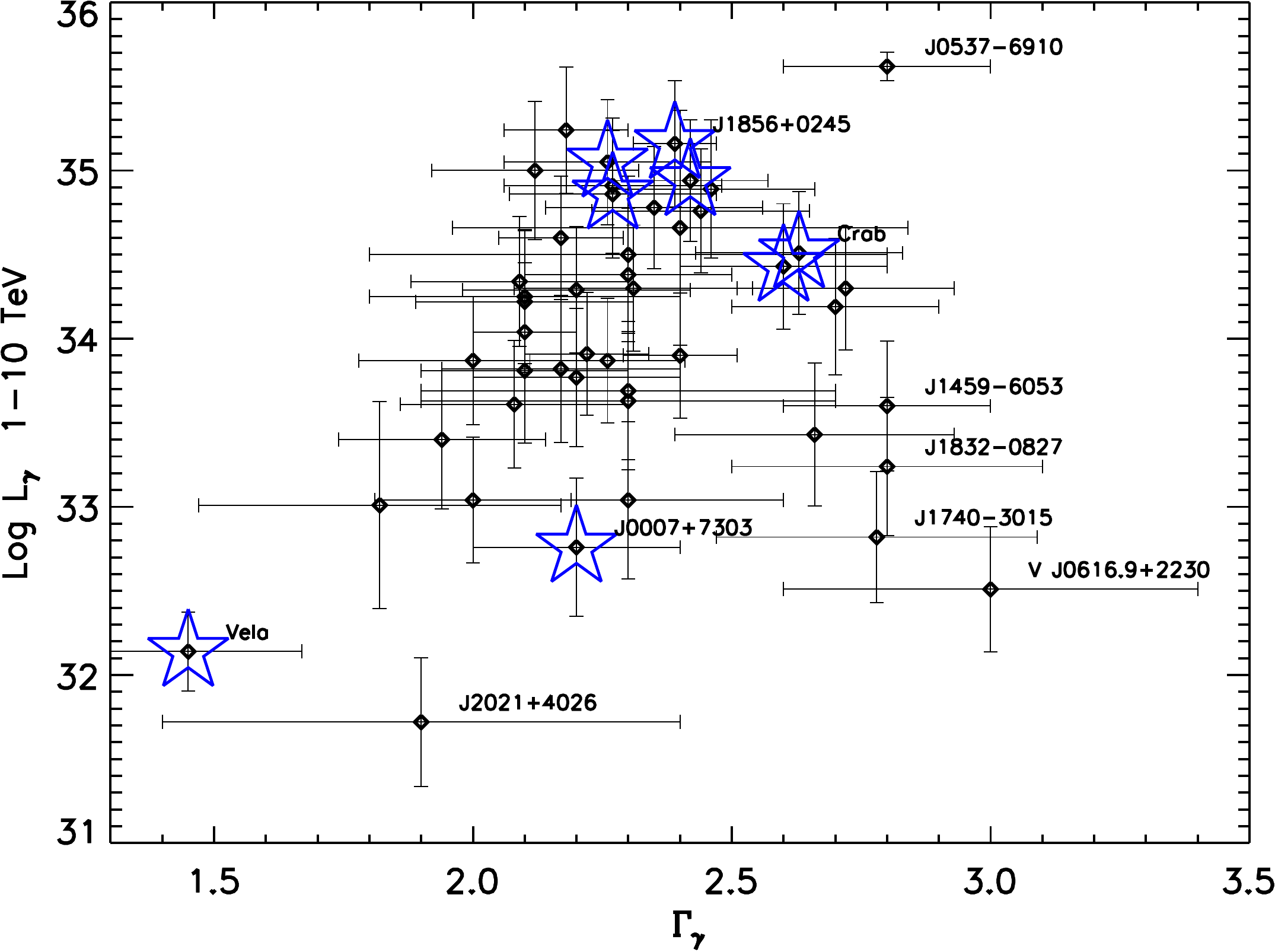}
\includegraphics[width=0.37\textwidth,angle=90]{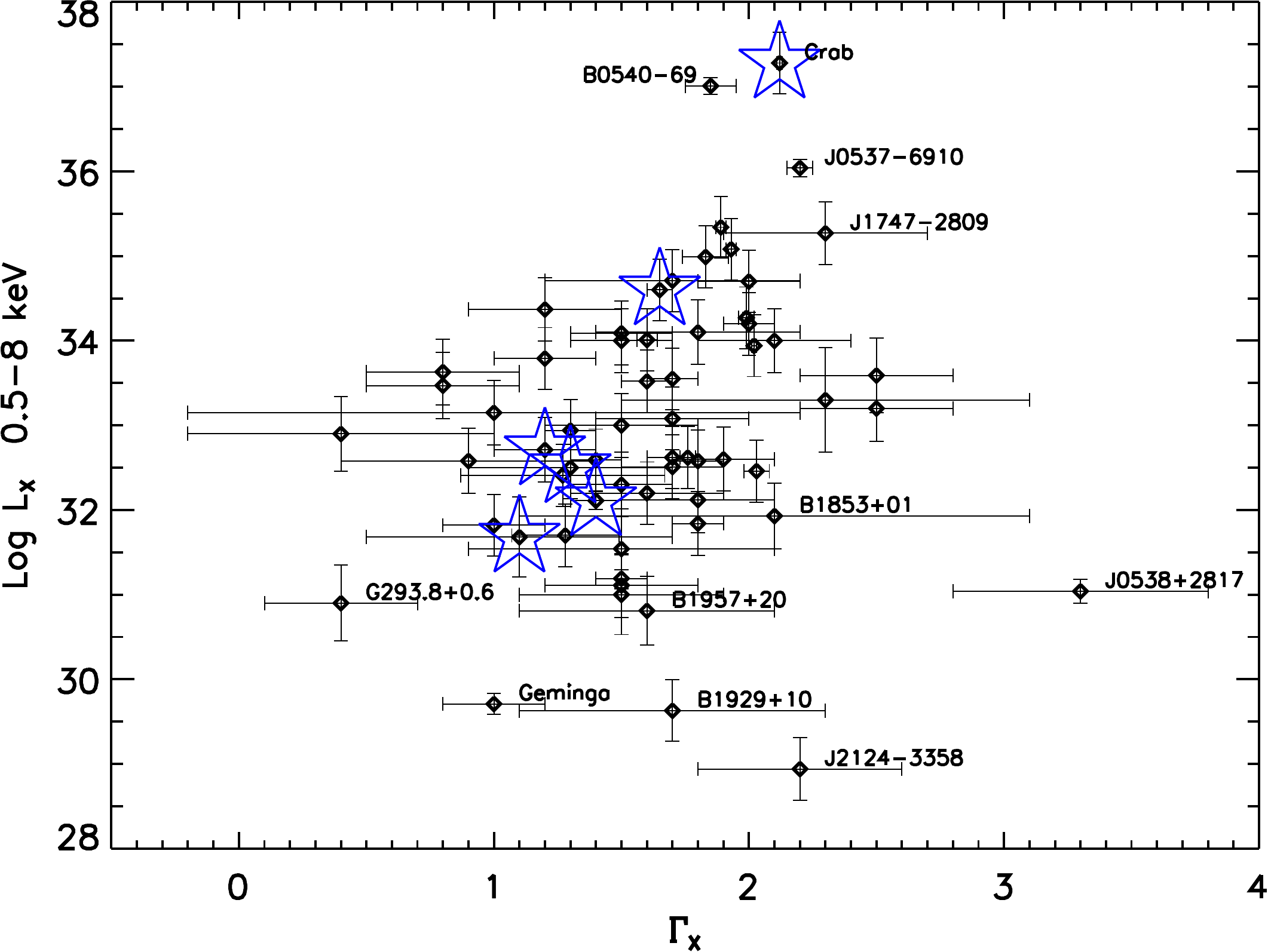}
 \caption{PWN luminosity vs.\ photon index in the TeV (\emph{left}) and X-ray (\emph{right}) bands. PWNe detected by {\sl Fermi} are marked by stars.}
\end{figure}

\begin{figure}
\centering
\includegraphics[width=0.5\textwidth]{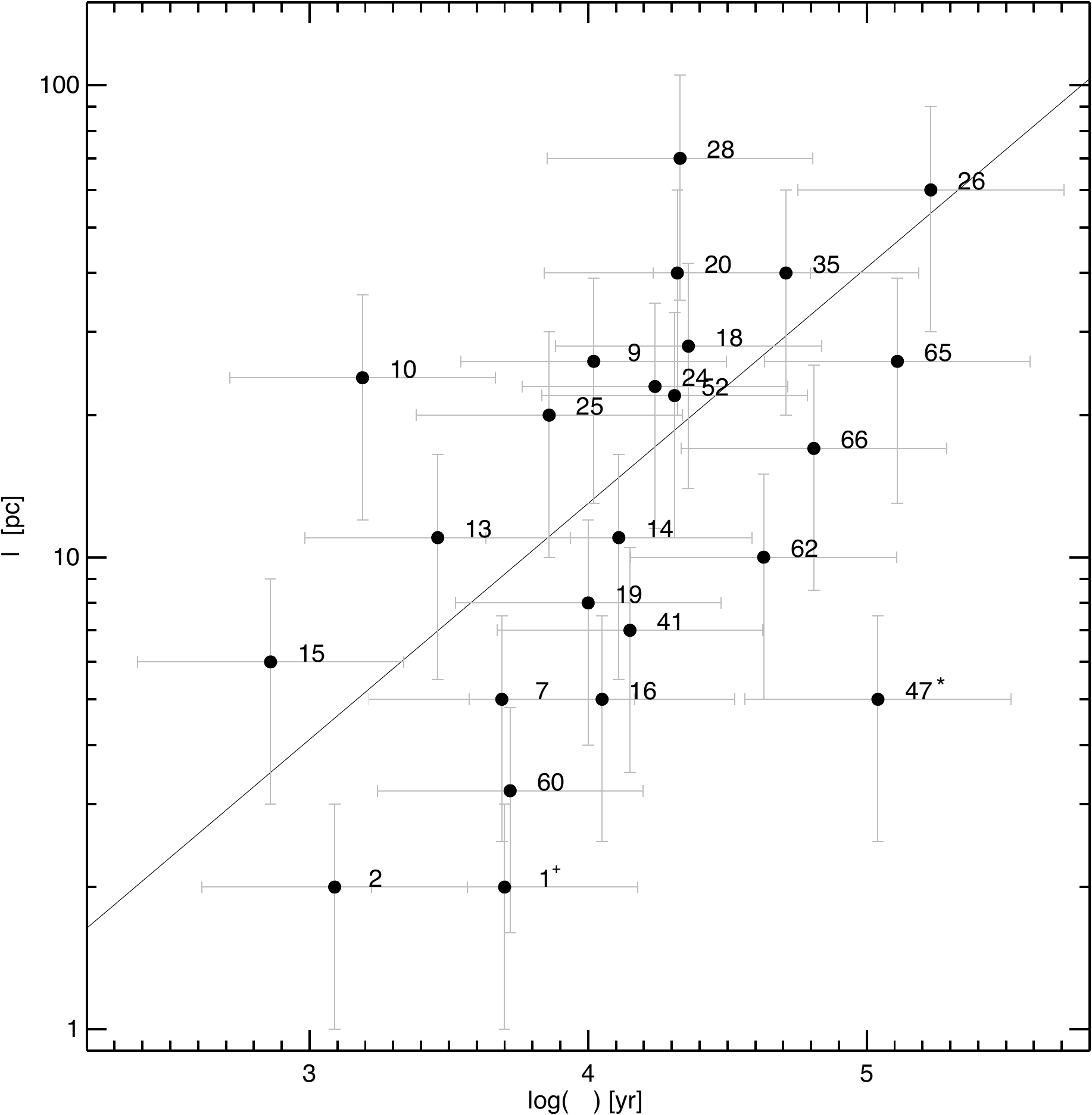}
\caption{Size of the TeV emission region (for firmly established TeV PWNe) as a function of pulsar's spin-down age. The straight line represent the characteristic size,  $\langle r^2 \rangle^{1/2}=(6Dt)^{1/2}$,  corresponding  to diffusion with the Bohm diffusion coefficient $D$  (see text). Source~1 (marked with  $^{+}$) is located in the Large Magellanic Cloud. Source~47 (marked with $^{\star}$) may have its distance underestimated by a factor of 2.}
\end{figure}

\begin{figure}
\centering
\renewcommand{\tabcolsep}{-0.2cm}
\begin{tabular}{cc}
\includegraphics[width=0.4\textwidth,angle=90,trim=0 0 0 15mm,clip]{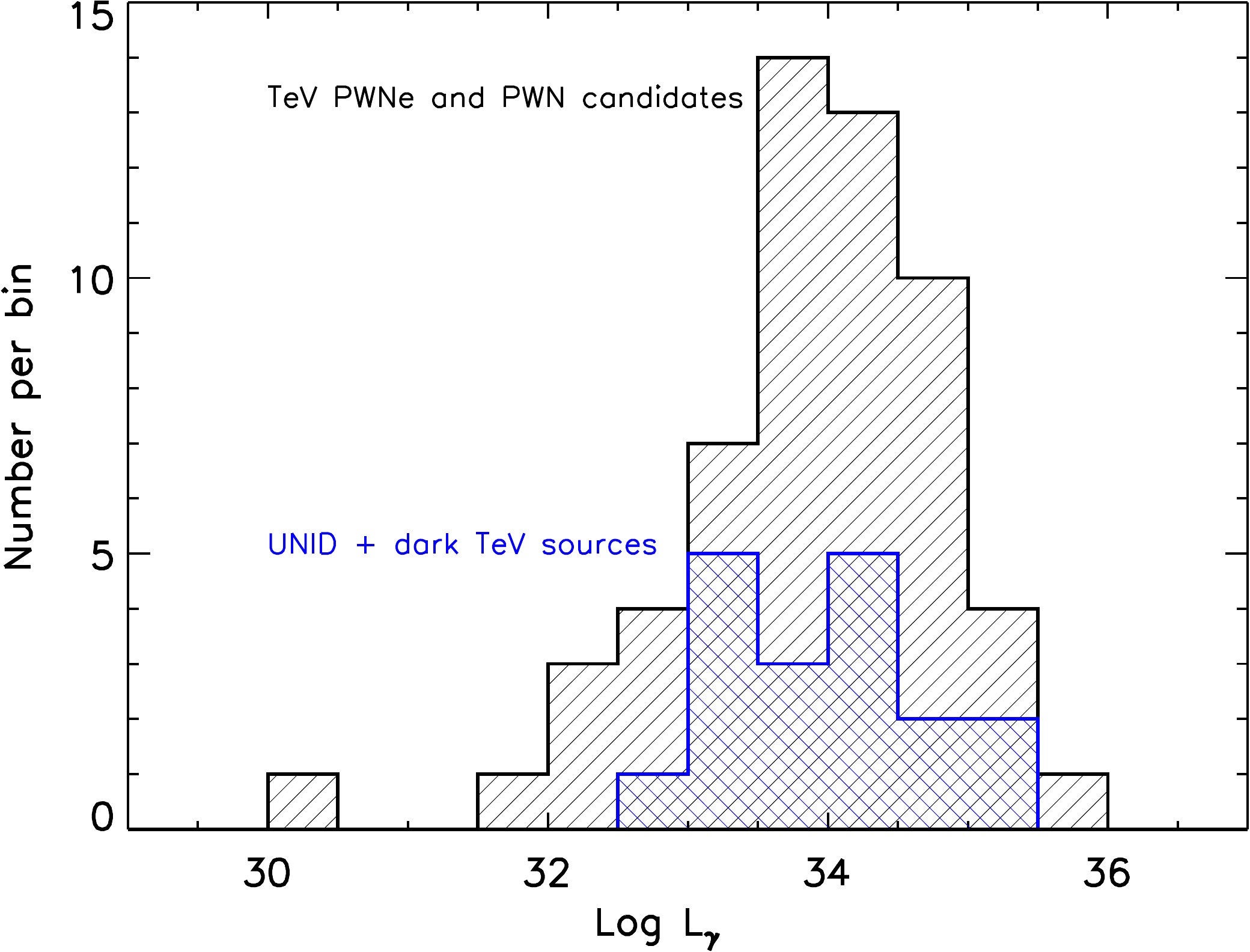}
\includegraphics[width=0.4\textwidth,angle=90,trim=0 0 0 15mm,clip]{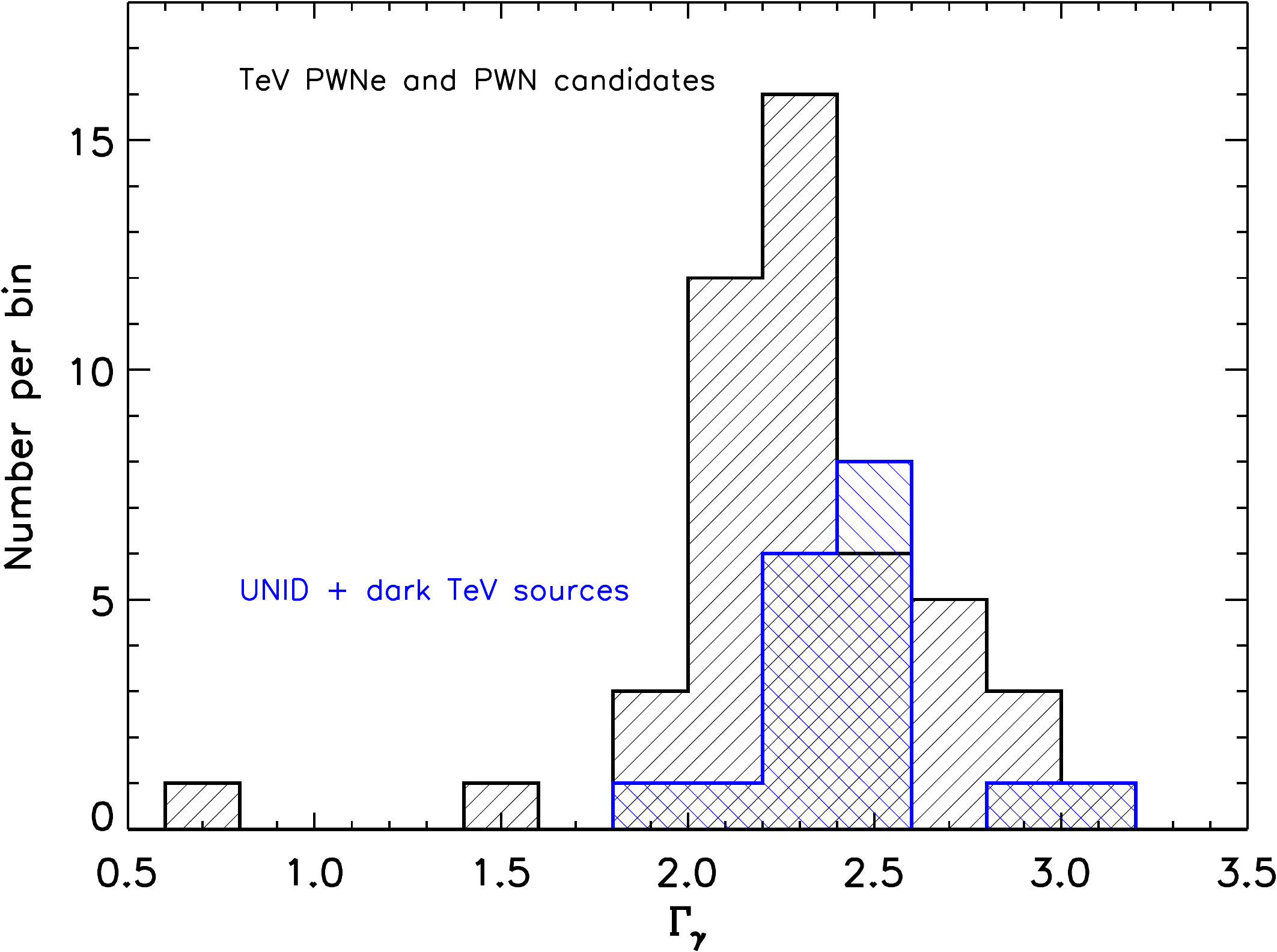}
\end{tabular}
\caption{TeV luminosities (in 1--10 TeV) and TeV power-law indices.}
\end{figure}

\begin{figure}
\centering
\renewcommand{\tabcolsep}{-0.2cm}
\begin{tabular}{cc}
\includegraphics[width=0.5\textwidth,angle=90]{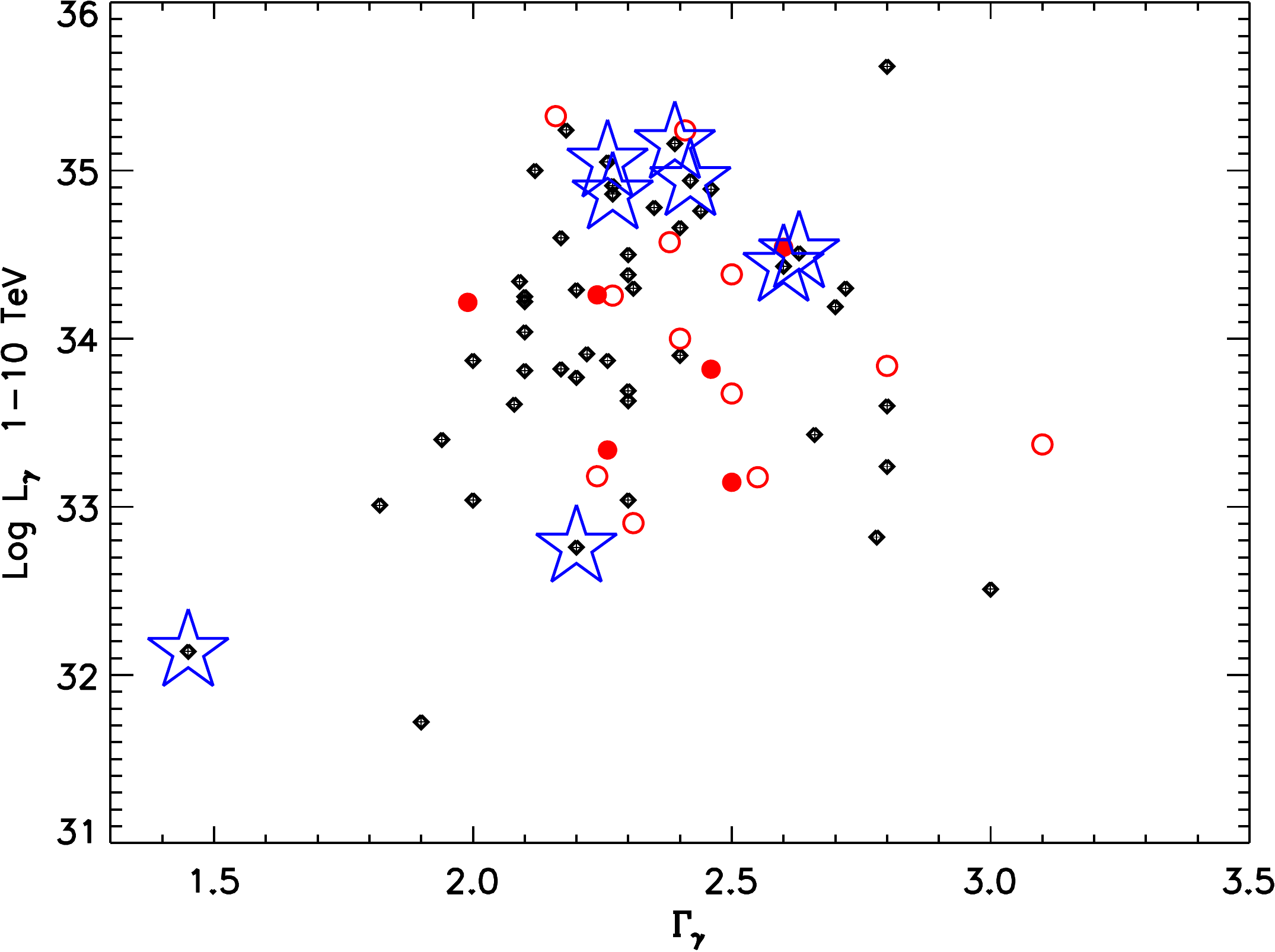}
\end{tabular}
\caption{Similar to Fig.~5 (left panel). The error bars are omitted for clarity. The unidentified and ``dark'' sources from Table~4 are shown with open filled red circles, respectively.}
\end{figure}

\section{Unidentified and dark VHE sources}

The properties of 20 unidentified Galactic VHE sources are summarized in Table~4. For some of these objects a tentative nature has been suggested based on either a positional coincidence (which may be a chance coincidence in the projection onto the sky) with some known object (such as a molecular cloud near an SNR or a star-forming region, or an offset pulsar); however, there is no sufficient evidence to establish a solid identification. To a certain extent this also applies to the 25 TeV PWN candidates marked by ``?'' in Table~1, although we consider most of these associations to be more secure because the chance coincidence is less likely. The bottom part of Table~4 contains sources which do not have any plausible association in X-rays (the images show multiple faint point sources) or radio. We note, however, that most of the TeV sources in Table~4 appear to have a 2FGL {\sl Fermi} LAT source within $20'$ radius, according to  the Second Fermi LAT source catalog \cite{2012ApJS..199...31N}. However, most of these 2FGL sources are unidentified as well. We also note that the $L_{\gamma}$ and $\Gamma_\gamma$ distributions for unidentified sources (shown in Figure~7) follow closely those for the TeV PWNe and PWN candidates from Table~1. This can also be seen from the plots shown in Figure~8. Therefore, it is plausible that most of the unidentified VHE sources in Table~4 are, in fact, relic PWNe produced by yet undetected pulsars.

\section{Summary}
 
We compiled and reviewed the emission properties of  Galactic VHE sources detected at TeV energies. The current population of identified VHE sources is  dominated by PWNe, and the locations of the TeV sources are well-correlated with the Galactic arms. It is also likely that a significant fraction (if not all) unidentified VHE sources are also relic PWNe whose synchrotron emission is too faint to be detected in  X-rays or it is shifted out of the X-ray band. Multiwavelength observations of unidentified VHE sources and dedicated pulsar searches within the extent of VHE sources can provide reliable identifications  and reveal the energetics and composition of pulsar winds.  In addition to  X-ray and TeV observations, the ICS PWN component can be detected with {\sl Fermi } LAT in the GeV band, where even old objects should exhibit uncooled ICS spectra, matching the radio synchrotron component. The data accumulation must be complemented by development of multi-zone models of PWN evolution \cite{2012ApJ...752...83T}, to understand the nature of pulsar winds and their role in seeding the Galaxy with energetic particles and magnetic fields. It seems that PWNe can be the dominant source of leptonic cosmic rays in the Galaxy, and a significant part of the diffuse Galactic background in GeV $\gamma$-rays could come from dissolved relic PWNe whose ages are even larger than those of TeV-identified PWNe.
  
 
\section*{Acknowledgements}
This material is  based upon work supported by NASA  grants NNX09AC84G and NNX09AC81G.

\begin{table}
 \setlength{\tabcolsep}{0.05in}
 \begin{tabular}{rlcccccc}
\hline

 \#& PSR & PWN$^*$ &VHE src.$^\dag$ &  $\log\dot{E}$ &  $\log\tau$ & $d^{**}$ & ${\rm Rad./H}_{\alpha}/{\rm GeV}^\ddag$\\
   &  & & & ${\rm [erg/s]}$& ${\rm [yr]}$ & ${\rm kpc}$ &\\

\hline
1   &   J0537--6910             &   N157B            & H~J0537-691              &   38.68          &   3.70    &     50~~~~~      &  Y/N/N     \\
2   &   B0531+21                   & Crab            &  H~J0534+220     &   38.66          &   3.09   &       2~~~        &  Y/N/Y      \\
3    &    J2022+3842              & G76.9+1.0       & ...               &   38.30          &   3.95    &     8~~~          &  Y/N/N    \\
4   &   B0540--69                 &    N158A           & ...            &   38.17         &   3.22   &       50~~~~~       &  Y/N/N     \\
5   &   J1813--1749             & G12.82--0.02     & H~J1813--178$^?$   &  37.75        &   3.75     &       4.5       &  N/N/N \\
6     &   J1400--6325             &   G310.6--1.6     & ...              & 37.71           &  4.10   &       6~~~~          &    Y/N/N  \\
7   &   J1833--1034           &   G21.50--0.89      & H~J1833--105       &  37.52      &   3.69    &    4.8              &  Y/N/N      \\
8   &   J0205+6449            &   3C\,58              & ...             &   37.43     &     3.73    &    3.2              &  Y/N/N      \\
9   &   J2229+6114          &  G106.65+2.96   & V~J2228+609       & 37.35  &   4.02   &      3~~~                &  Y/N/N     \\
10   &   B1509--58              &   Jellyfish             &  H~J1514--591       &   37.25     &   3.19  &    5~~~               &  P/N/Y      \\
11   &   J1617--5055           &  G332.50--0.28  & H~J1616-508$^?$   & 37.20      &   3.91  &     6.5             &  N/N/N      \\
12   &   J1124--5916           &   G292.04+1.75   & ...                          &  37.07   &     3.46   &      6~~~               &  Y/N/N     \\ 
13  &   J1930+1852           &  G54.10+0.27     & V~J1930+188      &   37.06   &   3.46   &    6.2~~             &  Y/N/N      \\
14  &   J1420--6048            &  G313.54+0.23  &  H~J1420--607    &   37.02   &   4.11   &      5.6              &  P/N/N     \\
 15  &   J1846--0258             &  Kes\,75           &   H~J1846--029     &  36.91   &   2.86    &     6$^{??}$~~                  &  Y/N/N      \\
16  &   B0833--45                & Vela                 &   H~J0835--455    &  36.84     &   4.05   &   ~~~~$0.29^{\rm p}$    &  Y/N/Y      \\
17  &   J1811--1925             &   G11.18--0.35  & ...                             &   36.81    &   4.37   &    5~~~            &  Y/N/N      \\
18     &   J1838--0655              &     G25.24--0.19 &  H~J1837--069     &  36.74    &   4.36   &      7~~~                 &  N/N/Y \\
19     &   J1418--6058             &  Rabbit               &  H~J1418--609    &    36.69   &   4.00    &           3$^?$~~              & Y/N/N   \\
20     &   J1856+0245             &   G36.01+0.06$^?$          &  H~J1857+026  & 36.66      & 4.32     &      9~~~               & N/N/Y \\
21  &   B1951+32$^{\ast}$                &  G68.77+2.82       &  ...                       &  36.57    &   5.03   &    2.5              &  Y/Y/N      \\
22  &   J1826--1256$^{\ast}$           & Eel                                       &  ...                   &  36.55         &  4.15   &      7$^?$~~                &  P/N/N   \\
23  &   J2021+3651           &   G75.23+0.12   &   M~J2019+37$^?$     &36.53   &   4.23   &    4~~~            &  N/N/N    \\
24  &   B1706--44               &   G343.10--2.69     & H~J1708--443 &   36.53   &   4.24   &     2~~~            &  Y/N/N     \\
25  &   J1357--6429           &  G309.92--2.51    & H~J1357--645  &   36.49   &   3.86    &    2.5              &  P/N/N      \\
26     &  J1913+1011            &   G44.48--0.17$^?$       & H~J1912+101  &  36.46   &   5.23     &    4.5              & N/N/N   \\
27     &   J1907+0602           &  G40.16--0.89$^?$        &  H~J1908+063$^?$ & 36.45   &  4.28     &       3.2             & N/N/N  \\
28  &   B1823--13              &  G18.00--0.69      &  H~J1825--137  &  36.45   &   4.33   &      4~~~            &  N/N/Y     \\
29  &   B1757--24$^{\ast}$             &            Duck                & ...                     &   36.41   &   4.19   &          5~~~  &  Y/N/N     \\
30  &   J1016--5857          &   G284.08--1.88    &    H~J1018-589B$^?$                      &   36.41   &   4.32  &        3~~~       &  N/N/N      \\
31  &   J1747--2958$^{\ast}$          &             Mouse          &                      &   36.40        &   4.41   &        5~~~   &  Y/N/N     \\
32  &   J1119--6127           &    G292.15--0.54   & H~J1119--615$^?$   & 36.37   &   3.21   &   8.4             &  N/N/N      \\
33  &   B1800--21              &  G8.40+0.15            &  H~J1804--216$^?$ & 36.34     &   4.20   &    4~~~            &  N/N/N      \\
34  &   B1046--58$^{\ast}$            &  G287.42+0.58      &   ...                         & 36.30           &   4.31      &   3~~~            &  N/N/N     \\
35  &   J1809--1917          &  G11.09+0.08       &   H~J1809--193        &  36.25  &   4.71    &    3.5              &  P/N/N       \\
36  &   J1301--6305           &  G304.10--0.24     & H~J1303--631$^?$  &  36.22   &   4.04   &    7~~~                &  N/N/N       \\
37  &   J1718--3825           &  G348.95--0.43    & H~J1718--385$^?$  &   36.11   &   4.95    &    4~~~             &  N/N/N     \\
38     & J1531-5610               &   G323.89+0.03    &   ...                    &   35.96  &   4.99    &    3~~~            &  N/N/N    \\
39  &   J1509--5850$^{\ast}$           &  G319.97--0.62     & ...                       &   35.71  &   5.19    &    4~~~           &  P/N/N       \\
40     &  J1857+0143            &       G35.17--0.57$^?$      & H~J1858+020$^?$  & 35.65    &   4.85    &   5.5        & N/N/N  \\
41     &   J0007+7303           &   CTA1                    &  V~J0006+729         &  35.65   &    4.15   &        1.4               &  N/N/Y   \\
42  &   B1853+01              &   G34.56-0.50      &   ...                        & 35.63      &   4.31   &   3~~~        &  Y/N/N       \\
43     &  J1809-2332             &       Taz      & ...                          &    35.60    &    4.36  &  2~~~~      &  Y/N/N      \\
44     &   J1958+2846           & G65.89--0.37$^?$      &   M~J1954+28       &  35.58    &   4.32  &   2$^?$~~       & N/N/N  \\
45  &   J1702--4128          &  G344.74+0.12      &  H~J1702--420$^?$ &  35.53    &   4.74   &    5~~~        &  N/N/N      \\
46  &   J0729--1448          &  G230.39--1.42     &  ...                                      &  35.45   &     4.54   &     4~~~      &  N/N/N       \\
47     &   J2032+4127           &     G80.22+1.02    &         V~J2032+415          &  35.43   &     5.04   &     1.7          &  N/N/N     \\
48  &   J1740+1000            &  G34.01+20.27    & ...                                       &   35.36   &   5.06    &    1.4              &  P/N/N       \\
49     &  J0631+1036            &  G201.22+0.45$^?$       &   M~J0630+10$^?$  &   35.24  &   4.64    &    3.6             & N/N/N \\
50  &   B1957+20$^{\ast}$               &  G59.20--4.70      &  ...                                       &  35.20    &   9.18    &    2.5              &  N/Y/N      \\
51     &   J0633+0632           &  G205.10--0.93         &    ...                               &  35.08   &  4.77     &    ~~1.5$^?$              &  N/N/N \\
52     &  J1740-3015              &  G358.29+0.24$^?$                &  H~J1741--302               &  34.91   &   4.31    &    3~~~                 &   N/N/N  \\

\hline
\end{tabular}
\caption{Pulsars with X-ray and/or TeV PWNe}
\label{tab:prop1}
\end{table}
\normalsize

\addtocounter{table}{-1}
\begin{table}
 \setlength{\tabcolsep}{0.05in}
 \begin{tabular}{rlcccccc}
\hline
 \#& PSR & PWN$^*$ &VHE src.$^\dag$ &  $\log\dot{E}$ &  $\log\tau$ & $d^{**}$ & ${\rm Rad./H}_{\alpha}/{\rm GeV}^\ddag$\\
   &  & & & ${\rm [erg/s]}$& ${\rm [yr]}$ & ${\rm kpc}$ &\\
\hline

53  &   J0538+2817            &  G179.72--1.69     &  ...                                   &  34.69  &   5.79   &   ~~~~$1.47^{\rm p}$    &  N/N/N      \\
54  &   B0355+54$^{\ast}$                 &       Mushroom     & ...                                      &   34.66      &   5.75   &   ~~~~$1.04^{\rm p}$    &  N/N/N       \\
55  &   J0633+1746$^{\ast}$             &      Geminga       & M~J0632+17$^?$      &   34.51     &   5.53   &     ~~~~$0.25^{\rm p}$   &  N/N/N       \\
56     &   J1745--3040              &   G358.55--0.96$^?$                  &   H~1745--305$^?$        &  33.93      &   5.74   &       2~~~     & N/N/N \\
57     & J1502--5828               & G319.39+0.13$^?$         &  H~J1503--582$^?$          &   33.68    &   5.46      &     8~~~                       &     N/N/N  \\          
58  &   B1929+10                &  G47.38--3.88  &      ...                                      &   33.59   &   6.49      &  ~~~~$0.36^{\rm p}$   &  P/N/N     \\
59  &   B2224+65$^{\ast}$              &          Guitar            & ...                                      &   33.07   &   6.05      &     1.5       &  N/Y/N      \\ 
\hline 
60     &   J1747--2809             & G0.9+0.1       &  H~1747--281                    &   37.63          &   3.72      &     8.5                         &     Y/N/N  \\          
61     &   J1023--5746             & G284.19-0.39$^?$         &  H~1023--575$^?$          &   37.03        &   3.66      &     8                       &     N/N/P  \\          
62    &  J1849--0001               &   G32.64+0.53          &      H~J1849--000      &  36.99               &           4.63        &         7      &   N/N/N \\
63    & J1437-5959$^{\ast}$                 &  G315.78+0.2       &    ...                   &       36.15  &      5.06     &        10           &   Y/N/N  \\
64    &  J0855-4644                & G266.97-1.00       &    H~J0852--463$^?$                     &     36.04   &     5.15     &        0.75          &       N/N/N  \\  
65   &   J1831--0952             &	G21.88-0.10     &       H~J1831--098     &    36.04    &       5.11    &      4.3~~~~    &   N/N/N \\
66     & J1459--6053               &   G317.89--1.79    &  H~J1458--608       &   35.96  &   4.81    &    4~~~            &  N/N/N    \\
67     &   J1028--5819             & G285.06-0.5         &  H~J1026--582$^?$          &   35.92          &   4.95      &     2.3                      &     N/N/N  \\          
68    &  B1830--08                  & G23.5+0.1                    &           ...                             &  35.76              &           5.17       &       5      &       P/N/N \\
69    &   J1853-0004              & ...                           &    H~J1852-000$^?$          &  35.32               &         5.46                &      7           & N/N/N   \\   
70    & J1648--4611                &  ...                          &       H~J1646--458B$^?$     &  35.22               &         5.04                &      5           &  N/N/N \\  
71     &   J2021+4026             & G78.23+2.09$^?$         &  V~J2019+407$^?$          &   35.08    &   4.88      &     1.5                       &     N/N/N  \\          
72     &    J1801-2304             &   ...                         &    H~1801--233$^?$           &    34.79      &  4.77       &      4                       &      N/N/N  \\
73    &   J1632-4757               &  ...                           &    H~J1632--478$^{??}$        &    34.70       &   5.38     &       7                       &   N/N/? \\   
74    & J1832--0827                 &  ...                           &   H~J1832--084$^?$        &  33.97               &         5.21             &        5        &  N/N/N   \\  
75    &  J1613-5211                 &  ...                            & H~J1614--518$^{?m}$       &    33.90              &    5.58                  &          6       &    N/N/N   \\  
76     &  J2124--3358              &  G10.92-45.43  &           ...                             &          33.83      &              9.58    &      0.25   &          N/Y/N  \\   
\hline
\end{tabular}
\caption{Pulsars with X-ray and/or TeV PWNe (continued)}
\footnotesize{$^*$ -- PWN name or Galactic coordinates. The superscript $^?$ mark the cases in which no X-ray PWN have been reported but TeV PWN candidates were found nearby. $^\dag$ -- TeV sources in the vicinity of the PSR/PWN. `H', `V' and `M' stand for HESS, VERITAS, and Milagro. The superscripts $^?$ and  $^??$ mark questionable associations, the superscript $^m$ is used if the TeV source may be a combination of multiple sources. $^{**}$ -- Our best guess for the pulsar distance, used to scale the distance-dependent parameters in Table~2. The superscript $^{\rm p}$ marks the distances determined from parallax measurements; the most uncertain distances (e.g., cases when even pulsar's dispersion measure is unknown) are marked $^?$. $^\ddag$ -- Is the PWN detected in radio/H$_{\alpha}$, and PSR/PWN in GeV $\gamma$-rays? Y=`yes', N=`No',  P = `possibly'.  $^{\ast}$ -- Pulsars with tails, which likely move supersonically through a relatively cold, low-density medium. The left portion of this table shows 59 sources presented in \cite{2010AIPC.1248...25K}, with some of the parameter values updated. }
\label{tab:prop1}
\end{table}

\begin{table}
 \setlength{\tabcolsep}{0.045in}
 \begin{tabular}{rccccclllc}
\hline

 \# &  $N_{\rm H,22}$$^\ast$ & $\log L_{\rm X}$$^\dagger$ & $\Gamma_{\rm X}$ & $\log L_\gamma$$^{\ast\ast}$ &  $\Gamma_{\gamma}$$^\ddagger$ &  $l_{X}$$^\S$ & $l_{\gamma}$$^\P$ & $\Delta$$^\parallel$ & ${\rm Refs.}$$^{\dagger\dagger}$ \\
     &   & ${\rm [erg/s]}$  &   & ${\rm [erg/s]}$  &    &    ${\rm pc}$  &  ${\rm pc}$  &      ${\rm arcmin}$ &   \\ 

\hline
1       &     0.5   &    $36.04  \pm   0.01$    &     $2.20\pm0.05$       &   $35.62\pm0.03$                            &     $2.8\pm0.2$           &  1.4  &    $<2$       &  $<1$     &  \cite{2006ApJ...651..237C,2012AnA...545L...2H}    \\
2       &     0.32  &    $37.28  \pm   0.01$    &     $2.12\pm0.01$       &   $34.51  \pm  0.06$   &      $2.63\pm0.20$          &  1.2  &   $  <2  $ &   $  <0.5   $    &  \cite{2004ApJ...609..186M, 2001AnA...365L.212W, 2006AnA...457..899A}    \\
3        &     1.3    &    $32.58  \pm  0.12$    &      $0.9\pm0.5$          &   ...                            &    ...                                &  0.4  &    ...     &     ...       &   \cite{2009_Arzoumanian_Boston}  \\
4       &     0.46  &    $37.01  \pm   0.01$    &     $1.85\pm0.10$       &   ...                          &      ...                             &  1.4  &   ...         &   ...    &  \cite{2001ApJ...546.1159K}    \\
5       &     10    &    $32.90\pm0.25$        &     $0.4^{+0.4}_{-0.7}$ &  $34.34 \pm 0.14$   &     $2.09\pm0.21$            &  2.0  &    $  6  $   &   $< 0.5$         & \cite{2007ApJ...665.1297H, 2006ApJ...636..777A,2012ApJ...753L..14H}\\  
6        &     2.1    &    $ 34.99\pm0.04$       &     $1.83\pm0.09$       &   ...                           &   ...                              &  2.3  &  ...             &  ...                  &  \cite{2010ApJ...716..663R} \\  
7       &     2.3   &    $35.36  \pm   0.01$    &     $1.89\pm0.02$       &   $33.63 \pm 0.12$    &      $2.08\pm0.22$          &  1.0  &   $<5$     &   $ < 1 $    &  \cite{2001ApJ...561..308S, 	2007arXiv0710.2247H}   \\
8       &     0.43  &    $33.94  \pm   0.01$    &     $2.02\pm0.01$       &   ...                          &    ...                              &  1.2 &   ...          &    ...            &  \cite{2004ApJ...616..403S}    \\
9       &     0.5   &    $32.94  \pm   0.01$    &     $1.3\pm0.1$         &   $33.63 \pm 0.20$     &      $2.3\pm0.4$             &  0.4  &   $ 26   $ &   $  24   $    &  \cite{2001ApJ...552L.125H, 2009_Aliu_Boston}    \\
10       &     0.8   &    $34.60  \pm   0.03$    &     $1.65\pm0.05$       &   $34.86  \pm  0.12$   &      $2.27\pm0.2$            &  4.5  &   $  24  $ &   $   2.4 $    &  \cite{2002ApJ...569..878G, 2005AnA...435L..17A}  \\
11       &     3.5   &    $33.79  \pm   0.02$    &     $1.2\pm0.2$         &   $34.78  \pm  0.06$   &      $2.35\pm0.21$          &  0.6  &   $60 $ &   $   10  $    & \cite{2009ApJ...690..891K, 2006ApJ...636..777A}     \\
12       &     0.37  &    $34.71  \pm   0.03$    &     $1.7\pm0.5$         &   ...                           &     ...                             &  0.2  & ...     &   ...    & \cite{2001ApJ...559L.153H} \\
13      &     1.9   &    $34.46  \pm   0.01$    &     $1.99\pm0.03$       &   $33.88  \pm  0.20$   &      $2.3\pm0.4$          &  1.5  &   $  <11 $ &   $   <3 $    & \cite{2002ApJ...568L..49L, 2009_Aliu_Boston}  \\
14      &     5.4   &    $33.15  \pm   0.11$    &     $1.0\pm 1.2$         &   $34.60  \pm  0.07$   &      $2.17\pm0.12$        &  0.4  &   $11$   &   $3.3$    &  \cite{2005ApJ...627..904N, 2006AnA...456..245A} \\
15      &     4.0   &    $35.08  \pm   0.03$    &     $1.93\pm0.02$       &   $33.87  \pm  0.09$   &      $2.26\pm0.15$          &  2.8  &   $<6 $ &   $<1$    & \cite{2003ApJ...582..783H}  \\
16      &     0.02  &    $32.11  \pm   0.03$    &     $1.4\pm0.1$         &   $32.14  \pm  0.22$   &      $1.45\pm0.22^?$  &  0.1  &   $ 5 $ &   $  30   $    &    \cite{2001ApJ...554L.189P, 2001ApJ...556..380H, 2006AnA...448L..43A} \\
17      &     3.1   &    $34.00  \pm   0.09$    &     $1.5\pm0.2$         &  ...                           &      ...                             &  1.0  &   ...        &  ...    & \cite{2003ApJ...588..992R}  \\
18         &      4.0   &    $33.47\pm0.14$        &      $0.8\pm0.3$          &  $34.91 \pm 0.18$       &      $2.27\pm0.21$           &   2.0  &  28       &    3    &\cite{2008ApJ...681..515G, 2006ApJ...636..777A} \\
19       &      2.3   &    $33.55\pm0.02$        &    $1.7\pm0.1$            &  $33.91\pm0.07$       &     $2.22\pm0.12$            &  1.5 &   8         &    3.5  &   \cite{2005ApJ...627..904N, 2006AnA...456..245A}  \\
20       &  $\lesssim4~~~$  &    $<32.8$                   &      ...                         &  $35.16\pm0.11$       &     $2.39\pm0.08$           &  1.7  &  40       &    6     & \cite{ 2008ApJ...682L..41H, 2008AnA...477..353A,2012AnA...544A...3R}\\  
21      &     0.34  &    $32.62  \pm   0.01$    &     $1.76\pm0.03$       &   ...                          &      ...                              &  0.4 &   ... &   ...    &  \cite{2004ApJ...610L..33M, 2005ApJ...628..931L}  \\
22        &       2.0  &   $32.41 \pm 0.05$       &     $1.27\pm0.40$      &  ...                            &      ...                               & 0.3   & ...   & ...   &  \cite{2007AAS...21114403R} \\
23      &     0.7   &    $33.08  \pm   0.07$    &     $1.7\pm0.3$         &   $33.09\pm0.50$       &     [2.6]   &  0.8  &   $ <140 $ &   $  18  $    & \cite{2004ApJ...612..389H, 2009ApJ...700L.127A} \\
24      &     0.5   &    $32.58  \pm   0.02$    &     $1.8\pm0.1$         &   $33.87  \pm  0.13$   &      $2.00\pm0.22$            &  0.2  &   $ 23 $ &   $   14  $    & \cite{2005ApJ...631..480R, 2009arXiv0906.5574H}  \\
25      &    0.23   &    $31.70   \pm  0.07$    &     $1.28\pm0.21$     &   $33.77   \pm 0.20$   &      $2.2\pm0.2$            &  0.1 &   $20$ &   $8   $    & \cite{2007ApJ...665L.143Z}   \\
26       & $\lesssim2$   &     $<31.22$           &    ...                         &  $34.19  \pm  0.19$    &      $2.7\pm0.2$            & ...      &   $60$ &   $12$     & \cite{2008AnA...484..435A} \\
27       & $\lesssim2$   &     ...                      &     ...                         &  $34.22 \pm 0.22$      &      $2.10\pm0.21$        & ...      &   $37$ &   $16$     & \cite{2009AnA...499..723A} \\        
28      &     1.0   &    $32.50  \pm   0.05$    &     $1.3\pm0.4$         &   $35.05  \pm  0.10$   &      $2.26\pm0.2^?$  &  0.2  &   $70$ &   $10$    &  \cite{2003ApJ...588..441G, 2008ApJ...675..683P, 2006AnA...460..365A}  \\
29      &     4.4   &    $33.20  \pm   0.14$    &     $2.5\pm0.3$         &   ...                            &     ...                             &  0.5  &   ...    &   ...   &  \cite{2001ApJ...562L.163K}   \\
30      &  $\!\!\!$[1.2] &    $32.30  \pm   0.11$    &     $1.5\pm0.2$  &   ...                         &         $2.9\pm0.4$                          &  0.1  & ...      &   ...     &  \cite{2004ApJ...616.1118C,2012AnA...541A...5H}   \\
31      &     3.0   &    $34.70  \pm   0.05$    &     $2.0\pm0.2$         &  ...                           &     ...                              &  0.5  &  ...     &   ...     &  \cite{2004ApJ...616..383G}   \\
32      &     1.6   &    $33.00  \pm   0.10$    &     $1.5\pm0.3$         &   $34.20  \pm  0.30$   &      $>2.2$                     &  0.5  &  $30$ &   $6$    &  \cite{2003ApJ...591L.143G, 2009_Djannati_Boston}   \\
33      &     1.4   &    $32.20  \pm   0.05$    &     $1.6\pm0.3$         &   $34.30  \pm  0.08$   &      $2.72\pm0.21$          &  0.2  &   $  58 $ &   $   10  $    &  \cite{2007ApJ...660.1413K, 2007ApJ...670..643K, 2006ApJ...636..777A}  \\
34      &  $\!\!\!$[0.4] &    $31.82  \pm   0.04$    &     $1.0\pm0.2$  &   ...                           &     ...                               &  0.2  &   ...    &   ...    &  \cite{2006ApJ...652..569G}   \\
35      &     0.7   &    $32.59  \pm   0.03$    &     $1.4\pm0.1$         &   $34.29  \pm  0.11$   &      $2.20\pm0.22$           &  0.2 &   $40$  &   $8$    &  \cite{2007ApJ...670..655K,  2007AnA...472..489A}    \\
36      &  $\!\!\!$[1.1] &    $32.16   \pm  0.50$    &     ...                 &   $34.76   \pm 0.08$   &      $2.44\pm0.21$           &  2.0  &  $39$  &   $11$      &  \cite{2005AnA...439.1013A}  \\
37      &     0.7   &    $32.60   \pm  0.10$    &     $1.9\pm0.2$         &   $33.74   \pm 0.30$   &      $0.7\pm0.6^?$   &  2.0    &  $7$    &     3       &  \cite{2007AnA...476L..25H, 2007AnA...472..489A}    \\
38        &  $\lesssim2$  &    $31.03 \pm  0.5$   &     ...                    &  ...                             &       ...                             &                                                          0.1  & ...        &   ...         & ...            \\
39      &     2.1   &    $32.12  \pm   0.13$    &     $1.8\pm0.3$         &   ...                            &      ...                             &  0.4  &  ...        &   ...          &  \cite{ 2008ApJ...684..542K}    \\
40       &   $\lesssim2$  &  ...                       &      ...                        &  $33.82 \pm 0.25$       &        $2.17\pm 0.23$    &  ...    &  $<13$    &   ...         & \cite{2008AnA...477..353A}\\
41       &  $\!\!\!$[0.3]  &    $31.68\pm0.30$ &      $1.1\pm0.6$        &      $32.76\pm 0.20$     &        $2.2\pm0.2$          &    0.1  &  7       &  $<5$  &\cite{2004ApJ...612..398H,2013ApJ...764...38A} \\
42      &     2.0   &    $31.93  \pm   0.13$    &     $2.1\pm1.0$         &   ...                             &     ...                             &  0.4  &   ...       &     ...    &  \cite{2002ApJ...579..404P}    \\
43       &   0.3      &    $31.54  \pm 0.30$       &      $1.5\pm0.6$         &  ...                             &    ...                             &  0.3  &   ...        &      ...    & ... \\
44       &    $\lesssim1$  &  ...                      &     ...                          &  $32.00\pm0.50$        &    $[2.6]$                         & ...    &   ...        &    $<40$  & \cite{2009ApJ...700L.127A} \\
45      & $\!\!\!$[1.1] &    $31.60   \pm   0.50$  &      ...                &   $34.30   \pm 0.11 $     &    $2.31\pm0.23$           &  0.2  &   $ 50   $ &   $ 30$    &  \cite{ 2006ApJ...636..777A}  \\
46      &   $\!\!\!$[0.3] &    $31.20\pm 0.50$       &     ...               &   ...                       &     ...                            &  0.05 &   ...        &   ...    &  ... \\
47       & $\!\!\!$0.5         &    $31.00\pm 0.10 $      &   $1.5\pm0.4$                 &   $33.04\pm0.09$         &   $2.0\pm0.2$               & 1.5   &   5         &   4     &  \cite{2009ApJ...705....1C, 2002AnA...393L..37A,2007ApnSS.309...29M}  \\  
48      & $\!\!\!$[0.1] &    $31.11  \pm   0.10$    & $1.5\pm0.3$         &  ...                             &  ...                              &  0.8  &  ... &   ...    &  \cite{ 2008ApJ...684..542K}  \\
49     &  $\lesssim0.6$ &   $<31.15$                         &   ...                       &     $32.64\pm0.50$     &   $[2.6]$                       & ...    &  $<180$   & ... &  \cite{2009ApJ...700L.127A, 2010ApJ...708.1426W}\\  
50      &     0.1   &    $30.81  \pm   0.18$    &     $1.6\pm0.5$         &  ...                               &    ...                              &  0.1  &   ... &  ...    &  \cite{2003Sci...299.1372S}    \\
51     &  $0.1$ &         $31.77\pm 0.12$                    &  $0.9\pm0.5$                &   ...           &...                   &  0.7     &   ... &  ...    &   \cite{2009_Kawai_Fermi} \\ 
52         &   $\lesssim1.5$ &         ...             &     ...                        &   $32.82\pm0.15$          &   $2.78\pm0.31$           & ...     &  22   & 12  & \cite{ 2009arXiv0907.0574T} \\

\hline
\end{tabular}
\caption{Properties of the X-ray/TeV PWNe listed in Table~1}\label{tab:prop2}
\end{table}

\addtocounter{table}{-1}
\begin{table}
 \setlength{\tabcolsep}{0.05in}
\begin{tabular}{rlcccclllc}
\hline

 \# &  $N_{\rm H,22}$$^\ast$ & $\log L_{\rm X}$$^\dagger$ & $\Gamma_{\rm X}$ & $\log L_\gamma$$^{\ast\ast}$ &  $\Gamma_{\gamma}$$^\ddagger$ &  $l_{X}$$^\S$ &  $l_{\gamma}$$^\P$ & $\Delta$$^\parallel$ & ${\rm Refs.}$$^{\dagger\dagger}$ \\

     &   & ${\rm [erg/s]}$  &   & ${\rm [erg/s]}$  &    &    ${\rm pc}$  &    ${\rm pc}$  &      ${\rm arcmin}$ &   \\ 
\hline

53     &     0.25  &    $31.04  \pm   0.10$    &     $3.3\pm0.5$         &  ...                               &   ...                              &  0.2 &           ... & ...   &  \cite{2003ApJ...585L..41R, 2007ApJ...654..487N} \\
54      &     0.6   &    $31.19  \pm   0.03$    &     $1.5\pm0.1$         &   ...                               &   ...                              &  0.1 &   ...          &  ...   &  \cite{2006ApJ...647.1300M}    \\
55      &     0.03  &    $29.71  \pm   0.07$    &     $1.0\pm0.2$         &   $30.22\pm0.50$         &      $[2.6]$                    &  0.02 &   $10$ &  $<30$    &  \cite{2006ApJ...643.1146P, 2009ApJ...700L.127A}    \\
56       &    $\lesssim1.3$ & $<31$              &       ...                       &    $33.01\pm0.50$             &   $1.82\pm0.35$            &  ...    &   $6$ &  $7$        &  \cite{2009arXiv0907.0574T}   \\  
57       &     $\lesssim2$ &           $<32.5$           &              ...                &   $34.66\pm0.50$          &   $2.40\pm0.44$              & ... &     $70$  &  ...          &   \cite{2009arXiv0907.0574T}  \\
58     &     0.17           &    $29.63  \pm   0.01$      &     $1.7\pm0.6$                  &   ...                  &   ...                             &  0.05 &   ...    &  ...  &  \cite{2008ApJ...685.1129M} \\
59      &     0.2            &            $29.5\pm0.5$     &         $[1.5]$                  &   ...                  &  ...                        &  0.07               &     ...                    &   $   ...  $    &  \cite{2007AnA...467.1209H}  \\
\hline
60         &  15                &           $35.27\pm0.07$      &  $2.3\pm0.4$         &     $33.90\pm0.10$               &  $2.4\pm0.1$   &  1.8      & 3.2  &  $<1$  &\cite{2001ApJ...556L.107G,2005AnA...432L..25A,2009ApJ...700L..34C}\\
61        &    1.5             &              $32.71\pm0.11$                           &         $1.2\pm0.2$                   &        $34.43\pm0.10$                          &   $2.6\pm0.2$        &   0.1    & 21  &    $<5$  & \cite{2011AnA...525A..46H,2010ApJ...725..571S}  \\
 62 & 4.3 &                     $34.00\pm0.10$                               &       $2.1\pm0.3$           &     $33.77\pm0.20$          &     ...            &     2   &  $<10$   & $<6$      &   \cite{2008AIPC.1085..312T,2011ApJ...729L..16G}     \\   
63     &1            &             ...                                          &              ...                                 &        ...                                 &       ...     &     ...   &   ...     & ...        &         \cite{2012ApJ...746..105N} \\
64    &   0.76              &             32.46                                           &              0.05                                 &        2.03                                &      0.05     &     ...   &   ...     & ...        &         \cite{2013AnA...551A...7A} \\
65    &  $<2$                &             $< 31.8$                                           &              ...                                 &        $33.90\pm0.12$                                 &       $2.1\pm0.1$     &     ...   &   26     & $<3$        &         \cite{2011ICRC....7..243S} \\
66   &       $<2$              &           $<32.43$                                               &          ...                           &          $33.36\pm0.14$                               &         $2.8\pm0.2$    &  ...      &
17  &  10    &     \cite{2012arXiv1205.0719D}\\
67        &      0.3              &               $31.50\pm0.3$                                   &              [1.5]                             &        $33.40\pm0.15$                         &    $1.9\pm0.2$        &  0.1     &  $<3$  &  10  &\cite{2011AnA...525A..46H, 2012arXiv1202.3838K} \\     
68  &  3.9   &              $33.30\pm0.5$                                                       &       $2.3\pm0.8$                                         &         ...                                                     &         ...                           &    4.4 
& ... &  ... &  \cite{2012ApJ...745...99K}     \\           
69  &   ...          &                            ...                                  &       ...                                         &         $34.03\pm0.15$                                                     &         ...                           &    
 ... & ... & 18 &  \cite{HESSColl2011}     \\     
70        &     1.2                 &                $<31.50$                                        &               ...                                &       $34.38\pm0.17$                     &    $2.3\pm0.2$        &    ...        &   ...     &
 40    &         \cite{2012AnA...537A.114A, 2012arXiv1202.3838K}        \\
71  &   1.6      &                      $31.0\pm0.5$                                                       &               [1.5]                                        &         $31.72\pm0.13$                          &    $1.9\pm0.5$                  &   ...& 
  $<6$      &   24&       \cite{2011arXiv1111.1034A} \\    
 72    &     ...      &             ...                                                   &                   ...                           &             $33.43\pm0.23$                            &       $2.66\pm0.27$              &   ...
&   17    &  20    & \cite{2008AnA...481..401A}     \\   
 73    &     2      &             $<32.0$                                                   &                   ...                           &             $35.00\pm0.20$                            &       $2.12\pm0.2$              &   ...
&   50    &  10    & \cite{2008AnA...481..401A}     \\   
74    &     ...      &             ...                                                   &                   ...                           &             $33.24\pm0.20$                            &       $2.8\pm0.3$              &   ...
&   $<10$    &  $<3$    & \cite{chav}     \\  
75    &     ...      &             ...                                                   &                   ...                           &             $34.89\pm0.20$                            &       $2.46\pm0.2$              &   ...
&   31    &  20    & \cite{2006ApJ...636..777A}     \\  

76  & 0.05 &                          $28.94\pm0.08$                                                         &         $2.2\pm0.4$                                &          ...                                               &          ...                           &   0.02   & 
... &  ... &  \cite{2006AnA...448L..13H}     \\

\hline
\end{tabular}
\caption{Properties of the X-ray/TeV PWNe listed in Table~1}
\footnotesize{$^\ast$ -- Hydrogen column density (in units of $10^{22}$ cm$^{-2}$) obtained from spectral fits to the PWN spectra or estimated  from the pulsar's dispersion measure assuming 10\% interstellar medium ionization (in square brackets for the latter case). In a few cases only the upper limits are given (based on the Galactic HI column density). $^\dagger$ -- Logarithm of PWN luminosity in the 0.5--8 keV band. The quoted errors are purely statistical, they do not include the distance errors. For bright PWNe (e.g., \#  2, 8, 16), we quote the luminosity of the PWN ``core'' restricted to the torus/arcs regions. For the PWNe with extended tails (\#  39, 42, 48, 53, 54, 55, 58) we quote only the luminosity of the bright ``bullet'' component. For \#  36, 45 and 46, faint extended emission is seen around the pulsar but its luminosity is very uncertain; we use $\pm 0.50$ dex as a conservative estimate for the uncertainty.  $^{\ast\ast}$ -- Logarithm of PWN  luminosity in the 1--10 TeV band. $^\ddagger$ -- Photon index of TeV spectrum determined from a power-law model. The fits are not good (e.g., an exponential cutoff is required or the spectral slope is nonuniform) in the cases marked by the superscript $^?$. The values of photon indices given in square brackets were assumed to estimate $L_\gamma$. $^\S$ -- Size of X-ray PWN ``core'' in which the PWN X-ray properties listed in this table were measured. $^\P$ -- Size of TeV source. If the source is unresolved, we quote the upper limit on $l_\gamma$. $^\parallel$ -- Offset between the X-ray and TeV components. $^{\dagger\dagger}$ -- The PWN/PSR X-ray properties listed here were measured by ourselves (except for \# 2, 5, 6, 10, 37, 47, 49 and 51), but we cite recent relevant papers when available. Top portion of this table shows 59 sources presented in \cite{2010AIPC.1248...25K} with some of the parameter values updated.}
\label{tab:prop3}
\end{table}

\begin{table}
\setlength{\tabcolsep}{0.03in} 
  {\scriptsize
\centering{
\begin{tabular}{lccccccccccc}
\hline
\# & SNR/PWN &  $d$ &  $\log L_{\rm X}$  & $\Gamma_{\rm X}$  &  $\log L_{\gamma}$  &  $\Gamma_{\gamma}$ & $l_{X}$ &   $l_{\gamma}$$^\P$ & $\Delta$$^\parallel$ & ${\rm Rad./H}_{\alpha}/{\rm GeV}$ & Refs. \\
    &                 & kpc  &                                                     &                                  &                                               &                                 &  pc &   pc &  amin &  & \\
\hline
77 & G15.4+0.1/HESS~J1818--154   &  ~~10          &    $<33.40$ &   ...         &  $34.25\pm0.17$   & $2.1\pm0.3$                        & ...  & 23& ...&  P/N/N    &  \cite{2011ICRC....7..247H}   \\
78 & G16.7+0.1/G16.73+0.08                  &   10         &    $34.37\pm0.07$ & $1.2\pm0.3$         &  $\lesssim 34.04$  & ...          &  2.2  & ... &  ... & P/N/N    &  \cite{2003ApJ...592..941H}  \\
79 &   G25.24--0.19                &   10           &    $33.63\pm0.14$ & $0.8\pm0.3$         &  ...    & ...                      &  2.2  & ... & ...& P/N/?    &  \cite{2008ApJ...681..515G, 2012ApJ...745...99K} \\
80 & G29.41+0.08/HESS~J1843-033C  &  ~~15         &  $35.12\pm0.10$  &  $2.3\pm0.5$         & $34.50\pm0.30$    &    ...          & $9$ & $<26$ &  $<2$ & P/N/?    & 
\cite{Terrier2011}   \\
81 & 3C\,396/G39.22--0.32              &  ~~8          &    $34.09\pm0.10$ & $1.5\pm0.2$         &  ...   & ...                          &  1.6  & ... & ... &  P/N/N    &  \cite{2003ApJ...592L..45O}   \\
82 & DA\,495/G65.73+1.18    &   ~~~~1.5       &    $31.84\pm0.09$ & $1.8\pm0.1$         & ...  & ...                        &  0.2  & ... & ... &  Y/N/N    & \cite{2008ApJ...687..505A} \\
83 & CTB\,87/G74.94+1.11              &  ~~6         &    $34.01\pm0.05$ & $1.60\pm0.04$         &  $33.04\pm0.30$   & $2.3\pm0.3$    &  5  & 50 & $<3$ &  P/N/N    &  ... \\
 & Veritas~J2016+372$^?$ \\
84 & IC\,443/G189.23+2.90      &   ~~~~1.5       &    $32.62\pm0.03$ & $1.7\pm0.1$         &  $32.51\pm0.10$    & $3.0\pm0.4$  &  0.3  &  4 & 18 & Y/P/N   &  \cite{2006ApJ...648.1037G,2009ApJ...698L.133A} \\
 & Veritas~J0616.9+2230$^?$ \\
85 & MSH 11--61A/G290.1--0.8 &  ~~10      &  $34.10\pm0.11$         &    $1.8\pm0.4$         &  $<34.21$   & ...               & 10    &   ... & ... &  Y/N/N    & 
\cite{2012ApJ...750L..39T}   \\ 
86 & MSH\,11--62/G291.02--0.11      &  ~~~~1.0        &    $33.52\pm0.05$ & $1.6\pm0.1$         &  $<32.21$    & ...                &  1.1   &  ... & ... & P/N/P    &  \cite{2004IAUS..218..203H,2012ApJ...749..131S} \\
87 & G293.8+0.6/G293.79+0.58      &   ~~2          &    $30.90\pm0.26$ & $0.4\pm0.3$         &  ...  & ...                            &  0.6   & ... & ... &  P/N/N    &  \cite{2003AAS...203.3907O} \\
88 & MSH\,15--56/G326.12--1.81      &   ~~4            &    $32.51\pm0.09$ & $1.7\pm0.2$         &  $<33.80$    & ...                    &  1.0    & ... & ... & Y/N/N    &  \cite{2002APS..APRN17037P}  \\
89 & G327.1--1.1/H~J1554--550                  &   ~~7            &    $34.20\pm0.07$ & $2.0\pm0.1$         &  $33.81\pm0.24$    & $2.1\pm0.2$            &  1.7  &  16 & 3 &  Y/N/?    &  \cite{2004AAS...205.8405S,2012arXiv1201.0481A} \\
90  & G334.79--0.15/H~J1626--490$^?$   &  ~~10     &  $33.22\pm0.50$      &   [1.5]                      &  $35.24\pm0.11$              & $2.18\pm0.12$  &  0.7    &  $<30$   & 10  &   N/N/?    &  \cite{2008AnA...477..353A}  \\
91 & G338.3-0.0/H~J1640-465 &  ~~10          &    $33.59\pm0.25$ & $2.5\pm0.3$         &  $34.94\pm0.05$   &  $2.42\pm0.15$         & 3   &  8   &  1  &  N/N/P    &  
\cite{2006ApJ...636..777A,2009ApJ...706.1269L}   \\
\hline
\end{tabular}
}}
\caption{Properties of X-ray PWNe without a known pulsar. \footnotesize{Parameters $\log L_{\rm X}$, $\Gamma_{\rm X}$ ,  $\log L_{\gamma}$ ,  $\Gamma_{\gamma}$, $l_{X}$,   $l_{\gamma}$, $\Delta$$^\parallel$ are defined in the caption  to Table~2.}}
\label{tab:prop3}
\end{table}

\begin{table}
\setlength{\tabcolsep}{0.03in} 
 {\scriptsize
\centering{
\begin{tabular}{clccccccccccc}
\hline
 \#& TeV Source & $l^a$& $b^a$& $d$&  $\log L_\gamma$$^b$&  $\Gamma_{\gamma}$& $\alpha$$^c$& $\Delta_{\rm 2FGL}$$^d$& $\Delta_{\rm ATNF}$$^{e}$& X-ray$^f$ & Suggested & Ref. \\
 & &  deg& deg& kpc& ${\rm [erg/s]}$&& deg& arcmin& arcmin & &type \\
\hline
\multicolumn{13}{c}{Unidentified Sources}\\
\hline
1 & HESS J1800-240C &   5.71 &  -0.06 &  2.0 & 32.90 & 2.31 & Unresolved &  0.53 &  ... & ... & SNR/MC & \cite{2008AnA...481..401A,2013arXiv1301.2437M}\\
2 & HESS J1800-240A &   6.14 &  -0.63 &  2.0 & 33.18 & 2.55 & 0.15 & 16.04 &  ... &  ... & SNR/MC & \cite{2008AnA...481..401A,2010ApJ...718..348A}\\
3 & HESS J1911+090 &  43.26 &  -0.19 &  8.0 & 33.37 & 3.10 & Unresolved &  0.97 &  ... & Ch:Diffuse & SNR/MC & \cite{2011arXiv1104.5003B}\\
4 & HESS J1729-345 & 353.44 &  -0.13 &  3.2 & 33.18 & 2.24 & 0.14 &  ... &  ... & ... & UNID & \cite{2011AnA...531A..81H}\\
5 & HESS J1808-204 &   9.93 &  -0.10 & 12.0 & 34.00 & 2.40 & ext & 15.40 &  9.39 & Ch\&XMM:mult.pt.src. & Magnetar & \cite{2012AIPC.1505..273R}\\
 6 & HESS J1841-055 &  26.79 &  -0.20 &  6.9 & 35.21 & 2.41 & 0.41/0.25 & 28.29 & 16.17 & Yes & PWN/PSR~J1838--0537? & \cite{2008AnA...477..353A,2009ApJ...697.1194S,2012xrb..confE..64W,2012ApJ...755L..20P}\\
 7 & HESS J1713-381 & 348.65 &   0.38 &  7.0 & 33.86 & 2.65 & Unresolved & 18.96 &  1.93 &  Ch:Diffuse+pt.src. & Shell/Magnetar & \cite{2008AnA...477..353A,2010ApJ...725.1384H}\\
8 & HESS J1834-087 &  23.26 &  -0.33 &  5.0 & 34.37 & 2.50 & Unresolved &  8.82 &  0.90 & XMM:Diffuse+pt.src. & PWN/Magnetar & \cite{2006ApJ...643L..53A,2011ApJ...735...33M}\\
9 & HESS J1427-608 & 314.41 &  -0.14 & 21.0 & 35.29 & 2.16 & Unresolved &  3.07 &  ... & XMM:mult.pt.src. & UNID & \cite{2008AnA...477..353A,2013arXiv1301.5274F}\\
10 & HESS J1747-248 & 3.78 & 1.72 &  ... &  ... & ... &  &  2.74 &  8.45 & Ch:mult.pt.src. & Terzan 5 & \cite{2011AnA...531L..18H}\\
11 & HESS J1634-472 & 337.00 &   0.22 &  8.0 & 34.57 & 2.38 & 0.11 &  9.53 &  ... &  ... & UNID & \cite{2006ApJ...636..777A}\\
12 & HESS J1848-018 &  31.00 &  -0.16 &  5.3 & 33.84 & 2.80 & 0.32 &  9.01 &  ... & ... & UNID & \cite{2008AIPC.1085..372C}\\
13 & HESS J1457-593 & 318.36 &  -0.43 &  ... &  ... & ... & 0.31/0.17 &  ... &  ... & ... & SNR/MC & \cite{2010tsra.confE.196H}\\
\hline
\multicolumn{13}{c}{Dark Sources}\\
\hline
14 & HESS J1923+141 &  49.13 &  -0.40 &  6.0 & 34.54 & 2.60 & ext &  3.79 &... & Ch:mult.pt.src. & SNR/MC & \cite{2009ApJ...700L.127A}\\
15 & HESS J1800-240B &   5.96 &  -0.38 &  2.0 & 32.89 & 2.50 & 0.15 &  2.76 &  ... & Ch:mult.pt.src. & SNR/MC & \cite{2008AnA...481..401A}\\
16 & HESS J1843-033B &  29.03 &   0.37 &  ... &  ... & ... & ext & 11.80 & 13.13 & Ch:mult.pt.src. & UNID & \cite{2008ICRC....2..579H}\\
17 & HESS J1614-518 & 331.51 &  -0.57 &  1.0 & 33.34 & 2.26 & 0.23/0.15 & 14.27 &  ... & Ch:mult.pt.src. & UNID & \cite{2006ApJ...636..777A}\\
18 & HESS J1507-622 & 317.95 &  -3.49 &  6.0 & 34.26 & 2.24 & 0.15 &  3.25 &  ... & Ch:Diffuse & UNID & \cite{2011AnA...525A..45H} \cite{2012AnA...545A..94D}\\
19 & HESS J1708-410 & 345.68 &  -0.47 &  3.0 & 33.80 & 2.46 & Unresolved &  ... &  ... & XMM:mult.pt.src. & UNID & \cite{2009ApJ...707.1717V}\\
20 & HESS J1641-463 & 338.52 &   0.09 & 11.0 & 34.22 & 1.99 & Unresolved & 15.56 &  ... & Ch:mult.pt.src. & UNID & \cite{2013arXiv1303.0979O}\\
\hline
\end{tabular}
}}
\caption{Unidentified and ``dark'' sources. $^a$ $-$ Galactic coordinates. $^b$ $-$ Logarithm of luminosity in the 1--10 TeV band. $^c$ $-$ Angular size of  TeV source. $^{d}$ $-$ Angular separation from the nearest 2FGL source with $\Delta_{\rm 2FGL}\lesssim30'$. $^{e}$ $-$ Angular separation from the nearest pulsar with $\Delta_{\rm ATNF}\lesssim20'$, $L_{\gamma}/\edot\lesssim10$,  $\tau_{\rm sd}\lesssim100$~kyr. $^{f}$ $-$ Existing X-ray observations (Ch. $=$ \emph{Chandra}; XMM $=$ \emph{XMM-Newton}; Diffuse $=$ diffuse X-ray emission detected; multi.pt.src $=$ multiple faint sources in the filed with no obvious counterpart; pt.src $=$ X-ray point source candidate).}
\label{tab:prop5}
\end{table}


\clearpage

\bibliography{bib}

\begin{thebibliography}{151}
\expandafter\ifx\csname natexlab\endcsname\relax\def\natexlab#1{#1}\fi
\expandafter\ifx\csname bibnamefont\endcsname\relax
  \def\bibnamefont#1{#1}\fi
\expandafter\ifx\csname bibfnamefont\endcsname\relax
  \def\bibfnamefont#1{#1}\fi
\expandafter\ifx\csname citenamefont\endcsname\relax
  \def\citenamefont#1{#1}\fi
\expandafter\ifx\csname url\endcsname\relax
  \def\url#1{\texttt{#1}}\fi
\expandafter\ifx\csname urlprefix\endcsname\relax\def\urlprefix{URL }\fi
\providecommand{\bibinfo}[2]{#2}
\providecommand{\eprint}[2][]{\url{#2}}

\bibitem[{\citenamefont{{Kennel} and {Coroniti}}(1984)}]{1984ApJ...283..694K}
\bibinfo{author}{\bibfnamefont{C.~F.} \bibnamefont{{Kennel}}} \bibnamefont{and}
  \bibinfo{author}{\bibfnamefont{F.~V.} \bibnamefont{{Coroniti}}},
  \bibinfo{journal}{\apj} \textbf{\bibinfo{volume}{283}}, \bibinfo{pages}{694}
  (\bibinfo{year}{1984}).

\bibitem[{\citenamefont{{Gaensler} and {Slane}}(2006)}]{2006ARAnA..44...17G}
\bibinfo{author}{\bibfnamefont{B.~M.} \bibnamefont{{Gaensler}}}
  \bibnamefont{and} \bibinfo{author}{\bibfnamefont{P.~O.}
  \bibnamefont{{Slane}}}, \bibinfo{journal}{\araa}
  \textbf{\bibinfo{volume}{44}}, \bibinfo{pages}{17} (\bibinfo{year}{2006}),
  \eprint{arXiv:astro-ph/0601081}.

\bibitem[{\citenamefont{{Kargaltsev} and {Pavlov}}(2008)}]{2008AIPC..983..171K}
\bibinfo{author}{\bibfnamefont{O.}~\bibnamefont{{Kargaltsev}}}
  \bibnamefont{and} \bibinfo{author}{\bibfnamefont{G.~G.}
  \bibnamefont{{Pavlov}}}, in \emph{\bibinfo{booktitle}{40 Years of Pulsars:
  Millisecond Pulsars, Magnetars and More}}, edited by
  \bibinfo{editor}{\bibfnamefont{C.}~\bibnamefont{{Bassa}}},
  \bibinfo{editor}{\bibfnamefont{Z.}~\bibnamefont{{Wang}}},
  \bibinfo{editor}{\bibfnamefont{A.}~\bibnamefont{{Cumming}}},
  \bibnamefont{and} \bibinfo{editor}{\bibfnamefont{V.~M.}
  \bibnamefont{{Kaspi}}} (\bibinfo{year}{2008}), vol. \bibinfo{volume}{983} of
  \emph{\bibinfo{series}{American Institute of Physics Conference Series}}, pp.
  \bibinfo{pages}{171--185}, \eprint{0801.2602}.

\bibitem[{\citenamefont{{Blondin} et~al.}(2001)\citenamefont{{Blondin},
  {Chevalier}, and {Frierson}}}]{2001ApJ...563..806B}
\bibinfo{author}{\bibfnamefont{J.~M.} \bibnamefont{{Blondin}}},
  \bibinfo{author}{\bibfnamefont{R.~A.} \bibnamefont{{Chevalier}}},
  \bibnamefont{and} \bibinfo{author}{\bibfnamefont{D.~M.}
  \bibnamefont{{Frierson}}}, \bibinfo{journal}{\apj}
  \textbf{\bibinfo{volume}{563}}, \bibinfo{pages}{806} (\bibinfo{year}{2001}),
  \eprint{arXiv:astro-ph/0107076}.

\bibitem[{\citenamefont{{de Jager} and
  {Djannati-Ata{\"i}}}(2009)}]{2009ASSL..357..451D}
\bibinfo{author}{\bibfnamefont{O.~C.} \bibnamefont{{de Jager}}}
  \bibnamefont{and}
  \bibinfo{author}{\bibfnamefont{A.}~\bibnamefont{{Djannati-Ata{\"i}}}}, in
  \emph{\bibinfo{booktitle}{Astrophysics and Space Science Library}}, edited by
  \bibinfo{editor}{\bibfnamefont{W.}~\bibnamefont{{Becker}}}
  (\bibinfo{year}{2009}), vol. \bibinfo{volume}{357} of
  \emph{\bibinfo{series}{Astrophysics and Space Science Library}}, p.
  \bibinfo{pages}{451}, \eprint{0803.0116}.

\bibitem[{\citenamefont{{Manchester} et~al.}(2005)\citenamefont{{Manchester},
  {Hobbs}, {Teoh}, and {Hobbs}}}]{2005AJ....129.1993M}
\bibinfo{author}{\bibfnamefont{R.~N.} \bibnamefont{{Manchester}}},
  \bibinfo{author}{\bibfnamefont{G.~B.} \bibnamefont{{Hobbs}}},
  \bibinfo{author}{\bibfnamefont{A.}~\bibnamefont{{Teoh}}}, \bibnamefont{and}
  \bibinfo{author}{\bibfnamefont{M.}~\bibnamefont{{Hobbs}}},
  \bibinfo{journal}{\aj} \textbf{\bibinfo{volume}{129}}, \bibinfo{pages}{1993}
  (\bibinfo{year}{2005}), \eprint{arXiv:astro-ph/0412641}.

\bibitem[{\citenamefont{{Kargaltsev}
  et~al.}(2012{\natexlab{a}})\citenamefont{{Kargaltsev}, {Durant}, {Pavlov},
  and {Garmire}}}]{2012arXiv1202.3838K}
\bibinfo{author}{\bibfnamefont{O.}~\bibnamefont{{Kargaltsev}}},
  \bibinfo{author}{\bibfnamefont{M.}~\bibnamefont{{Durant}}},
  \bibinfo{author}{\bibfnamefont{G.~G.} \bibnamefont{{Pavlov}}},
  \bibnamefont{and}
  \bibinfo{author}{\bibfnamefont{G.}~\bibnamefont{{Garmire}}},
  \bibinfo{journal}{ApJS} \textbf{\bibinfo{volume}{201}}, \bibinfo{eid}{37}
  (\bibinfo{year}{2012}{\natexlab{a}}), \eprint{1202.3838}.

\bibitem[{\citenamefont{{Goldreich} and {Julian}}(1969)}]{1969ApJ...157..869G}
\bibinfo{author}{\bibfnamefont{P.}~\bibnamefont{{Goldreich}}} \bibnamefont{and}
  \bibinfo{author}{\bibfnamefont{W.~H.} \bibnamefont{{Julian}}},
  \bibinfo{journal}{\apj} \textbf{\bibinfo{volume}{157}}, \bibinfo{pages}{869}
  (\bibinfo{year}{1969}).

\bibitem[{\citenamefont{{de Jager} et~al.}(2009)\citenamefont{{de Jager},
  {Ferreira}, {Djannati-Ata{\"i}}, {Dalton}, {Deil}, {Kosack}, {Renaud},
  {Schwanke}, and {Tibolla}}}]{2009arXiv0906.2644D}
\bibinfo{author}{\bibfnamefont{O.~C.} \bibnamefont{{de Jager}}},
  \bibinfo{author}{\bibfnamefont{S.~E.~S.} \bibnamefont{{Ferreira}}},
  \bibinfo{author}{\bibfnamefont{A.}~\bibnamefont{{Djannati-Ata{\"i}}}},
  \bibinfo{author}{\bibfnamefont{M.}~\bibnamefont{{Dalton}}},
  \bibinfo{author}{\bibfnamefont{C.}~\bibnamefont{{Deil}}},
  \bibinfo{author}{\bibfnamefont{K.}~\bibnamefont{{Kosack}}},
  \bibinfo{author}{\bibfnamefont{M.}~\bibnamefont{{Renaud}}},
  \bibinfo{author}{\bibfnamefont{U.}~\bibnamefont{{Schwanke}}},
  \bibnamefont{and}
  \bibinfo{author}{\bibfnamefont{O.}~\bibnamefont{{Tibolla}}},
  \bibinfo{journal}{ArXiv e-prints}  (\bibinfo{year}{2009}),
  \eprint{0906.2644}.

\bibitem[{\citenamefont{{Mattana} et~al.}(2009)\citenamefont{{Mattana},
  {Falanga}, {G{\"o}tz}, {Terrier}, {Esposito}, {Pellizzoni}, {De Luca},
  {Marandon}, {Goldwurm}, and {Caraveo}}}]{2009ApJ...694...12M}
\bibinfo{author}{\bibfnamefont{F.}~\bibnamefont{{Mattana}}},
  \bibinfo{author}{\bibfnamefont{M.}~\bibnamefont{{Falanga}}},
  \bibinfo{author}{\bibfnamefont{D.}~\bibnamefont{{G{\"o}tz}}},
  \bibinfo{author}{\bibfnamefont{R.}~\bibnamefont{{Terrier}}},
  \bibinfo{author}{\bibfnamefont{P.}~\bibnamefont{{Esposito}}},
  \bibinfo{author}{\bibfnamefont{A.}~\bibnamefont{{Pellizzoni}}},
  \bibinfo{author}{\bibfnamefont{A.}~\bibnamefont{{De Luca}}},
  \bibinfo{author}{\bibfnamefont{V.}~\bibnamefont{{Marandon}}},
  \bibinfo{author}{\bibfnamefont{A.}~\bibnamefont{{Goldwurm}}},
  \bibnamefont{and} \bibinfo{author}{\bibfnamefont{P.~A.}
  \bibnamefont{{Caraveo}}}, \bibinfo{journal}{\apj}
  \textbf{\bibinfo{volume}{694}}, \bibinfo{pages}{12} (\bibinfo{year}{2009}),
  \eprint{0811.0327}.

\bibitem[{\citenamefont{{Aharonian} et~al.}(1997)\citenamefont{{Aharonian},
  {Atoyan}, and {Kifune}}}]{1997MNRAS.291..162A}
\bibinfo{author}{\bibfnamefont{F.~A.} \bibnamefont{{Aharonian}}},
  \bibinfo{author}{\bibfnamefont{A.~M.} \bibnamefont{{Atoyan}}},
  \bibnamefont{and} \bibinfo{author}{\bibfnamefont{T.}~\bibnamefont{{Kifune}}},
  \bibinfo{journal}{\mnras} \textbf{\bibinfo{volume}{291}},
  \bibinfo{pages}{162} (\bibinfo{year}{1997}).

\bibitem[{\citenamefont{{de Jager} et~al.}(2008)\citenamefont{{de Jager},
  {Ferreira}, and {Djannati-Ata{\"i}}}}]{2008AIPC.1085..199D}
\bibinfo{author}{\bibfnamefont{O.~C.} \bibnamefont{{de Jager}}},
  \bibinfo{author}{\bibfnamefont{S.~E.~S.} \bibnamefont{{Ferreira}}},
  \bibnamefont{and}
  \bibinfo{author}{\bibfnamefont{A.}~\bibnamefont{{Djannati-Ata{\"i}}}}, in
  \emph{\bibinfo{booktitle}{American Institute of Physics Conference Series}},
  edited by \bibinfo{editor}{\bibfnamefont{F.~A.} \bibnamefont{{Aharonian}}},
  \bibinfo{editor}{\bibfnamefont{W.}~\bibnamefont{{Hofmann}}},
  \bibnamefont{and} \bibinfo{editor}{\bibfnamefont{F.}~\bibnamefont{{Rieger}}}
  (\bibinfo{year}{2008}), vol. \bibinfo{volume}{1085} of
  \emph{\bibinfo{series}{American Institute of Physics Conference Series}}, pp.
  \bibinfo{pages}{199--202}.

\bibitem[{\citenamefont{{Gelfand} et~al.}(2009)\citenamefont{{Gelfand},
  {Slane}, and {Zhang}}}]{2009ApJ...703.2051G}
\bibinfo{author}{\bibfnamefont{J.~D.} \bibnamefont{{Gelfand}}},
  \bibinfo{author}{\bibfnamefont{P.~O.} \bibnamefont{{Slane}}},
  \bibnamefont{and} \bibinfo{author}{\bibfnamefont{W.}~\bibnamefont{{Zhang}}},
  \bibinfo{journal}{\apj} \textbf{\bibinfo{volume}{703}}, \bibinfo{pages}{2051}
  (\bibinfo{year}{2009}), \eprint{0904.4053}.

\bibitem[{\citenamefont{{Bucciantini} et~al.}(2011)\citenamefont{{Bucciantini},
  {Arons}, and {Amato}}}]{2011MNRAS.410..381B}
\bibinfo{author}{\bibfnamefont{N.}~\bibnamefont{{Bucciantini}}},
  \bibinfo{author}{\bibfnamefont{J.}~\bibnamefont{{Arons}}}, \bibnamefont{and}
  \bibinfo{author}{\bibfnamefont{E.}~\bibnamefont{{Amato}}},
  \bibinfo{journal}{\mnras} \textbf{\bibinfo{volume}{410}},
  \bibinfo{pages}{381} (\bibinfo{year}{2011}).

\bibitem[{\citenamefont{{Tanaka} and {Takahara}}(2013)}]{2013MNRAS.429.2945T}
\bibinfo{author}{\bibfnamefont{S.~J.} \bibnamefont{{Tanaka}}} \bibnamefont{and}
  \bibinfo{author}{\bibfnamefont{F.}~\bibnamefont{{Takahara}}},
  \bibinfo{journal}{\mnras} \textbf{\bibinfo{volume}{429}},
  \bibinfo{pages}{2945} (\bibinfo{year}{2013}), \eprint{1211.7266}.

\bibitem[{\citenamefont{{Van Etten} and {Romani}}(2011)}]{2011ApJ...742...62V}
\bibinfo{author}{\bibfnamefont{A.}~\bibnamefont{{Van Etten}}} \bibnamefont{and}
  \bibinfo{author}{\bibfnamefont{R.~W.} \bibnamefont{{Romani}}},
  \bibinfo{journal}{\apj} \textbf{\bibinfo{volume}{742}}, \bibinfo{eid}{62}
  (\bibinfo{year}{2011}), \eprint{1108.3573}.

\bibitem[{\citenamefont{{Mart{\'{\i}}n}
  et~al.}(2012)\citenamefont{{Mart{\'{\i}}n}, {Torres}, and
  {Rea}}}]{2012MNRAS.427..415M}
\bibinfo{author}{\bibfnamefont{J.}~\bibnamefont{{Mart{\'{\i}}n}}},
  \bibinfo{author}{\bibfnamefont{D.~F.} \bibnamefont{{Torres}}},
  \bibnamefont{and} \bibinfo{author}{\bibfnamefont{N.}~\bibnamefont{{Rea}}},
  \bibinfo{journal}{\mnras} \textbf{\bibinfo{volume}{427}},
  \bibinfo{pages}{415} (\bibinfo{year}{2012}), \eprint{1209.0300}.

\bibitem[{\citenamefont{{Vorster} and {Moraal}}(2013)}]{2013ApJ...765...30V}
\bibinfo{author}{\bibfnamefont{M.~J.} \bibnamefont{{Vorster}}}
  \bibnamefont{and} \bibinfo{author}{\bibfnamefont{H.}~\bibnamefont{{Moraal}}},
  \bibinfo{journal}{\apj} \textbf{\bibinfo{volume}{765}}, \bibinfo{eid}{30}
  (\bibinfo{year}{2013}).

\bibitem[{\citenamefont{{Horns} et~al.}(2007)\citenamefont{{Horns},
  {Aharonian}, {Hoffmann}, and {Santangelo}}}]{2007ApnSS.309..189H}
\bibinfo{author}{\bibfnamefont{D.}~\bibnamefont{{Horns}}},
  \bibinfo{author}{\bibfnamefont{F.}~\bibnamefont{{Aharonian}}},
  \bibinfo{author}{\bibfnamefont{A.~I.~D.} \bibnamefont{{Hoffmann}}},
  \bibnamefont{and}
  \bibinfo{author}{\bibfnamefont{A.}~\bibnamefont{{Santangelo}}},
  \bibinfo{journal}{\apss} \textbf{\bibinfo{volume}{309}}, \bibinfo{pages}{189}
  (\bibinfo{year}{2007}), \eprint{arXiv:astro-ph/0609386}.

\bibitem[{\citenamefont{{Arons} and {Tavani}}(1994)}]{1994ApJS...90..797A}
\bibinfo{author}{\bibfnamefont{J.}~\bibnamefont{{Arons}}} \bibnamefont{and}
  \bibinfo{author}{\bibfnamefont{M.}~\bibnamefont{{Tavani}}},
  \bibinfo{journal}{ApJS} \textbf{\bibinfo{volume}{90}}, \bibinfo{pages}{797}
  (\bibinfo{year}{1994}).

\bibitem[{\citenamefont{{Bartko} and {Bednarek}}(2008)}]{2008MNRAS.385.1105B}
\bibinfo{author}{\bibfnamefont{H.}~\bibnamefont{{Bartko}}} \bibnamefont{and}
  \bibinfo{author}{\bibfnamefont{W.}~\bibnamefont{{Bednarek}}},
  \bibinfo{journal}{\mnras} \textbf{\bibinfo{volume}{385}},
  \bibinfo{pages}{1105} (\bibinfo{year}{2008}), \eprint{0712.2964}.

\bibitem[{\citenamefont{{Nolan} et~al.}(2012)\citenamefont{{Nolan}, {Abdo},
  {Ackermann}, {Ajello}, {Allafort}, {Antolini}, {Atwood}, {Axelsson},
  {Baldini}, {Ballet} et~al.}}]{2012ApJS..199...31N}
\bibinfo{author}{\bibfnamefont{P.~L.} \bibnamefont{{Nolan}}},
  \bibinfo{author}{\bibfnamefont{A.~A.} \bibnamefont{{Abdo}}},
  \bibinfo{author}{\bibfnamefont{M.}~\bibnamefont{{Ackermann}}},
  \bibinfo{author}{\bibfnamefont{M.}~\bibnamefont{{Ajello}}},
  \bibinfo{author}{\bibfnamefont{A.}~\bibnamefont{{Allafort}}},
  \bibinfo{author}{\bibfnamefont{E.}~\bibnamefont{{Antolini}}},
  \bibinfo{author}{\bibfnamefont{W.~B.} \bibnamefont{{Atwood}}},
  \bibinfo{author}{\bibfnamefont{M.}~\bibnamefont{{Axelsson}}},
  \bibinfo{author}{\bibfnamefont{L.}~\bibnamefont{{Baldini}}},
  \bibinfo{author}{\bibfnamefont{J.}~\bibnamefont{{Ballet}}},
  \bibnamefont{et~al.}, \bibinfo{journal}{ApJS} \textbf{\bibinfo{volume}{199}},
  \bibinfo{eid}{31} (\bibinfo{year}{2012}), \eprint{1108.1435}.

\bibitem[{\citenamefont{{Tang} and {Chevalier}}(2012)}]{2012ApJ...752...83T}
\bibinfo{author}{\bibfnamefont{X.}~\bibnamefont{{Tang}}} \bibnamefont{and}
  \bibinfo{author}{\bibfnamefont{R.~A.} \bibnamefont{{Chevalier}}},
  \bibinfo{journal}{\apj} \textbf{\bibinfo{volume}{752}}, \bibinfo{eid}{83}
  (\bibinfo{year}{2012}), \eprint{1204.3913}.

\bibitem[{\citenamefont{{Kargaltsev} and {Pavlov}}(2010)}]{2010AIPC.1248...25K}
\bibinfo{author}{\bibfnamefont{O.}~\bibnamefont{{Kargaltsev}}}
  \bibnamefont{and} \bibinfo{author}{\bibfnamefont{G.~G.}
  \bibnamefont{{Pavlov}}}, \bibinfo{journal}{X-ray Astronomy 2009; Present
  Status, Multi-Wavelength Approach and Future Perspectives}
  \textbf{\bibinfo{volume}{1248}}, \bibinfo{pages}{25} (\bibinfo{year}{2010}),
  \eprint{1002.0885}.

\bibitem[{\citenamefont{{Chen} et~al.}(2006)\citenamefont{{Chen}, {Wang},
  {Gotthelf}, {Jiang}, {Chu}, and {Gruendl}}}]{2006ApJ...651..237C}
\bibinfo{author}{\bibfnamefont{Y.}~\bibnamefont{{Chen}}},
  \bibinfo{author}{\bibfnamefont{Q.~D.} \bibnamefont{{Wang}}},
  \bibinfo{author}{\bibfnamefont{E.~V.} \bibnamefont{{Gotthelf}}},
  \bibinfo{author}{\bibfnamefont{B.}~\bibnamefont{{Jiang}}},
  \bibinfo{author}{\bibfnamefont{Y.-H.} \bibnamefont{{Chu}}}, \bibnamefont{and}
  \bibinfo{author}{\bibfnamefont{R.}~\bibnamefont{{Gruendl}}},
  \bibinfo{journal}{\apj} \textbf{\bibinfo{volume}{651}}, \bibinfo{pages}{237}
  (\bibinfo{year}{2006}), \eprint{arXiv:astro-ph/0603123}.

\bibitem[{\citenamefont{{H.E.S.S.~Collaboration}
  et~al.}(2012)\citenamefont{{H.E.S.S.~Collaboration}, {Abramowski}, {Acero},
  {Aharonian}, {Akhperjanian}, {Anton}, {Balenderan}, {Balzer}, {Barnacka},
  {Becherini} et~al.}}]{2012AnA...545L...2H}
\bibinfo{author}{\bibnamefont{{H.E.S.S.~Collaboration}}},
  \bibinfo{author}{\bibfnamefont{A.}~\bibnamefont{{Abramowski}}},
  \bibinfo{author}{\bibfnamefont{F.}~\bibnamefont{{Acero}}},
  \bibinfo{author}{\bibfnamefont{F.}~\bibnamefont{{Aharonian}}},
  \bibinfo{author}{\bibfnamefont{A.~G.} \bibnamefont{{Akhperjanian}}},
  \bibinfo{author}{\bibfnamefont{G.}~\bibnamefont{{Anton}}},
  \bibinfo{author}{\bibfnamefont{S.}~\bibnamefont{{Balenderan}}},
  \bibinfo{author}{\bibfnamefont{A.}~\bibnamefont{{Balzer}}},
  \bibinfo{author}{\bibfnamefont{A.}~\bibnamefont{{Barnacka}}},
  \bibinfo{author}{\bibfnamefont{Y.}~\bibnamefont{{Becherini}}},
  \bibnamefont{et~al.}, \bibinfo{journal}{\aap} \textbf{\bibinfo{volume}{545}},
  \bibinfo{eid}{L2} (\bibinfo{year}{2012}), \eprint{1208.1636}.

\bibitem[{\citenamefont{{Mori} et~al.}(2004)\citenamefont{{Mori}, {Burrows},
  {Hester}, {Pavlov}, {Shibata}, and {Tsunemi}}}]{2004ApJ...609..186M}
\bibinfo{author}{\bibfnamefont{K.}~\bibnamefont{{Mori}}},
  \bibinfo{author}{\bibfnamefont{D.~N.} \bibnamefont{{Burrows}}},
  \bibinfo{author}{\bibfnamefont{J.~J.} \bibnamefont{{Hester}}},
  \bibinfo{author}{\bibfnamefont{G.~G.} \bibnamefont{{Pavlov}}},
  \bibinfo{author}{\bibfnamefont{S.}~\bibnamefont{{Shibata}}},
  \bibnamefont{and}
  \bibinfo{author}{\bibfnamefont{H.}~\bibnamefont{{Tsunemi}}},
  \bibinfo{journal}{\apj} \textbf{\bibinfo{volume}{609}}, \bibinfo{pages}{186}
  (\bibinfo{year}{2004}), \eprint{arXiv:astro-ph/0403287}.

\bibitem[{\citenamefont{{Willingale} et~al.}(2001)\citenamefont{{Willingale},
  {Aschenbach}, {Griffiths}, {Sembay}, {Warwick}, {Becker}, {Abbey}, and
  {Bonnet-Bidaud}}}]{2001AnA...365L.212W}
\bibinfo{author}{\bibfnamefont{R.}~\bibnamefont{{Willingale}}},
  \bibinfo{author}{\bibfnamefont{B.}~\bibnamefont{{Aschenbach}}},
  \bibinfo{author}{\bibfnamefont{R.~G.} \bibnamefont{{Griffiths}}},
  \bibinfo{author}{\bibfnamefont{S.}~\bibnamefont{{Sembay}}},
  \bibinfo{author}{\bibfnamefont{R.~S.} \bibnamefont{{Warwick}}},
  \bibinfo{author}{\bibfnamefont{W.}~\bibnamefont{{Becker}}},
  \bibinfo{author}{\bibfnamefont{A.~F.} \bibnamefont{{Abbey}}},
  \bibnamefont{and} \bibinfo{author}{\bibfnamefont{J.-M.}
  \bibnamefont{{Bonnet-Bidaud}}}, \bibinfo{journal}{\aap}
  \textbf{\bibinfo{volume}{365}}, \bibinfo{pages}{L212} (\bibinfo{year}{2001}).

\bibitem[{\citenamefont{{Aharonian}
  et~al.}(2006{\natexlab{a}})\citenamefont{{Aharonian}, {Akhperjanian},
  {Bazer-Bachi}, {Beilicke}, {Benbow}, {Berge}, {Bernl{\"o}hr}, {Boisson},
  {Bolz}, {Borrel} et~al.}}]{2006AnA...457..899A}
\bibinfo{author}{\bibfnamefont{F.}~\bibnamefont{{Aharonian}}},
  \bibinfo{author}{\bibfnamefont{A.~G.} \bibnamefont{{Akhperjanian}}},
  \bibinfo{author}{\bibfnamefont{A.~R.} \bibnamefont{{Bazer-Bachi}}},
  \bibinfo{author}{\bibfnamefont{M.}~\bibnamefont{{Beilicke}}},
  \bibinfo{author}{\bibfnamefont{W.}~\bibnamefont{{Benbow}}},
  \bibinfo{author}{\bibfnamefont{D.}~\bibnamefont{{Berge}}},
  \bibinfo{author}{\bibfnamefont{K.}~\bibnamefont{{Bernl{\"o}hr}}},
  \bibinfo{author}{\bibfnamefont{C.}~\bibnamefont{{Boisson}}},
  \bibinfo{author}{\bibfnamefont{O.}~\bibnamefont{{Bolz}}},
  \bibinfo{author}{\bibfnamefont{V.}~\bibnamefont{{Borrel}}},
  \bibnamefont{et~al.}, \bibinfo{journal}{\aap} \textbf{\bibinfo{volume}{457}},
  \bibinfo{pages}{899} (\bibinfo{year}{2006}{\natexlab{a}}),
  \eprint{arXiv:astro-ph/0607333}.

\bibitem[{\citenamefont{Arzoumanian}(2009)}]{2009_Arzoumanian_Boston}
\bibinfo{author}{\bibfnamefont{Z.}~\bibnamefont{Arzoumanian}}, in
  \emph{\bibinfo{booktitle}{at Supernova Remnants and PWNe in the Chandra Era}}
  (\bibinfo{year}{2009}).

\bibitem[{\citenamefont{{Kaaret} et~al.}(2001)\citenamefont{{Kaaret},
  {Marshall}, {Aldcroft}, {Graessle}, {Karovska}, {Murray}, {Rots}, {Schulz},
  and {Seward}}}]{2001ApJ...546.1159K}
\bibinfo{author}{\bibfnamefont{P.}~\bibnamefont{{Kaaret}}},
  \bibinfo{author}{\bibfnamefont{H.~L.} \bibnamefont{{Marshall}}},
  \bibinfo{author}{\bibfnamefont{T.~L.} \bibnamefont{{Aldcroft}}},
  \bibinfo{author}{\bibfnamefont{D.~E.} \bibnamefont{{Graessle}}},
  \bibinfo{author}{\bibfnamefont{M.}~\bibnamefont{{Karovska}}},
  \bibinfo{author}{\bibfnamefont{S.~S.} \bibnamefont{{Murray}}},
  \bibinfo{author}{\bibfnamefont{A.~H.} \bibnamefont{{Rots}}},
  \bibinfo{author}{\bibfnamefont{N.~S.} \bibnamefont{{Schulz}}},
  \bibnamefont{and} \bibinfo{author}{\bibfnamefont{F.~D.}
  \bibnamefont{{Seward}}}, \bibinfo{journal}{\apj}
  \textbf{\bibinfo{volume}{546}}, \bibinfo{pages}{1159} (\bibinfo{year}{2001}),
  \eprint{arXiv:astro-ph/0008388}.

\bibitem[{\citenamefont{{Helfand} et~al.}(2007)\citenamefont{{Helfand},
  {Gotthelf}, {Halpern}, {Camilo}, {Semler}, {Becker}, and
  {White}}}]{2007ApJ...665.1297H}
\bibinfo{author}{\bibfnamefont{D.~J.} \bibnamefont{{Helfand}}},
  \bibinfo{author}{\bibfnamefont{E.~V.} \bibnamefont{{Gotthelf}}},
  \bibinfo{author}{\bibfnamefont{J.~P.} \bibnamefont{{Halpern}}},
  \bibinfo{author}{\bibfnamefont{F.}~\bibnamefont{{Camilo}}},
  \bibinfo{author}{\bibfnamefont{D.~R.} \bibnamefont{{Semler}}},
  \bibinfo{author}{\bibfnamefont{R.~H.} \bibnamefont{{Becker}}},
  \bibnamefont{and} \bibinfo{author}{\bibfnamefont{R.~L.}
  \bibnamefont{{White}}}, \bibinfo{journal}{\apj}
  \textbf{\bibinfo{volume}{665}}, \bibinfo{pages}{1297} (\bibinfo{year}{2007}),
  \eprint{0705.0065}.

\bibitem[{\citenamefont{{Aharonian}
  et~al.}(2006{\natexlab{b}})\citenamefont{{Aharonian}, {Akhperjanian},
  {Bazer-Bachi}, {Beilicke}, {Benbow}, {Berge}, {Bernl{\"o}hr}, {Boisson},
  {Bolz}, {Borrel} et~al.}}]{2006ApJ...636..777A}
\bibinfo{author}{\bibfnamefont{F.}~\bibnamefont{{Aharonian}}},
  \bibinfo{author}{\bibfnamefont{A.~G.} \bibnamefont{{Akhperjanian}}},
  \bibinfo{author}{\bibfnamefont{A.~R.} \bibnamefont{{Bazer-Bachi}}},
  \bibinfo{author}{\bibfnamefont{M.}~\bibnamefont{{Beilicke}}},
  \bibinfo{author}{\bibfnamefont{W.}~\bibnamefont{{Benbow}}},
  \bibinfo{author}{\bibfnamefont{D.}~\bibnamefont{{Berge}}},
  \bibinfo{author}{\bibfnamefont{K.}~\bibnamefont{{Bernl{\"o}hr}}},
  \bibinfo{author}{\bibfnamefont{C.}~\bibnamefont{{Boisson}}},
  \bibinfo{author}{\bibfnamefont{O.}~\bibnamefont{{Bolz}}},
  \bibinfo{author}{\bibfnamefont{V.}~\bibnamefont{{Borrel}}},
  \bibnamefont{et~al.}, \bibinfo{journal}{\apj} \textbf{\bibinfo{volume}{636}},
  \bibinfo{pages}{777} (\bibinfo{year}{2006}{\natexlab{b}}),
  \eprint{arXiv:astro-ph/0510397}.

\bibitem[{\citenamefont{{Halpern} et~al.}(2012)\citenamefont{{Halpern},
  {Gotthelf}, and {Camilo}}}]{2012ApJ...753L..14H}
\bibinfo{author}{\bibfnamefont{J.~P.} \bibnamefont{{Halpern}}},
  \bibinfo{author}{\bibfnamefont{E.~V.} \bibnamefont{{Gotthelf}}},
  \bibnamefont{and} \bibinfo{author}{\bibfnamefont{F.}~\bibnamefont{{Camilo}}},
  \bibinfo{journal}{\apjl} \textbf{\bibinfo{volume}{753}}, \bibinfo{eid}{L14}
  (\bibinfo{year}{2012}), \eprint{1206.2338}.

\bibitem[{\citenamefont{{Renaud} et~al.}(2010)\citenamefont{{Renaud},
  {Marandon}, {Gotthelf}, {Rodriguez}, {Terrier}, {Mattana}, {Lebrun},
  {Tomsick}, and {Manchester}}}]{2010ApJ...716..663R}
\bibinfo{author}{\bibfnamefont{M.}~\bibnamefont{{Renaud}}},
  \bibinfo{author}{\bibfnamefont{V.}~\bibnamefont{{Marandon}}},
  \bibinfo{author}{\bibfnamefont{E.~V.} \bibnamefont{{Gotthelf}}},
  \bibinfo{author}{\bibfnamefont{J.}~\bibnamefont{{Rodriguez}}},
  \bibinfo{author}{\bibfnamefont{R.}~\bibnamefont{{Terrier}}},
  \bibinfo{author}{\bibfnamefont{F.}~\bibnamefont{{Mattana}}},
  \bibinfo{author}{\bibfnamefont{F.}~\bibnamefont{{Lebrun}}},
  \bibinfo{author}{\bibfnamefont{J.~A.} \bibnamefont{{Tomsick}}},
  \bibnamefont{and} \bibinfo{author}{\bibfnamefont{R.~N.}
  \bibnamefont{{Manchester}}}, \bibinfo{journal}{\apj}
  \textbf{\bibinfo{volume}{716}}, \bibinfo{pages}{663} (\bibinfo{year}{2010}),
  \eprint{0910.3074}.

\bibitem[{\citenamefont{{Safi-Harb} et~al.}(2001)\citenamefont{{Safi-Harb},
  {Harrus}, {Petre}, {Pavlov}, {Koptsevich}, and
  {Sanwal}}}]{2001ApJ...561..308S}
\bibinfo{author}{\bibfnamefont{S.}~\bibnamefont{{Safi-Harb}}},
  \bibinfo{author}{\bibfnamefont{I.~M.} \bibnamefont{{Harrus}}},
  \bibinfo{author}{\bibfnamefont{R.}~\bibnamefont{{Petre}}},
  \bibinfo{author}{\bibfnamefont{G.~G.} \bibnamefont{{Pavlov}}},
  \bibinfo{author}{\bibfnamefont{A.~B.} \bibnamefont{{Koptsevich}}},
  \bibnamefont{and} \bibinfo{author}{\bibfnamefont{D.}~\bibnamefont{{Sanwal}}},
  \bibinfo{journal}{\apj} \textbf{\bibinfo{volume}{561}}, \bibinfo{pages}{308}
  (\bibinfo{year}{2001}), \eprint{arXiv:astro-ph/0107175}.

\bibitem[{\citenamefont{{H.~E.~S.~S.~Collaboration: A.~Djannati-Atai}
  et~al.}(2007)\citenamefont{{H.~E.~S.~S.~Collaboration: A.~Djannati-Atai}, {De
  Jager}, {Terrier}, {Gallant}, and {Hoppe}}}]{2007arXiv0710.2247H}
\bibinfo{author}{\bibnamefont{{H.~E.~S.~S.~Collaboration: A.~Djannati-Atai}}},
  \bibinfo{author}{\bibfnamefont{O.~C.} \bibnamefont{{De Jager}}},
  \bibinfo{author}{\bibfnamefont{R.}~\bibnamefont{{Terrier}}},
  \bibinfo{author}{\bibfnamefont{Y.~A.} \bibnamefont{{Gallant}}},
  \bibnamefont{and} \bibinfo{author}{\bibfnamefont{S.}~\bibnamefont{{Hoppe}}},
  \bibinfo{journal}{ArXiv e-prints}  (\bibinfo{year}{2007}),
  \eprint{0710.2247}.

\bibitem[{\citenamefont{{Slane}
  et~al.}(2004{\natexlab{a}})\citenamefont{{Slane}, {Helfand}, {van der
  Swaluw}, and {Murray}}}]{2004ApJ...616..403S}
\bibinfo{author}{\bibfnamefont{P.}~\bibnamefont{{Slane}}},
  \bibinfo{author}{\bibfnamefont{D.~J.} \bibnamefont{{Helfand}}},
  \bibinfo{author}{\bibfnamefont{E.}~\bibnamefont{{van der Swaluw}}},
  \bibnamefont{and} \bibinfo{author}{\bibfnamefont{S.~S.}
  \bibnamefont{{Murray}}}, \bibinfo{journal}{\apj}
  \textbf{\bibinfo{volume}{616}}, \bibinfo{pages}{403}
  (\bibinfo{year}{2004}{\natexlab{a}}), \eprint{arXiv:astro-ph/0405380}.

\bibitem[{\citenamefont{{Halpern} et~al.}(2001)\citenamefont{{Halpern},
  {Camilo}, {Gotthelf}, {Helfand}, {Kramer}, {Lyne}, {Leighly}, and
  {Eracleous}}}]{2001ApJ...552L.125H}
\bibinfo{author}{\bibfnamefont{J.~P.} \bibnamefont{{Halpern}}},
  \bibinfo{author}{\bibfnamefont{F.}~\bibnamefont{{Camilo}}},
  \bibinfo{author}{\bibfnamefont{E.~V.} \bibnamefont{{Gotthelf}}},
  \bibinfo{author}{\bibfnamefont{D.~J.} \bibnamefont{{Helfand}}},
  \bibinfo{author}{\bibfnamefont{M.}~\bibnamefont{{Kramer}}},
  \bibinfo{author}{\bibfnamefont{A.~G.} \bibnamefont{{Lyne}}},
  \bibinfo{author}{\bibfnamefont{K.~M.} \bibnamefont{{Leighly}}},
  \bibnamefont{and}
  \bibinfo{author}{\bibfnamefont{M.}~\bibnamefont{{Eracleous}}},
  \bibinfo{journal}{\apjl} \textbf{\bibinfo{volume}{552}},
  \bibinfo{pages}{L125} (\bibinfo{year}{2001}),
  \eprint{arXiv:astro-ph/0104109}.

\bibitem[{\citenamefont{Aliu}(2009)}]{2009_Aliu_Boston}
\bibinfo{author}{\bibfnamefont{E.}~\bibnamefont{Aliu}}, in
  \emph{\bibinfo{booktitle}{at Supernova Remnants and PWNe in the Chandra Era}}
  (\bibinfo{year}{2009}).

\bibitem[{\citenamefont{{Gaensler} et~al.}(2002)\citenamefont{{Gaensler},
  {Arons}, {Kaspi}, {Pivovaroff}, {Kawai}, and {Tamura}}}]{2002ApJ...569..878G}
\bibinfo{author}{\bibfnamefont{B.~M.} \bibnamefont{{Gaensler}}},
  \bibinfo{author}{\bibfnamefont{J.}~\bibnamefont{{Arons}}},
  \bibinfo{author}{\bibfnamefont{V.~M.} \bibnamefont{{Kaspi}}},
  \bibinfo{author}{\bibfnamefont{M.~J.} \bibnamefont{{Pivovaroff}}},
  \bibinfo{author}{\bibfnamefont{N.}~\bibnamefont{{Kawai}}}, \bibnamefont{and}
  \bibinfo{author}{\bibfnamefont{K.}~\bibnamefont{{Tamura}}},
  \bibinfo{journal}{\apj} \textbf{\bibinfo{volume}{569}}, \bibinfo{pages}{878}
  (\bibinfo{year}{2002}), \eprint{arXiv:astro-ph/0110454}.

\bibitem[{\citenamefont{{Aharonian}
  et~al.}(2005{\natexlab{a}})\citenamefont{{Aharonian}, {Akhperjanian}, {Aye},
  {Bazer-Bachi}, {Beilicke}, {Benbow}, {Berge}, {Berghaus}, {Bernl{\"o}hr},
  {Boisson} et~al.}}]{2005AnA...435L..17A}
\bibinfo{author}{\bibfnamefont{F.}~\bibnamefont{{Aharonian}}},
  \bibinfo{author}{\bibfnamefont{A.~G.} \bibnamefont{{Akhperjanian}}},
  \bibinfo{author}{\bibfnamefont{K.-M.} \bibnamefont{{Aye}}},
  \bibinfo{author}{\bibfnamefont{A.~R.} \bibnamefont{{Bazer-Bachi}}},
  \bibinfo{author}{\bibfnamefont{M.}~\bibnamefont{{Beilicke}}},
  \bibinfo{author}{\bibfnamefont{W.}~\bibnamefont{{Benbow}}},
  \bibinfo{author}{\bibfnamefont{D.}~\bibnamefont{{Berge}}},
  \bibinfo{author}{\bibfnamefont{P.}~\bibnamefont{{Berghaus}}},
  \bibinfo{author}{\bibfnamefont{K.}~\bibnamefont{{Bernl{\"o}hr}}},
  \bibinfo{author}{\bibfnamefont{C.}~\bibnamefont{{Boisson}}},
  \bibnamefont{et~al.}, \bibinfo{journal}{\aap} \textbf{\bibinfo{volume}{435}},
  \bibinfo{pages}{L17} (\bibinfo{year}{2005}{\natexlab{a}}),
  \eprint{arXiv:astro-ph/0504120}.

\bibitem[{\citenamefont{{Kargaltsev} et~al.}(2009)\citenamefont{{Kargaltsev},
  {Pavlov}, and {Wong}}}]{2009ApJ...690..891K}
\bibinfo{author}{\bibfnamefont{O.}~\bibnamefont{{Kargaltsev}}},
  \bibinfo{author}{\bibfnamefont{G.~G.} \bibnamefont{{Pavlov}}},
  \bibnamefont{and} \bibinfo{author}{\bibfnamefont{J.~A.}
  \bibnamefont{{Wong}}}, \bibinfo{journal}{\apj}
  \textbf{\bibinfo{volume}{690}}, \bibinfo{pages}{891} (\bibinfo{year}{2009}).

\bibitem[{\citenamefont{{Hughes} et~al.}(2001)\citenamefont{{Hughes}, {Slane},
  {Burrows}, {Garmire}, {Nousek}, {Olbert}, and
  {Keohane}}}]{2001ApJ...559L.153H}
\bibinfo{author}{\bibfnamefont{J.~P.} \bibnamefont{{Hughes}}},
  \bibinfo{author}{\bibfnamefont{P.~O.} \bibnamefont{{Slane}}},
  \bibinfo{author}{\bibfnamefont{D.~N.} \bibnamefont{{Burrows}}},
  \bibinfo{author}{\bibfnamefont{G.}~\bibnamefont{{Garmire}}},
  \bibinfo{author}{\bibfnamefont{J.~A.} \bibnamefont{{Nousek}}},
  \bibinfo{author}{\bibfnamefont{C.~M.} \bibnamefont{{Olbert}}},
  \bibnamefont{and} \bibinfo{author}{\bibfnamefont{J.~W.}
  \bibnamefont{{Keohane}}}, \bibinfo{journal}{\apjl}
  \textbf{\bibinfo{volume}{559}}, \bibinfo{pages}{L153} (\bibinfo{year}{2001}),
  \eprint{arXiv:astro-ph/0106031}.

\bibitem[{\citenamefont{{Lu} et~al.}(2002)\citenamefont{{Lu}, {Wang},
  {Aschenbach}, {Durouchoux}, and {Song}}}]{2002ApJ...568L..49L}
\bibinfo{author}{\bibfnamefont{F.~J.} \bibnamefont{{Lu}}},
  \bibinfo{author}{\bibfnamefont{Q.~D.} \bibnamefont{{Wang}}},
  \bibinfo{author}{\bibfnamefont{B.}~\bibnamefont{{Aschenbach}}},
  \bibinfo{author}{\bibfnamefont{P.}~\bibnamefont{{Durouchoux}}},
  \bibnamefont{and} \bibinfo{author}{\bibfnamefont{L.~M.}
  \bibnamefont{{Song}}}, \bibinfo{journal}{\apjl}
  \textbf{\bibinfo{volume}{568}}, \bibinfo{pages}{L49} (\bibinfo{year}{2002}),
  \eprint{arXiv:astro-ph/0202169}.

\bibitem[{\citenamefont{{Ng} et~al.}(2005)\citenamefont{{Ng}, {Roberts}, and
  {Romani}}}]{2005ApJ...627..904N}
\bibinfo{author}{\bibfnamefont{C.-Y.} \bibnamefont{{Ng}}},
  \bibinfo{author}{\bibfnamefont{M.~S.~E.} \bibnamefont{{Roberts}}},
  \bibnamefont{and} \bibinfo{author}{\bibfnamefont{R.~W.}
  \bibnamefont{{Romani}}}, \bibinfo{journal}{\apj}
  \textbf{\bibinfo{volume}{627}}, \bibinfo{pages}{904} (\bibinfo{year}{2005}),
  \eprint{arXiv:astro-ph/0503684}.

\bibitem[{\citenamefont{{Aharonian}
  et~al.}(2006{\natexlab{c}})\citenamefont{{Aharonian}, {Akhperjanian},
  {Bazer-Bachi}, {Beilicke}, {Benbow}, {Berge}, {Bernl{\"o}hr}, {Boisson},
  {Bolz}, {Borrel} et~al.}}]{2006AnA...456..245A}
\bibinfo{author}{\bibfnamefont{F.}~\bibnamefont{{Aharonian}}},
  \bibinfo{author}{\bibfnamefont{A.~G.} \bibnamefont{{Akhperjanian}}},
  \bibinfo{author}{\bibfnamefont{A.~R.} \bibnamefont{{Bazer-Bachi}}},
  \bibinfo{author}{\bibfnamefont{M.}~\bibnamefont{{Beilicke}}},
  \bibinfo{author}{\bibfnamefont{W.}~\bibnamefont{{Benbow}}},
  \bibinfo{author}{\bibfnamefont{D.}~\bibnamefont{{Berge}}},
  \bibinfo{author}{\bibfnamefont{K.}~\bibnamefont{{Bernl{\"o}hr}}},
  \bibinfo{author}{\bibfnamefont{C.}~\bibnamefont{{Boisson}}},
  \bibinfo{author}{\bibfnamefont{O.}~\bibnamefont{{Bolz}}},
  \bibinfo{author}{\bibfnamefont{V.}~\bibnamefont{{Borrel}}},
  \bibnamefont{et~al.}, \bibinfo{journal}{\aap} \textbf{\bibinfo{volume}{456}},
  \bibinfo{pages}{245} (\bibinfo{year}{2006}{\natexlab{c}}),
  \eprint{arXiv:astro-ph/0606311}.

\bibitem[{\citenamefont{{Helfand}
  et~al.}(2003{\natexlab{a}})\citenamefont{{Helfand}, {Collins}, and
  {Gotthelf}}}]{2003ApJ...582..783H}
\bibinfo{author}{\bibfnamefont{D.~J.} \bibnamefont{{Helfand}}},
  \bibinfo{author}{\bibfnamefont{B.~F.} \bibnamefont{{Collins}}},
  \bibnamefont{and} \bibinfo{author}{\bibfnamefont{E.~V.}
  \bibnamefont{{Gotthelf}}}, \bibinfo{journal}{\apj}
  \textbf{\bibinfo{volume}{582}}, \bibinfo{pages}{783}
  (\bibinfo{year}{2003}{\natexlab{a}}), \eprint{arXiv:astro-ph/0209348}.

\bibitem[{\citenamefont{{Pavlov} et~al.}(2001)\citenamefont{{Pavlov},
  {Kargaltsev}, {Sanwal}, and {Garmire}}}]{2001ApJ...554L.189P}
\bibinfo{author}{\bibfnamefont{G.~G.} \bibnamefont{{Pavlov}}},
  \bibinfo{author}{\bibfnamefont{O.~Y.} \bibnamefont{{Kargaltsev}}},
  \bibinfo{author}{\bibfnamefont{D.}~\bibnamefont{{Sanwal}}}, \bibnamefont{and}
  \bibinfo{author}{\bibfnamefont{G.~P.} \bibnamefont{{Garmire}}},
  \bibinfo{journal}{\apjl} \textbf{\bibinfo{volume}{554}},
  \bibinfo{pages}{L189} (\bibinfo{year}{2001}),
  \eprint{arXiv:astro-ph/0104264}.

\bibitem[{\citenamefont{{Helfand} et~al.}(2001)\citenamefont{{Helfand},
  {Gotthelf}, and {Halpern}}}]{2001ApJ...556..380H}
\bibinfo{author}{\bibfnamefont{D.~J.} \bibnamefont{{Helfand}}},
  \bibinfo{author}{\bibfnamefont{E.~V.} \bibnamefont{{Gotthelf}}},
  \bibnamefont{and} \bibinfo{author}{\bibfnamefont{J.~P.}
  \bibnamefont{{Halpern}}}, \bibinfo{journal}{\apj}
  \textbf{\bibinfo{volume}{556}}, \bibinfo{pages}{380} (\bibinfo{year}{2001}),
  \eprint{arXiv:astro-ph/0007310}.

\bibitem[{\citenamefont{{Aharonian}
  et~al.}(2006{\natexlab{d}})\citenamefont{{Aharonian}, {Akhperjanian},
  {Bazer-Bachi}, {Beilicke}, {Benbow}, {Berge}, {Bernl{\"o}hr}, {Boisson},
  {Bolz}, {Borrel} et~al.}}]{2006AnA...448L..43A}
\bibinfo{author}{\bibfnamefont{F.}~\bibnamefont{{Aharonian}}},
  \bibinfo{author}{\bibfnamefont{A.~G.} \bibnamefont{{Akhperjanian}}},
  \bibinfo{author}{\bibfnamefont{A.~R.} \bibnamefont{{Bazer-Bachi}}},
  \bibinfo{author}{\bibfnamefont{M.}~\bibnamefont{{Beilicke}}},
  \bibinfo{author}{\bibfnamefont{W.}~\bibnamefont{{Benbow}}},
  \bibinfo{author}{\bibfnamefont{D.}~\bibnamefont{{Berge}}},
  \bibinfo{author}{\bibfnamefont{K.}~\bibnamefont{{Bernl{\"o}hr}}},
  \bibinfo{author}{\bibfnamefont{C.}~\bibnamefont{{Boisson}}},
  \bibinfo{author}{\bibfnamefont{O.}~\bibnamefont{{Bolz}}},
  \bibinfo{author}{\bibfnamefont{V.}~\bibnamefont{{Borrel}}},
  \bibnamefont{et~al.}, \bibinfo{journal}{\aap} \textbf{\bibinfo{volume}{448}},
  \bibinfo{pages}{L43} (\bibinfo{year}{2006}{\natexlab{d}}),
  \eprint{arXiv:astro-ph/0601575}.

\bibitem[{\citenamefont{{Roberts} et~al.}(2003)\citenamefont{{Roberts}, {Tam},
  {Kaspi}, {Lyutikov}, {Vasisht}, {Pivovaroff}, {Gotthelf}, and
  {Kawai}}}]{2003ApJ...588..992R}
\bibinfo{author}{\bibfnamefont{M.~S.~E.} \bibnamefont{{Roberts}}},
  \bibinfo{author}{\bibfnamefont{C.~R.} \bibnamefont{{Tam}}},
  \bibinfo{author}{\bibfnamefont{V.~M.} \bibnamefont{{Kaspi}}},
  \bibinfo{author}{\bibfnamefont{M.}~\bibnamefont{{Lyutikov}}},
  \bibinfo{author}{\bibfnamefont{G.}~\bibnamefont{{Vasisht}}},
  \bibinfo{author}{\bibfnamefont{M.}~\bibnamefont{{Pivovaroff}}},
  \bibinfo{author}{\bibfnamefont{E.~V.} \bibnamefont{{Gotthelf}}},
  \bibnamefont{and} \bibinfo{author}{\bibfnamefont{N.}~\bibnamefont{{Kawai}}},
  \bibinfo{journal}{\apj} \textbf{\bibinfo{volume}{588}}, \bibinfo{pages}{992}
  (\bibinfo{year}{2003}), \eprint{arXiv:astro-ph/0206450}.

\bibitem[{\citenamefont{{Gotthelf} and {Halpern}}(2008)}]{2008ApJ...681..515G}
\bibinfo{author}{\bibfnamefont{E.~V.} \bibnamefont{{Gotthelf}}}
  \bibnamefont{and} \bibinfo{author}{\bibfnamefont{J.~P.}
  \bibnamefont{{Halpern}}}, \bibinfo{journal}{\apj}
  \textbf{\bibinfo{volume}{681}}, \bibinfo{pages}{515} (\bibinfo{year}{2008}),
  \eprint{0803.1361}.

\bibitem[{\citenamefont{{Hessels} et~al.}(2008)\citenamefont{{Hessels}, {Nice},
  {Gaensler}, {Kaspi}, {Lorimer}, {Champion}, {Lyne}, {Kramer}, {Cordes},
  {Freire} et~al.}}]{2008ApJ...682L..41H}
\bibinfo{author}{\bibfnamefont{J.~W.~T.} \bibnamefont{{Hessels}}},
  \bibinfo{author}{\bibfnamefont{D.~J.} \bibnamefont{{Nice}}},
  \bibinfo{author}{\bibfnamefont{B.~M.} \bibnamefont{{Gaensler}}},
  \bibinfo{author}{\bibfnamefont{V.~M.} \bibnamefont{{Kaspi}}},
  \bibinfo{author}{\bibfnamefont{D.~R.} \bibnamefont{{Lorimer}}},
  \bibinfo{author}{\bibfnamefont{D.~J.} \bibnamefont{{Champion}}},
  \bibinfo{author}{\bibfnamefont{A.~G.} \bibnamefont{{Lyne}}},
  \bibinfo{author}{\bibfnamefont{M.}~\bibnamefont{{Kramer}}},
  \bibinfo{author}{\bibfnamefont{J.~M.} \bibnamefont{{Cordes}}},
  \bibinfo{author}{\bibfnamefont{P.~C.~C.} \bibnamefont{{Freire}}},
  \bibnamefont{et~al.}, \bibinfo{journal}{\apjl}
  \textbf{\bibinfo{volume}{682}}, \bibinfo{pages}{L41} (\bibinfo{year}{2008}),
  \eprint{0806.1200}.

\bibitem[{\citenamefont{{Aharonian}
  et~al.}(2008{\natexlab{a}})\citenamefont{{Aharonian}, {Akhperjanian}, {Barres
  de Almeida}, {Bazer-Bachi}, {Behera}, {Beilicke}, {Benbow}, {Bernl{\"o}hr},
  {Boisson}, {Bolz} et~al.}}]{2008AnA...477..353A}
\bibinfo{author}{\bibfnamefont{F.}~\bibnamefont{{Aharonian}}},
  \bibinfo{author}{\bibfnamefont{A.~G.} \bibnamefont{{Akhperjanian}}},
  \bibinfo{author}{\bibfnamefont{U.}~\bibnamefont{{Barres de Almeida}}},
  \bibinfo{author}{\bibfnamefont{A.~R.} \bibnamefont{{Bazer-Bachi}}},
  \bibinfo{author}{\bibfnamefont{B.}~\bibnamefont{{Behera}}},
  \bibinfo{author}{\bibfnamefont{M.}~\bibnamefont{{Beilicke}}},
  \bibinfo{author}{\bibfnamefont{W.}~\bibnamefont{{Benbow}}},
  \bibinfo{author}{\bibfnamefont{K.}~\bibnamefont{{Bernl{\"o}hr}}},
  \bibinfo{author}{\bibfnamefont{C.}~\bibnamefont{{Boisson}}},
  \bibinfo{author}{\bibfnamefont{O.}~\bibnamefont{{Bolz}}},
  \bibnamefont{et~al.}, \bibinfo{journal}{\aap} \textbf{\bibinfo{volume}{477}},
  \bibinfo{pages}{353} (\bibinfo{year}{2008}{\natexlab{a}}),
  \eprint{0712.1173}.

\bibitem[{\citenamefont{{Rousseau} et~al.}(2012)\citenamefont{{Rousseau},
  {Grondin}, {Van Etten}, {Lemoine-Goumard}, {Bogdanov}, {Hessels}, {Kaspi},
  {Arzoumanian}, {Camilo}, {Casandjian} et~al.}}]{2012AnA...544A...3R}
\bibinfo{author}{\bibfnamefont{R.}~\bibnamefont{{Rousseau}}},
  \bibinfo{author}{\bibfnamefont{M.-H.} \bibnamefont{{Grondin}}},
  \bibinfo{author}{\bibfnamefont{A.}~\bibnamefont{{Van Etten}}},
  \bibinfo{author}{\bibfnamefont{M.}~\bibnamefont{{Lemoine-Goumard}}},
  \bibinfo{author}{\bibfnamefont{S.}~\bibnamefont{{Bogdanov}}},
  \bibinfo{author}{\bibfnamefont{J.~W.~T.} \bibnamefont{{Hessels}}},
  \bibinfo{author}{\bibfnamefont{V.~M.} \bibnamefont{{Kaspi}}},
  \bibinfo{author}{\bibfnamefont{Z.}~\bibnamefont{{Arzoumanian}}},
  \bibinfo{author}{\bibfnamefont{F.}~\bibnamefont{{Camilo}}},
  \bibinfo{author}{\bibfnamefont{J.~M.} \bibnamefont{{Casandjian}}},
  \bibnamefont{et~al.}, \bibinfo{journal}{\aap} \textbf{\bibinfo{volume}{544}},
  \bibinfo{eid}{A3} (\bibinfo{year}{2012}), \eprint{1206.3324}.

\bibitem[{\citenamefont{{Moon} et~al.}(2004)\citenamefont{{Moon}, {Lee},
  {Eikenberry}, {Koo}, {Chatterjee}, {Kaplan}, {Hester}, {Cordes}, {Gallant},
  and {Koch-Miramond}}}]{2004ApJ...610L..33M}
\bibinfo{author}{\bibfnamefont{D.-S.} \bibnamefont{{Moon}}},
  \bibinfo{author}{\bibfnamefont{J.-J.} \bibnamefont{{Lee}}},
  \bibinfo{author}{\bibfnamefont{S.~S.} \bibnamefont{{Eikenberry}}},
  \bibinfo{author}{\bibfnamefont{B.-C.} \bibnamefont{{Koo}}},
  \bibinfo{author}{\bibfnamefont{S.}~\bibnamefont{{Chatterjee}}},
  \bibinfo{author}{\bibfnamefont{D.~L.} \bibnamefont{{Kaplan}}},
  \bibinfo{author}{\bibfnamefont{J.~J.} \bibnamefont{{Hester}}},
  \bibinfo{author}{\bibfnamefont{J.~M.} \bibnamefont{{Cordes}}},
  \bibinfo{author}{\bibfnamefont{Y.~A.} \bibnamefont{{Gallant}}},
  \bibnamefont{and}
  \bibinfo{author}{\bibfnamefont{L.}~\bibnamefont{{Koch-Miramond}}},
  \bibinfo{journal}{\apjl} \textbf{\bibinfo{volume}{610}}, \bibinfo{pages}{L33}
  (\bibinfo{year}{2004}), \eprint{arXiv:astro-ph/0406240}.

\bibitem[{\citenamefont{{Li} et~al.}(2005)\citenamefont{{Li}, {Lu}, and
  {Li}}}]{2005ApJ...628..931L}
\bibinfo{author}{\bibfnamefont{X.~H.} \bibnamefont{{Li}}},
  \bibinfo{author}{\bibfnamefont{F.~J.} \bibnamefont{{Lu}}}, \bibnamefont{and}
  \bibinfo{author}{\bibfnamefont{T.~P.} \bibnamefont{{Li}}},
  \bibinfo{journal}{\apj} \textbf{\bibinfo{volume}{628}}, \bibinfo{pages}{931}
  (\bibinfo{year}{2005}), \eprint{arXiv:astro-ph/0504293}.

\bibitem[{\citenamefont{{Roberts} et~al.}(2007)\citenamefont{{Roberts},
  {Brogan}, and {Lyutikov}}}]{2007AAS...21114403R}
\bibinfo{author}{\bibfnamefont{M.}~\bibnamefont{{Roberts}}},
  \bibinfo{author}{\bibfnamefont{C.}~\bibnamefont{{Brogan}}}, \bibnamefont{and}
  \bibinfo{author}{\bibfnamefont{M.}~\bibnamefont{{Lyutikov}}}, in
  \emph{\bibinfo{booktitle}{American Astronomical Society Meeting Abstracts}}
  (\bibinfo{year}{2007}), vol.~\bibinfo{volume}{39} of
  \emph{\bibinfo{series}{Bulletin of the American Astronomical Society}}, p.
  \bibinfo{pages}{144.03}.

\bibitem[{\citenamefont{{Hessels} et~al.}(2004)\citenamefont{{Hessels},
  {Roberts}, {Ransom}, {Kaspi}, {Romani}, {Ng}, {Freire}, and
  {Gaensler}}}]{2004ApJ...612..389H}
\bibinfo{author}{\bibfnamefont{J.~W.~T.} \bibnamefont{{Hessels}}},
  \bibinfo{author}{\bibfnamefont{M.~S.~E.} \bibnamefont{{Roberts}}},
  \bibinfo{author}{\bibfnamefont{S.~M.} \bibnamefont{{Ransom}}},
  \bibinfo{author}{\bibfnamefont{V.~M.} \bibnamefont{{Kaspi}}},
  \bibinfo{author}{\bibfnamefont{R.~W.} \bibnamefont{{Romani}}},
  \bibinfo{author}{\bibfnamefont{C.-Y.} \bibnamefont{{Ng}}},
  \bibinfo{author}{\bibfnamefont{P.~C.~C.} \bibnamefont{{Freire}}},
  \bibnamefont{and} \bibinfo{author}{\bibfnamefont{B.~M.}
  \bibnamefont{{Gaensler}}}, \bibinfo{journal}{\apj}
  \textbf{\bibinfo{volume}{612}}, \bibinfo{pages}{389} (\bibinfo{year}{2004}),
  \eprint{arXiv:astro-ph/0403632}.

\bibitem[{\citenamefont{{Abdo} et~al.}(2009)\citenamefont{{Abdo}, {Allen},
  {Aune}, {Berley}, {Chen}, {Christopher}, {DeYoung}, {Dingus}, {Ellsworth},
  {Gonzalez} et~al.}}]{2009ApJ...700L.127A}
\bibinfo{author}{\bibfnamefont{A.~A.} \bibnamefont{{Abdo}}},
  \bibinfo{author}{\bibfnamefont{B.~T.} \bibnamefont{{Allen}}},
  \bibinfo{author}{\bibfnamefont{T.}~\bibnamefont{{Aune}}},
  \bibinfo{author}{\bibfnamefont{D.}~\bibnamefont{{Berley}}},
  \bibinfo{author}{\bibfnamefont{C.}~\bibnamefont{{Chen}}},
  \bibinfo{author}{\bibfnamefont{G.~E.} \bibnamefont{{Christopher}}},
  \bibinfo{author}{\bibfnamefont{T.}~\bibnamefont{{DeYoung}}},
  \bibinfo{author}{\bibfnamefont{B.~L.} \bibnamefont{{Dingus}}},
  \bibinfo{author}{\bibfnamefont{R.~W.} \bibnamefont{{Ellsworth}}},
  \bibinfo{author}{\bibfnamefont{M.~M.} \bibnamefont{{Gonzalez}}},
  \bibnamefont{et~al.}, \bibinfo{journal}{\apjl}
  \textbf{\bibinfo{volume}{700}}, \bibinfo{pages}{L127} (\bibinfo{year}{2009}),
  \eprint{0904.1018}.

\bibitem[{\citenamefont{{Romani} et~al.}(2005)\citenamefont{{Romani}, {Ng},
  {Dodson}, and {Brisken}}}]{2005ApJ...631..480R}
\bibinfo{author}{\bibfnamefont{R.~W.} \bibnamefont{{Romani}}},
  \bibinfo{author}{\bibfnamefont{C.-Y.} \bibnamefont{{Ng}}},
  \bibinfo{author}{\bibfnamefont{R.}~\bibnamefont{{Dodson}}}, \bibnamefont{and}
  \bibinfo{author}{\bibfnamefont{W.}~\bibnamefont{{Brisken}}},
  \bibinfo{journal}{\apj} \textbf{\bibinfo{volume}{631}}, \bibinfo{pages}{480}
  (\bibinfo{year}{2005}), \eprint{arXiv:astro-ph/0506089}.

\bibitem[{\citenamefont{{Hoppe} et~al.}(2009)\citenamefont{{Hoppe}, {de
  O{\~n}a-Wilhemi}, {Kh{\'e}lifi}, {Chaves}, {de Jager}, {Stegmann}, {Terrier},
  and {for the H.~E.~S.~S.~Collaboration}}}]{2009arXiv0906.5574H}
\bibinfo{author}{\bibfnamefont{S.}~\bibnamefont{{Hoppe}}},
  \bibinfo{author}{\bibfnamefont{E.}~\bibnamefont{{de O{\~n}a-Wilhemi}}},
  \bibinfo{author}{\bibfnamefont{B.}~\bibnamefont{{Kh{\'e}lifi}}},
  \bibinfo{author}{\bibfnamefont{R.~C.~G.} \bibnamefont{{Chaves}}},
  \bibinfo{author}{\bibfnamefont{O.~C.} \bibnamefont{{de Jager}}},
  \bibinfo{author}{\bibfnamefont{C.}~\bibnamefont{{Stegmann}}},
  \bibinfo{author}{\bibfnamefont{R.}~\bibnamefont{{Terrier}}},
  \bibnamefont{and} \bibinfo{author}{\bibnamefont{{for the
  H.~E.~S.~S.~Collaboration}}}, \bibinfo{journal}{ArXiv e-prints}
  (\bibinfo{year}{2009}), \eprint{0906.5574}.

\bibitem[{\citenamefont{{Zavlin}}(2007)}]{2007ApJ...665L.143Z}
\bibinfo{author}{\bibfnamefont{V.~E.} \bibnamefont{{Zavlin}}},
  \bibinfo{journal}{\apjl} \textbf{\bibinfo{volume}{665}},
  \bibinfo{pages}{L143} (\bibinfo{year}{2007}),
  \eprint{arXiv:astro-ph/0703802}.

\bibitem[{\citenamefont{{Aharonian}
  et~al.}(2008{\natexlab{b}})\citenamefont{{Aharonian}, {Akhperjanian}, {Barres
  de Almeida}, {Bazer-Bachi}, {Behera}, {Beilicke}, {Benbow}, {Bernl{\"o}hr},
  {Boisson}, {Bolz} et~al.}}]{2008AnA...484..435A}
\bibinfo{author}{\bibfnamefont{F.}~\bibnamefont{{Aharonian}}},
  \bibinfo{author}{\bibfnamefont{A.~G.} \bibnamefont{{Akhperjanian}}},
  \bibinfo{author}{\bibfnamefont{U.}~\bibnamefont{{Barres de Almeida}}},
  \bibinfo{author}{\bibfnamefont{A.~R.} \bibnamefont{{Bazer-Bachi}}},
  \bibinfo{author}{\bibfnamefont{B.}~\bibnamefont{{Behera}}},
  \bibinfo{author}{\bibfnamefont{M.}~\bibnamefont{{Beilicke}}},
  \bibinfo{author}{\bibfnamefont{W.}~\bibnamefont{{Benbow}}},
  \bibinfo{author}{\bibfnamefont{K.}~\bibnamefont{{Bernl{\"o}hr}}},
  \bibinfo{author}{\bibfnamefont{C.}~\bibnamefont{{Boisson}}},
  \bibinfo{author}{\bibfnamefont{O.}~\bibnamefont{{Bolz}}},
  \bibnamefont{et~al.}, \bibinfo{journal}{\aap} \textbf{\bibinfo{volume}{484}},
  \bibinfo{pages}{435} (\bibinfo{year}{2008}{\natexlab{b}}),
  \eprint{0802.3841}.

\bibitem[{\citenamefont{{Aharonian} et~al.}(2009)\citenamefont{{Aharonian},
  {Akhperjanian}, {Anton}, {Barres de Almeida}, {Bazer-Bachi}, {Becherini},
  {Behera}, {Benbow}, {Bernl{\"o}hr}, {Boisson} et~al.}}]{2009AnA...499..723A}
\bibinfo{author}{\bibfnamefont{F.}~\bibnamefont{{Aharonian}}},
  \bibinfo{author}{\bibfnamefont{A.~G.} \bibnamefont{{Akhperjanian}}},
  \bibinfo{author}{\bibfnamefont{G.}~\bibnamefont{{Anton}}},
  \bibinfo{author}{\bibfnamefont{U.}~\bibnamefont{{Barres de Almeida}}},
  \bibinfo{author}{\bibfnamefont{A.~R.} \bibnamefont{{Bazer-Bachi}}},
  \bibinfo{author}{\bibfnamefont{Y.}~\bibnamefont{{Becherini}}},
  \bibinfo{author}{\bibfnamefont{B.}~\bibnamefont{{Behera}}},
  \bibinfo{author}{\bibfnamefont{W.}~\bibnamefont{{Benbow}}},
  \bibinfo{author}{\bibfnamefont{K.}~\bibnamefont{{Bernl{\"o}hr}}},
  \bibinfo{author}{\bibfnamefont{C.}~\bibnamefont{{Boisson}}},
  \bibnamefont{et~al.}, \bibinfo{journal}{\aap} \textbf{\bibinfo{volume}{499}},
  \bibinfo{pages}{723} (\bibinfo{year}{2009}), \eprint{0904.3409}.

\bibitem[{\citenamefont{{Gaensler} et~al.}(2003)\citenamefont{{Gaensler},
  {Schulz}, {Kaspi}, {Pivovaroff}, and {Becker}}}]{2003ApJ...588..441G}
\bibinfo{author}{\bibfnamefont{B.~M.} \bibnamefont{{Gaensler}}},
  \bibinfo{author}{\bibfnamefont{N.~S.} \bibnamefont{{Schulz}}},
  \bibinfo{author}{\bibfnamefont{V.~M.} \bibnamefont{{Kaspi}}},
  \bibinfo{author}{\bibfnamefont{M.~J.} \bibnamefont{{Pivovaroff}}},
  \bibnamefont{and} \bibinfo{author}{\bibfnamefont{W.~E.}
  \bibnamefont{{Becker}}}, \bibinfo{journal}{\apj}
  \textbf{\bibinfo{volume}{588}}, \bibinfo{pages}{441} (\bibinfo{year}{2003}),
  \eprint{arXiv:astro-ph/0211359}.

\bibitem[{\citenamefont{{Pavlov} et~al.}(2008)\citenamefont{{Pavlov},
  {Kargaltsev}, and {Brisken}}}]{2008ApJ...675..683P}
\bibinfo{author}{\bibfnamefont{G.~G.} \bibnamefont{{Pavlov}}},
  \bibinfo{author}{\bibfnamefont{O.}~\bibnamefont{{Kargaltsev}}},
  \bibnamefont{and} \bibinfo{author}{\bibfnamefont{W.~F.}
  \bibnamefont{{Brisken}}}, \bibinfo{journal}{\apj}
  \textbf{\bibinfo{volume}{675}}, \bibinfo{pages}{683} (\bibinfo{year}{2008}),
  \eprint{0707.3529}.

\bibitem[{\citenamefont{{Aharonian}
  et~al.}(2006{\natexlab{e}})\citenamefont{{Aharonian}, {Akhperjanian},
  {Bazer-Bachi}, {Beilicke}, {Benbow}, {Berge}, {Bernl{\"o}hr}, {Boisson},
  {Bolz}, {Borrel} et~al.}}]{2006AnA...460..365A}
\bibinfo{author}{\bibfnamefont{F.}~\bibnamefont{{Aharonian}}},
  \bibinfo{author}{\bibfnamefont{A.~G.} \bibnamefont{{Akhperjanian}}},
  \bibinfo{author}{\bibfnamefont{A.~R.} \bibnamefont{{Bazer-Bachi}}},
  \bibinfo{author}{\bibfnamefont{M.}~\bibnamefont{{Beilicke}}},
  \bibinfo{author}{\bibfnamefont{W.}~\bibnamefont{{Benbow}}},
  \bibinfo{author}{\bibfnamefont{D.}~\bibnamefont{{Berge}}},
  \bibinfo{author}{\bibfnamefont{K.}~\bibnamefont{{Bernl{\"o}hr}}},
  \bibinfo{author}{\bibfnamefont{C.}~\bibnamefont{{Boisson}}},
  \bibinfo{author}{\bibfnamefont{O.}~\bibnamefont{{Bolz}}},
  \bibinfo{author}{\bibfnamefont{V.}~\bibnamefont{{Borrel}}},
  \bibnamefont{et~al.}, \bibinfo{journal}{\aap} \textbf{\bibinfo{volume}{460}},
  \bibinfo{pages}{365} (\bibinfo{year}{2006}{\natexlab{e}}),
  \eprint{arXiv:astro-ph/0607548}.

\bibitem[{\citenamefont{{Kaspi} et~al.}(2001)\citenamefont{{Kaspi}, {Gotthelf},
  {Gaensler}, and {Lyutikov}}}]{2001ApJ...562L.163K}
\bibinfo{author}{\bibfnamefont{V.~M.} \bibnamefont{{Kaspi}}},
  \bibinfo{author}{\bibfnamefont{E.~V.} \bibnamefont{{Gotthelf}}},
  \bibinfo{author}{\bibfnamefont{B.~M.} \bibnamefont{{Gaensler}}},
  \bibnamefont{and}
  \bibinfo{author}{\bibfnamefont{M.}~\bibnamefont{{Lyutikov}}},
  \bibinfo{journal}{\apjl} \textbf{\bibinfo{volume}{562}},
  \bibinfo{pages}{L163} (\bibinfo{year}{2001}),
  \eprint{arXiv:astro-ph/0110188}.

\bibitem[{\citenamefont{{Camilo} et~al.}(2004)\citenamefont{{Camilo},
  {Gaensler}, {Gotthelf}, {Halpern}, and {Manchester}}}]{2004ApJ...616.1118C}
\bibinfo{author}{\bibfnamefont{F.}~\bibnamefont{{Camilo}}},
  \bibinfo{author}{\bibfnamefont{B.~M.} \bibnamefont{{Gaensler}}},
  \bibinfo{author}{\bibfnamefont{E.~V.} \bibnamefont{{Gotthelf}}},
  \bibinfo{author}{\bibfnamefont{J.~P.} \bibnamefont{{Halpern}}},
  \bibnamefont{and} \bibinfo{author}{\bibfnamefont{R.~N.}
  \bibnamefont{{Manchester}}}, \bibinfo{journal}{\apj}
  \textbf{\bibinfo{volume}{616}}, \bibinfo{pages}{1118} (\bibinfo{year}{2004}),
  \eprint{arXiv:astro-ph/0407593}.

\bibitem[{\citenamefont{{H.~E.~S.~S.~Collaboration}
  et~al.}(2012)\citenamefont{{H.~E.~S.~S.~Collaboration}, {Abramowski},
  {Acero}, {Aharonian}, {Akhperjanian}, {Anton}, {Balzer}, {Barnacka},
  {Becherini}, {Becker} et~al.}}]{2012AnA...541A...5H}
\bibinfo{author}{\bibnamefont{{H.~E.~S.~S.~Collaboration}}},
  \bibinfo{author}{\bibfnamefont{A.}~\bibnamefont{{Abramowski}}},
  \bibinfo{author}{\bibfnamefont{F.}~\bibnamefont{{Acero}}},
  \bibinfo{author}{\bibfnamefont{F.}~\bibnamefont{{Aharonian}}},
  \bibinfo{author}{\bibfnamefont{A.~G.} \bibnamefont{{Akhperjanian}}},
  \bibinfo{author}{\bibfnamefont{G.}~\bibnamefont{{Anton}}},
  \bibinfo{author}{\bibfnamefont{A.}~\bibnamefont{{Balzer}}},
  \bibinfo{author}{\bibfnamefont{A.}~\bibnamefont{{Barnacka}}},
  \bibinfo{author}{\bibfnamefont{Y.}~\bibnamefont{{Becherini}}},
  \bibinfo{author}{\bibfnamefont{J.}~\bibnamefont{{Becker}}},
  \bibnamefont{et~al.}, \bibinfo{journal}{\aap} \textbf{\bibinfo{volume}{541}},
  \bibinfo{eid}{A5} (\bibinfo{year}{2012}).

\bibitem[{\citenamefont{{Gaensler} et~al.}(2004)\citenamefont{{Gaensler}, {van
  der Swaluw}, {Camilo}, {Kaspi}, {Baganoff}, {Yusef-Zadeh}, and
  {Manchester}}}]{2004ApJ...616..383G}
\bibinfo{author}{\bibfnamefont{B.~M.} \bibnamefont{{Gaensler}}},
  \bibinfo{author}{\bibfnamefont{E.}~\bibnamefont{{van der Swaluw}}},
  \bibinfo{author}{\bibfnamefont{F.}~\bibnamefont{{Camilo}}},
  \bibinfo{author}{\bibfnamefont{V.~M.} \bibnamefont{{Kaspi}}},
  \bibinfo{author}{\bibfnamefont{F.~K.} \bibnamefont{{Baganoff}}},
  \bibinfo{author}{\bibfnamefont{F.}~\bibnamefont{{Yusef-Zadeh}}},
  \bibnamefont{and} \bibinfo{author}{\bibfnamefont{R.~N.}
  \bibnamefont{{Manchester}}}, \bibinfo{journal}{\apj}
  \textbf{\bibinfo{volume}{616}}, \bibinfo{pages}{383} (\bibinfo{year}{2004}),
  \eprint{arXiv:astro-ph/0312362}.

\bibitem[{\citenamefont{{Gonzalez} and
  {Safi-Harb}}(2003)}]{2003ApJ...591L.143G}
\bibinfo{author}{\bibfnamefont{M.}~\bibnamefont{{Gonzalez}}} \bibnamefont{and}
  \bibinfo{author}{\bibfnamefont{S.}~\bibnamefont{{Safi-Harb}}},
  \bibinfo{journal}{\apjl} \textbf{\bibinfo{volume}{591}},
  \bibinfo{pages}{L143} (\bibinfo{year}{2003}),
  \eprint{arXiv:astro-ph/0305497}.

\bibitem[{\citenamefont{{Djannati-Atai}}(2009)}]{2009_Djannati_Boston}
\bibinfo{author}{\bibfnamefont{A.}~\bibnamefont{{Djannati-Atai}}}, in
  \emph{\bibinfo{booktitle}{at Supernova Remnants and PWNe in the Chandra Era,
  Boston, MA}} (\bibinfo{year}{2009}).

\bibitem[{\citenamefont{{Kargaltsev}
  et~al.}(2007{\natexlab{a}})\citenamefont{{Kargaltsev}, {Pavlov}, and
  {Garmire}}}]{2007ApJ...660.1413K}
\bibinfo{author}{\bibfnamefont{O.}~\bibnamefont{{Kargaltsev}}},
  \bibinfo{author}{\bibfnamefont{G.~G.} \bibnamefont{{Pavlov}}},
  \bibnamefont{and} \bibinfo{author}{\bibfnamefont{G.~P.}
  \bibnamefont{{Garmire}}}, \bibinfo{journal}{\apj}
  \textbf{\bibinfo{volume}{660}}, \bibinfo{pages}{1413}
  (\bibinfo{year}{2007}{\natexlab{a}}), \eprint{arXiv:astro-ph/0611599}.

\bibitem[{\citenamefont{{Kargaltsev}
  et~al.}(2007{\natexlab{b}})\citenamefont{{Kargaltsev}, {Pavlov}, and
  {Garmire}}}]{2007ApJ...670..643K}
\bibinfo{author}{\bibfnamefont{O.}~\bibnamefont{{Kargaltsev}}},
  \bibinfo{author}{\bibfnamefont{G.~G.} \bibnamefont{{Pavlov}}},
  \bibnamefont{and} \bibinfo{author}{\bibfnamefont{G.~P.}
  \bibnamefont{{Garmire}}}, \bibinfo{journal}{\apj}
  \textbf{\bibinfo{volume}{670}}, \bibinfo{pages}{643}
  (\bibinfo{year}{2007}{\natexlab{b}}), \eprint{arXiv:astro-ph/0701069}.

\bibitem[{\citenamefont{{Gonzalez} et~al.}(2006)\citenamefont{{Gonzalez},
  {Kaspi}, {Pivovaroff}, and {Gaensler}}}]{2006ApJ...652..569G}
\bibinfo{author}{\bibfnamefont{M.~E.} \bibnamefont{{Gonzalez}}},
  \bibinfo{author}{\bibfnamefont{V.~M.} \bibnamefont{{Kaspi}}},
  \bibinfo{author}{\bibfnamefont{M.~J.} \bibnamefont{{Pivovaroff}}},
  \bibnamefont{and} \bibinfo{author}{\bibfnamefont{B.~M.}
  \bibnamefont{{Gaensler}}}, \bibinfo{journal}{\apj}
  \textbf{\bibinfo{volume}{652}}, \bibinfo{pages}{569} (\bibinfo{year}{2006}),
  \eprint{arXiv:astro-ph/0610523}.

\bibitem[{\citenamefont{{Kargaltsev} and {Pavlov}}(2007)}]{2007ApJ...670..655K}
\bibinfo{author}{\bibfnamefont{O.}~\bibnamefont{{Kargaltsev}}}
  \bibnamefont{and} \bibinfo{author}{\bibfnamefont{G.~G.}
  \bibnamefont{{Pavlov}}}, \bibinfo{journal}{\apj}
  \textbf{\bibinfo{volume}{670}}, \bibinfo{pages}{655} (\bibinfo{year}{2007}),
  \eprint{0705.2378}.

\bibitem[{\citenamefont{{Aharonian} et~al.}(2007)\citenamefont{{Aharonian},
  {Akhperjanian}, {Bazer-Bachi}, {Behera}, {Beilicke}, {Benbow}, {Berge},
  {Bernl{\"o}hr}, {Boisson}, {Bolz} et~al.}}]{2007AnA...472..489A}
\bibinfo{author}{\bibfnamefont{F.}~\bibnamefont{{Aharonian}}},
  \bibinfo{author}{\bibfnamefont{A.~G.} \bibnamefont{{Akhperjanian}}},
  \bibinfo{author}{\bibfnamefont{A.~R.} \bibnamefont{{Bazer-Bachi}}},
  \bibinfo{author}{\bibfnamefont{B.}~\bibnamefont{{Behera}}},
  \bibinfo{author}{\bibfnamefont{M.}~\bibnamefont{{Beilicke}}},
  \bibinfo{author}{\bibfnamefont{W.}~\bibnamefont{{Benbow}}},
  \bibinfo{author}{\bibfnamefont{D.}~\bibnamefont{{Berge}}},
  \bibinfo{author}{\bibfnamefont{K.}~\bibnamefont{{Bernl{\"o}hr}}},
  \bibinfo{author}{\bibfnamefont{C.}~\bibnamefont{{Boisson}}},
  \bibinfo{author}{\bibfnamefont{O.}~\bibnamefont{{Bolz}}},
  \bibnamefont{et~al.}, \bibinfo{journal}{\aap} \textbf{\bibinfo{volume}{472}},
  \bibinfo{pages}{489} (\bibinfo{year}{2007}), \eprint{0705.1605}.

\bibitem[{\citenamefont{{Aharonian}
  et~al.}(2005{\natexlab{b}})\citenamefont{{Aharonian}, {Akhperjanian}, {Aye},
  {Bazer-Bachi}, {Beilicke}, {Benbow}, {Berge}, {Berghaus}, {Bernl{\"o}hr},
  {Boisson} et~al.}}]{2005AnA...439.1013A}
\bibinfo{author}{\bibfnamefont{F.}~\bibnamefont{{Aharonian}}},
  \bibinfo{author}{\bibfnamefont{A.~G.} \bibnamefont{{Akhperjanian}}},
  \bibinfo{author}{\bibfnamefont{K.-M.} \bibnamefont{{Aye}}},
  \bibinfo{author}{\bibfnamefont{A.~R.} \bibnamefont{{Bazer-Bachi}}},
  \bibinfo{author}{\bibfnamefont{M.}~\bibnamefont{{Beilicke}}},
  \bibinfo{author}{\bibfnamefont{W.}~\bibnamefont{{Benbow}}},
  \bibinfo{author}{\bibfnamefont{D.}~\bibnamefont{{Berge}}},
  \bibinfo{author}{\bibfnamefont{P.}~\bibnamefont{{Berghaus}}},
  \bibinfo{author}{\bibfnamefont{K.}~\bibnamefont{{Bernl{\"o}hr}}},
  \bibinfo{author}{\bibfnamefont{C.}~\bibnamefont{{Boisson}}},
  \bibnamefont{et~al.}, \bibinfo{journal}{\aap} \textbf{\bibinfo{volume}{439}},
  \bibinfo{pages}{1013} (\bibinfo{year}{2005}{\natexlab{b}}),
  \eprint{arXiv:astro-ph/0505219}.

\bibitem[{\citenamefont{{Hinton} et~al.}(2007)\citenamefont{{Hinton}, {Funk},
  {Carrigan}, {Gallant}, {de Jager}, {Kosack}, {Lemi{\`e}re}, and
  {P{\"u}hlhofer}}}]{2007AnA...476L..25H}
\bibinfo{author}{\bibfnamefont{J.~A.} \bibnamefont{{Hinton}}},
  \bibinfo{author}{\bibfnamefont{S.}~\bibnamefont{{Funk}}},
  \bibinfo{author}{\bibfnamefont{S.}~\bibnamefont{{Carrigan}}},
  \bibinfo{author}{\bibfnamefont{Y.~A.} \bibnamefont{{Gallant}}},
  \bibinfo{author}{\bibfnamefont{O.~C.} \bibnamefont{{de Jager}}},
  \bibinfo{author}{\bibfnamefont{K.}~\bibnamefont{{Kosack}}},
  \bibinfo{author}{\bibfnamefont{A.}~\bibnamefont{{Lemi{\`e}re}}},
  \bibnamefont{and}
  \bibinfo{author}{\bibfnamefont{G.}~\bibnamefont{{P{\"u}hlhofer}}},
  \bibinfo{journal}{\aap} \textbf{\bibinfo{volume}{476}}, \bibinfo{pages}{L25}
  (\bibinfo{year}{2007}), \eprint{0710.0367}.

\bibitem[{\citenamefont{{Kargaltsev} et~al.}(2008)\citenamefont{{Kargaltsev},
  {Misanovic}, {Pavlov}, {Wong}, and {Garmire}}}]{2008ApJ...684..542K}
\bibinfo{author}{\bibfnamefont{O.}~\bibnamefont{{Kargaltsev}}},
  \bibinfo{author}{\bibfnamefont{Z.}~\bibnamefont{{Misanovic}}},
  \bibinfo{author}{\bibfnamefont{G.~G.} \bibnamefont{{Pavlov}}},
  \bibinfo{author}{\bibfnamefont{J.~A.} \bibnamefont{{Wong}}},
  \bibnamefont{and} \bibinfo{author}{\bibfnamefont{G.~P.}
  \bibnamefont{{Garmire}}}, \bibinfo{journal}{\apj}
  \textbf{\bibinfo{volume}{684}}, \bibinfo{pages}{542} (\bibinfo{year}{2008}),
  \eprint{0802.2963}.

\bibitem[{\citenamefont{{Halpern} et~al.}(2004)\citenamefont{{Halpern},
  {Gotthelf}, {Camilo}, {Helfand}, and {Ransom}}}]{2004ApJ...612..398H}
\bibinfo{author}{\bibfnamefont{J.~P.} \bibnamefont{{Halpern}}},
  \bibinfo{author}{\bibfnamefont{E.~V.} \bibnamefont{{Gotthelf}}},
  \bibinfo{author}{\bibfnamefont{F.}~\bibnamefont{{Camilo}}},
  \bibinfo{author}{\bibfnamefont{D.~J.} \bibnamefont{{Helfand}}},
  \bibnamefont{and} \bibinfo{author}{\bibfnamefont{S.~M.}
  \bibnamefont{{Ransom}}}, \bibinfo{journal}{\apj}
  \textbf{\bibinfo{volume}{612}}, \bibinfo{pages}{398} (\bibinfo{year}{2004}),
  \eprint{arXiv:astro-ph/0404312}.

\bibitem[{\citenamefont{{Aliu} et~al.}(2013)\citenamefont{{Aliu},
  {Archambault}, {Arlen}, {Aune}, {Beilicke}, {Benbow}, {Bouvier}, {Buckley},
  {Bugaev}, {Cesarini} et~al.}}]{2013ApJ...764...38A}
\bibinfo{author}{\bibfnamefont{E.}~\bibnamefont{{Aliu}}},
  \bibinfo{author}{\bibfnamefont{S.}~\bibnamefont{{Archambault}}},
  \bibinfo{author}{\bibfnamefont{T.}~\bibnamefont{{Arlen}}},
  \bibinfo{author}{\bibfnamefont{T.}~\bibnamefont{{Aune}}},
  \bibinfo{author}{\bibfnamefont{M.}~\bibnamefont{{Beilicke}}},
  \bibinfo{author}{\bibfnamefont{W.}~\bibnamefont{{Benbow}}},
  \bibinfo{author}{\bibfnamefont{A.}~\bibnamefont{{Bouvier}}},
  \bibinfo{author}{\bibfnamefont{J.~H.} \bibnamefont{{Buckley}}},
  \bibinfo{author}{\bibfnamefont{V.}~\bibnamefont{{Bugaev}}},
  \bibinfo{author}{\bibfnamefont{A.}~\bibnamefont{{Cesarini}}},
  \bibnamefont{et~al.}, \bibinfo{journal}{\apj} \textbf{\bibinfo{volume}{764}},
  \bibinfo{eid}{38} (\bibinfo{year}{2013}), \eprint{1212.4739}.

\bibitem[{\citenamefont{{Petre} et~al.}(2002)\citenamefont{{Petre}, {Kuntz},
  and {Shelton}}}]{2002ApJ...579..404P}
\bibinfo{author}{\bibfnamefont{R.}~\bibnamefont{{Petre}}},
  \bibinfo{author}{\bibfnamefont{K.~D.} \bibnamefont{{Kuntz}}},
  \bibnamefont{and} \bibinfo{author}{\bibfnamefont{R.~L.}
  \bibnamefont{{Shelton}}}, \bibinfo{journal}{\apj}
  \textbf{\bibinfo{volume}{579}}, \bibinfo{pages}{404} (\bibinfo{year}{2002}),
  \eprint{arXiv:astro-ph/0207092}.

\bibitem[{\citenamefont{{Camilo}
  et~al.}(2009{\natexlab{a}})\citenamefont{{Camilo}, {Ray}, {Ransom}, {Burgay},
  {Johnson}, {Kerr}, {Gotthelf}, {Halpern}, {Reynolds}, {Romani}
  et~al.}}]{2009ApJ...705....1C}
\bibinfo{author}{\bibfnamefont{F.}~\bibnamefont{{Camilo}}},
  \bibinfo{author}{\bibfnamefont{P.~S.} \bibnamefont{{Ray}}},
  \bibinfo{author}{\bibfnamefont{S.~M.} \bibnamefont{{Ransom}}},
  \bibinfo{author}{\bibfnamefont{M.}~\bibnamefont{{Burgay}}},
  \bibinfo{author}{\bibfnamefont{T.~J.} \bibnamefont{{Johnson}}},
  \bibinfo{author}{\bibfnamefont{M.}~\bibnamefont{{Kerr}}},
  \bibinfo{author}{\bibfnamefont{E.~V.} \bibnamefont{{Gotthelf}}},
  \bibinfo{author}{\bibfnamefont{J.~P.} \bibnamefont{{Halpern}}},
  \bibinfo{author}{\bibfnamefont{J.}~\bibnamefont{{Reynolds}}},
  \bibinfo{author}{\bibfnamefont{R.~W.} \bibnamefont{{Romani}}},
  \bibnamefont{et~al.}, \bibinfo{journal}{\apj} \textbf{\bibinfo{volume}{705}},
  \bibinfo{pages}{1} (\bibinfo{year}{2009}{\natexlab{a}}), \eprint{0908.2626}.

\bibitem[{\citenamefont{{Aharonian} et~al.}(2002)\citenamefont{{Aharonian},
  {Akhperjanian}, {Beilicke}, {Bernl{\"o}hr}, {B{\"o}rst}, {Bojahr}, {Bolz},
  {Coarasa}, {Contreras}, {Cortina} et~al.}}]{2002AnA...393L..37A}
\bibinfo{author}{\bibfnamefont{F.}~\bibnamefont{{Aharonian}}},
  \bibinfo{author}{\bibfnamefont{A.}~\bibnamefont{{Akhperjanian}}},
  \bibinfo{author}{\bibfnamefont{M.}~\bibnamefont{{Beilicke}}},
  \bibinfo{author}{\bibfnamefont{K.}~\bibnamefont{{Bernl{\"o}hr}}},
  \bibinfo{author}{\bibfnamefont{H.}~\bibnamefont{{B{\"o}rst}}},
  \bibinfo{author}{\bibfnamefont{H.}~\bibnamefont{{Bojahr}}},
  \bibinfo{author}{\bibfnamefont{O.}~\bibnamefont{{Bolz}}},
  \bibinfo{author}{\bibfnamefont{T.}~\bibnamefont{{Coarasa}}},
  \bibinfo{author}{\bibfnamefont{J.}~\bibnamefont{{Contreras}}},
  \bibinfo{author}{\bibfnamefont{J.}~\bibnamefont{{Cortina}}},
  \bibnamefont{et~al.}, \bibinfo{journal}{\aap} \textbf{\bibinfo{volume}{393}},
  \bibinfo{pages}{L37} (\bibinfo{year}{2002}), \eprint{arXiv:astro-ph/0207528}.

\bibitem[{\citenamefont{{Mukherjee} et~al.}(2007)\citenamefont{{Mukherjee},
  {Gotthelf}, and {Halpern}}}]{2007ApnSS.309...29M}
\bibinfo{author}{\bibfnamefont{R.}~\bibnamefont{{Mukherjee}}},
  \bibinfo{author}{\bibfnamefont{E.~V.} \bibnamefont{{Gotthelf}}},
  \bibnamefont{and} \bibinfo{author}{\bibfnamefont{J.~P.}
  \bibnamefont{{Halpern}}}, \bibinfo{journal}{\apss}
  \textbf{\bibinfo{volume}{309}}, \bibinfo{pages}{29} (\bibinfo{year}{2007}),
  \eprint{arXiv:astro-ph/0610299}.

\bibitem[{\citenamefont{{Weltevrede} et~al.}(2010)\citenamefont{{Weltevrede},
  {Abdo}, {Ackermann}, {Ajello}, {Axelsson}, {Baldini}, {Ballet},
  {Barbiellini}, {Bastieri}, {Baughman} et~al.}}]{2010ApJ...708.1426W}
\bibinfo{author}{\bibfnamefont{P.}~\bibnamefont{{Weltevrede}}},
  \bibinfo{author}{\bibfnamefont{A.~A.} \bibnamefont{{Abdo}}},
  \bibinfo{author}{\bibfnamefont{M.}~\bibnamefont{{Ackermann}}},
  \bibinfo{author}{\bibfnamefont{M.}~\bibnamefont{{Ajello}}},
  \bibinfo{author}{\bibfnamefont{M.}~\bibnamefont{{Axelsson}}},
  \bibinfo{author}{\bibfnamefont{L.}~\bibnamefont{{Baldini}}},
  \bibinfo{author}{\bibfnamefont{J.}~\bibnamefont{{Ballet}}},
  \bibinfo{author}{\bibfnamefont{G.}~\bibnamefont{{Barbiellini}}},
  \bibinfo{author}{\bibfnamefont{D.}~\bibnamefont{{Bastieri}}},
  \bibinfo{author}{\bibfnamefont{B.~M.} \bibnamefont{{Baughman}}},
  \bibnamefont{et~al.}, \bibinfo{journal}{\apj} \textbf{\bibinfo{volume}{708}},
  \bibinfo{pages}{1426} (\bibinfo{year}{2010}), \eprint{0911.3063}.

\bibitem[{\citenamefont{{Stappers} et~al.}(2003)\citenamefont{{Stappers},
  {Gaensler}, {Kaspi}, {van der Klis}, and {Lewin}}}]{2003Sci...299.1372S}
\bibinfo{author}{\bibfnamefont{B.~W.} \bibnamefont{{Stappers}}},
  \bibinfo{author}{\bibfnamefont{B.~M.} \bibnamefont{{Gaensler}}},
  \bibinfo{author}{\bibfnamefont{V.~M.} \bibnamefont{{Kaspi}}},
  \bibinfo{author}{\bibfnamefont{M.}~\bibnamefont{{van der Klis}}},
  \bibnamefont{and} \bibinfo{author}{\bibfnamefont{W.~H.~G.}
  \bibnamefont{{Lewin}}}, \bibinfo{journal}{Science}
  \textbf{\bibinfo{volume}{299}}, \bibinfo{pages}{1372} (\bibinfo{year}{2003}),
  \eprint{arXiv:astro-ph/0302588}.

\bibitem[{\citenamefont{{Kawai}}(2009)}]{2009_Kawai_Fermi}
\bibinfo{author}{\bibfnamefont{N.}~\bibnamefont{{Kawai}}}, in
  \emph{\bibinfo{booktitle}{talk at Fermi Symposium, Washington, DC}}
  (\bibinfo{year}{2009}).

\bibitem[{\citenamefont{{Tibolla} et~al.}(2009)\citenamefont{{Tibolla},
  {Chaves}, {de Jager}, {Domainko}, {Fiasson}, {Komin}, and
  {Kosack}}}]{2009arXiv0907.0574T}
\bibinfo{author}{\bibfnamefont{O.}~\bibnamefont{{Tibolla}}},
  \bibinfo{author}{\bibfnamefont{R.~C.~G.} \bibnamefont{{Chaves}}},
  \bibinfo{author}{\bibfnamefont{O.}~\bibnamefont{{de Jager}}},
  \bibinfo{author}{\bibfnamefont{W.}~\bibnamefont{{Domainko}}},
  \bibinfo{author}{\bibfnamefont{A.}~\bibnamefont{{Fiasson}}},
  \bibinfo{author}{\bibfnamefont{N.}~\bibnamefont{{Komin}}}, \bibnamefont{and}
  \bibinfo{author}{\bibfnamefont{K.}~\bibnamefont{{Kosack}}},
  \bibinfo{journal}{ArXiv e-prints}  (\bibinfo{year}{2009}),
  \eprint{0907.0574}.

\bibitem[{\citenamefont{{Romani} and {Ng}}(2003)}]{2003ApJ...585L..41R}
\bibinfo{author}{\bibfnamefont{R.~W.} \bibnamefont{{Romani}}} \bibnamefont{and}
  \bibinfo{author}{\bibfnamefont{C.-Y.} \bibnamefont{{Ng}}},
  \bibinfo{journal}{\apjl} \textbf{\bibinfo{volume}{585}}, \bibinfo{pages}{L41}
  (\bibinfo{year}{2003}), \eprint{arXiv:astro-ph/0301506}.

\bibitem[{\citenamefont{{Ng} et~al.}(2007)\citenamefont{{Ng}, {Romani},
  {Brisken}, {Chatterjee}, and {Kramer}}}]{2007ApJ...654..487N}
\bibinfo{author}{\bibfnamefont{C.-Y.} \bibnamefont{{Ng}}},
  \bibinfo{author}{\bibfnamefont{R.~W.} \bibnamefont{{Romani}}},
  \bibinfo{author}{\bibfnamefont{W.~F.} \bibnamefont{{Brisken}}},
  \bibinfo{author}{\bibfnamefont{S.}~\bibnamefont{{Chatterjee}}},
  \bibnamefont{and} \bibinfo{author}{\bibfnamefont{M.}~\bibnamefont{{Kramer}}},
  \bibinfo{journal}{\apj} \textbf{\bibinfo{volume}{654}}, \bibinfo{pages}{487}
  (\bibinfo{year}{2007}), \eprint{arXiv:astro-ph/0611068}.

\bibitem[{\citenamefont{{McGowan} et~al.}(2006)\citenamefont{{McGowan},
  {Vestrand}, {Kennea}, {Zane}, {Cropper}, and
  {C{\'o}rdova}}}]{2006ApJ...647.1300M}
\bibinfo{author}{\bibfnamefont{K.~E.} \bibnamefont{{McGowan}}},
  \bibinfo{author}{\bibfnamefont{W.~T.} \bibnamefont{{Vestrand}}},
  \bibinfo{author}{\bibfnamefont{J.~A.} \bibnamefont{{Kennea}}},
  \bibinfo{author}{\bibfnamefont{S.}~\bibnamefont{{Zane}}},
  \bibinfo{author}{\bibfnamefont{M.}~\bibnamefont{{Cropper}}},
  \bibnamefont{and} \bibinfo{author}{\bibfnamefont{F.~A.}
  \bibnamefont{{C{\'o}rdova}}}, \bibinfo{journal}{\apj}
  \textbf{\bibinfo{volume}{647}}, \bibinfo{pages}{1300} (\bibinfo{year}{2006}),
  \eprint{arXiv:astro-ph/0605087}.

\bibitem[{\citenamefont{{Pavlov} et~al.}(2006)\citenamefont{{Pavlov}, {Sanwal},
  and {Zavlin}}}]{2006ApJ...643.1146P}
\bibinfo{author}{\bibfnamefont{G.~G.} \bibnamefont{{Pavlov}}},
  \bibinfo{author}{\bibfnamefont{D.}~\bibnamefont{{Sanwal}}}, \bibnamefont{and}
  \bibinfo{author}{\bibfnamefont{V.~E.} \bibnamefont{{Zavlin}}},
  \bibinfo{journal}{\apj} \textbf{\bibinfo{volume}{643}}, \bibinfo{pages}{1146}
  (\bibinfo{year}{2006}), \eprint{arXiv:astro-ph/0511364}.

\bibitem[{\citenamefont{{Misanovic} et~al.}(2008)\citenamefont{{Misanovic},
  {Pavlov}, and {Garmire}}}]{2008ApJ...685.1129M}
\bibinfo{author}{\bibfnamefont{Z.}~\bibnamefont{{Misanovic}}},
  \bibinfo{author}{\bibfnamefont{G.~G.} \bibnamefont{{Pavlov}}},
  \bibnamefont{and} \bibinfo{author}{\bibfnamefont{G.~P.}
  \bibnamefont{{Garmire}}}, \bibinfo{journal}{\apj}
  \textbf{\bibinfo{volume}{685}}, \bibinfo{pages}{1129} (\bibinfo{year}{2008}),
  \eprint{0711.4171}.

\bibitem[{\citenamefont{{Hui} and {Becker}}(2007)}]{2007AnA...467.1209H}
\bibinfo{author}{\bibfnamefont{C.~Y.} \bibnamefont{{Hui}}} \bibnamefont{and}
  \bibinfo{author}{\bibfnamefont{W.}~\bibnamefont{{Becker}}},
  \bibinfo{journal}{\aap} \textbf{\bibinfo{volume}{467}}, \bibinfo{pages}{1209}
  (\bibinfo{year}{2007}), \eprint{arXiv:astro-ph/0610505}.

\bibitem[{\citenamefont{{Gaensler} et~al.}(2001)\citenamefont{{Gaensler},
  {Pivovaroff}, and {Garmire}}}]{2001ApJ...556L.107G}
\bibinfo{author}{\bibfnamefont{B.~M.} \bibnamefont{{Gaensler}}},
  \bibinfo{author}{\bibfnamefont{M.~J.} \bibnamefont{{Pivovaroff}}},
  \bibnamefont{and} \bibinfo{author}{\bibfnamefont{G.~P.}
  \bibnamefont{{Garmire}}}, \bibinfo{journal}{\apjl}
  \textbf{\bibinfo{volume}{556}}, \bibinfo{pages}{L107} (\bibinfo{year}{2001}),
  \eprint{arXiv:astro-ph/0106486}.

\bibitem[{\citenamefont{{Aharonian}
  et~al.}(2005{\natexlab{c}})\citenamefont{{Aharonian}, {Akhperjanian}, {Aye},
  {Bazer-Bachi}, {Beilicke}, {Benbow}, {Berge}, {Berghaus}, {Bernl{\"o}hr},
  {Boisson} et~al.}}]{2005AnA...432L..25A}
\bibinfo{author}{\bibfnamefont{F.}~\bibnamefont{{Aharonian}}},
  \bibinfo{author}{\bibfnamefont{A.~G.} \bibnamefont{{Akhperjanian}}},
  \bibinfo{author}{\bibfnamefont{K.-M.} \bibnamefont{{Aye}}},
  \bibinfo{author}{\bibfnamefont{A.~R.} \bibnamefont{{Bazer-Bachi}}},
  \bibinfo{author}{\bibfnamefont{M.}~\bibnamefont{{Beilicke}}},
  \bibinfo{author}{\bibfnamefont{W.}~\bibnamefont{{Benbow}}},
  \bibinfo{author}{\bibfnamefont{D.}~\bibnamefont{{Berge}}},
  \bibinfo{author}{\bibfnamefont{P.}~\bibnamefont{{Berghaus}}},
  \bibinfo{author}{\bibfnamefont{K.}~\bibnamefont{{Bernl{\"o}hr}}},
  \bibinfo{author}{\bibfnamefont{C.}~\bibnamefont{{Boisson}}},
  \bibnamefont{et~al.}, \bibinfo{journal}{\aap} \textbf{\bibinfo{volume}{432}},
  \bibinfo{pages}{L25} (\bibinfo{year}{2005}{\natexlab{c}}),
  \eprint{arXiv:astro-ph/0501265}.

\bibitem[{\citenamefont{{Camilo}
  et~al.}(2009{\natexlab{b}})\citenamefont{{Camilo}, {Ransom}, {Gaensler}, and
  {Lorimer}}}]{2009ApJ...700L..34C}
\bibinfo{author}{\bibfnamefont{F.}~\bibnamefont{{Camilo}}},
  \bibinfo{author}{\bibfnamefont{S.~M.} \bibnamefont{{Ransom}}},
  \bibinfo{author}{\bibfnamefont{B.~M.} \bibnamefont{{Gaensler}}},
  \bibnamefont{and} \bibinfo{author}{\bibfnamefont{D.~R.}
  \bibnamefont{{Lorimer}}}, \bibinfo{journal}{\apjl}
  \textbf{\bibinfo{volume}{700}}, \bibinfo{pages}{L34}
  (\bibinfo{year}{2009}{\natexlab{b}}), \eprint{0906.3289}.

\bibitem[{\citenamefont{{H.E.S.S.~Collaboration}
  et~al.}(2011{\natexlab{a}})\citenamefont{{H.E.S.S.~Collaboration},
  {Abramowski}, {Acero}, {Aharonian}, {Akhperjanian}, {Anton}, {Barnacka},
  {Barres de Almeida}, {Bazer-Bachi}, {Becherini}
  et~al.}}]{2011AnA...525A..46H}
\bibinfo{author}{\bibnamefont{{H.E.S.S.~Collaboration}}},
  \bibinfo{author}{\bibfnamefont{A.}~\bibnamefont{{Abramowski}}},
  \bibinfo{author}{\bibfnamefont{F.}~\bibnamefont{{Acero}}},
  \bibinfo{author}{\bibfnamefont{F.}~\bibnamefont{{Aharonian}}},
  \bibinfo{author}{\bibfnamefont{A.~G.} \bibnamefont{{Akhperjanian}}},
  \bibinfo{author}{\bibfnamefont{G.}~\bibnamefont{{Anton}}},
  \bibinfo{author}{\bibfnamefont{A.}~\bibnamefont{{Barnacka}}},
  \bibinfo{author}{\bibfnamefont{U.}~\bibnamefont{{Barres de Almeida}}},
  \bibinfo{author}{\bibfnamefont{A.~R.} \bibnamefont{{Bazer-Bachi}}},
  \bibinfo{author}{\bibfnamefont{Y.}~\bibnamefont{{Becherini}}},
  \bibnamefont{et~al.}, \bibinfo{journal}{\aap} \textbf{\bibinfo{volume}{525}},
  \bibinfo{eid}{A46} (\bibinfo{year}{2011}{\natexlab{a}}), \eprint{1009.3012}.

\bibitem[{\citenamefont{{Saz Parkinson} et~al.}(2010)\citenamefont{{Saz
  Parkinson}, {Dormody}, {Ziegler}, {Ray}, {Abdo}, {Ballet}, {Baring},
  {Belfiore}, {Burnett}, {Caliandro} et~al.}}]{2010ApJ...725..571S}
\bibinfo{author}{\bibfnamefont{P.~M.} \bibnamefont{{Saz Parkinson}}},
  \bibinfo{author}{\bibfnamefont{M.}~\bibnamefont{{Dormody}}},
  \bibinfo{author}{\bibfnamefont{M.}~\bibnamefont{{Ziegler}}},
  \bibinfo{author}{\bibfnamefont{P.~S.} \bibnamefont{{Ray}}},
  \bibinfo{author}{\bibfnamefont{A.~A.} \bibnamefont{{Abdo}}},
  \bibinfo{author}{\bibfnamefont{J.}~\bibnamefont{{Ballet}}},
  \bibinfo{author}{\bibfnamefont{M.~G.} \bibnamefont{{Baring}}},
  \bibinfo{author}{\bibfnamefont{A.}~\bibnamefont{{Belfiore}}},
  \bibinfo{author}{\bibfnamefont{T.~H.} \bibnamefont{{Burnett}}},
  \bibinfo{author}{\bibfnamefont{G.~A.} \bibnamefont{{Caliandro}}},
  \bibnamefont{et~al.}, \bibinfo{journal}{\apj} \textbf{\bibinfo{volume}{725}},
  \bibinfo{pages}{571} (\bibinfo{year}{2010}), \eprint{1006.2134}.

\bibitem[{\citenamefont{{Terrier} et~al.}(2008)\citenamefont{{Terrier},
  {Mattana}, {Djannati-Atai}, {Marandon}, {Renaud}, and
  {Dubois}}}]{2008AIPC.1085..312T}
\bibinfo{author}{\bibfnamefont{R.}~\bibnamefont{{Terrier}}},
  \bibinfo{author}{\bibfnamefont{F.}~\bibnamefont{{Mattana}}},
  \bibinfo{author}{\bibfnamefont{A.}~\bibnamefont{{Djannati-Atai}}},
  \bibinfo{author}{\bibfnamefont{V.}~\bibnamefont{{Marandon}}},
  \bibinfo{author}{\bibfnamefont{M.}~\bibnamefont{{Renaud}}}, \bibnamefont{and}
  \bibinfo{author}{\bibfnamefont{F.}~\bibnamefont{{Dubois}}}, in
  \emph{\bibinfo{booktitle}{American Institute of Physics Conference Series}},
  edited by \bibinfo{editor}{\bibfnamefont{F.~A.} \bibnamefont{{Aharonian}}},
  \bibinfo{editor}{\bibfnamefont{W.}~\bibnamefont{{Hofmann}}},
  \bibnamefont{and} \bibinfo{editor}{\bibfnamefont{F.}~\bibnamefont{{Rieger}}}
  (\bibinfo{year}{2008}), vol. \bibinfo{volume}{1085} of
  \emph{\bibinfo{series}{American Institute of Physics Conference Series}}, pp.
  \bibinfo{pages}{312--315}.

\bibitem[{\citenamefont{{Gotthelf} et~al.}(2011)\citenamefont{{Gotthelf},
  {Halpern}, {Terrier}, and {Mattana}}}]{2011ApJ...729L..16G}
\bibinfo{author}{\bibfnamefont{E.~V.} \bibnamefont{{Gotthelf}}},
  \bibinfo{author}{\bibfnamefont{J.~P.} \bibnamefont{{Halpern}}},
  \bibinfo{author}{\bibfnamefont{R.}~\bibnamefont{{Terrier}}},
  \bibnamefont{and}
  \bibinfo{author}{\bibfnamefont{F.}~\bibnamefont{{Mattana}}},
  \bibinfo{journal}{\apjl} \textbf{\bibinfo{volume}{729}}, \bibinfo{eid}{L16}
  (\bibinfo{year}{2011}), \eprint{1012.2121}.

\bibitem[{\citenamefont{{Ng} et~al.}(2012)\citenamefont{{Ng}, {Bucciantini},
  {Gaensler}, {Camilo}, {Chatterjee}, and {Bouchard}}}]{2012ApJ...746..105N}
\bibinfo{author}{\bibfnamefont{C.-Y.} \bibnamefont{{Ng}}},
  \bibinfo{author}{\bibfnamefont{N.}~\bibnamefont{{Bucciantini}}},
  \bibinfo{author}{\bibfnamefont{B.~M.} \bibnamefont{{Gaensler}}},
  \bibinfo{author}{\bibfnamefont{F.}~\bibnamefont{{Camilo}}},
  \bibinfo{author}{\bibfnamefont{S.}~\bibnamefont{{Chatterjee}}},
  \bibnamefont{and}
  \bibinfo{author}{\bibfnamefont{A.}~\bibnamefont{{Bouchard}}},
  \bibinfo{journal}{\apj} \textbf{\bibinfo{volume}{746}}, \bibinfo{eid}{105}
  (\bibinfo{year}{2012}), \eprint{1109.2233}.

\bibitem[{\citenamefont{{Acero} et~al.}(2013)\citenamefont{{Acero}, {Gallant},
  {Ballet}, {Renaud}, and {Terrier}}}]{2013AnA...551A...7A}
\bibinfo{author}{\bibfnamefont{F.}~\bibnamefont{{Acero}}},
  \bibinfo{author}{\bibfnamefont{Y.}~\bibnamefont{{Gallant}}},
  \bibinfo{author}{\bibfnamefont{J.}~\bibnamefont{{Ballet}}},
  \bibinfo{author}{\bibfnamefont{M.}~\bibnamefont{{Renaud}}}, \bibnamefont{and}
  \bibinfo{author}{\bibfnamefont{R.}~\bibnamefont{{Terrier}}},
  \bibinfo{journal}{\aap} \textbf{\bibinfo{volume}{551}}, \bibinfo{eid}{A7}
  (\bibinfo{year}{2013}), \eprint{1212.4156}.

\bibitem[{\citenamefont{{Sheidaei}}(2011)}]{2011ICRC....7..243S}
\bibinfo{author}{\bibfnamefont{F.}~\bibnamefont{{Sheidaei}}}, in
  \emph{\bibinfo{booktitle}{International Cosmic Ray Conference}}
  (\bibinfo{year}{2011}), vol.~\bibinfo{volume}{7} of
  \emph{\bibinfo{series}{International Cosmic Ray Conference}}, p.
  \bibinfo{pages}{243}, \eprint{1110.6837}.

\bibitem[{\citenamefont{{de los Reyes} et~al.}(2012)\citenamefont{{de los
  Reyes}, {Zajczyk}, {Chaves}, and {for the
  H.~E.~S.~S.~collaboration}}}]{2012arXiv1205.0719D}
\bibinfo{author}{\bibfnamefont{R.}~\bibnamefont{{de los Reyes}}},
  \bibinfo{author}{\bibfnamefont{A.}~\bibnamefont{{Zajczyk}}},
  \bibinfo{author}{\bibfnamefont{R.~C.~G.} \bibnamefont{{Chaves}}},
  \bibnamefont{and} \bibinfo{author}{\bibnamefont{{for the
  H.~E.~S.~S.~collaboration}}}, \bibinfo{journal}{ArXiv e-prints}
  (\bibinfo{year}{2012}), \eprint{1205.0719}.

\bibitem[{\citenamefont{{Kargaltsev}
  et~al.}(2012{\natexlab{b}})\citenamefont{{Kargaltsev}, {Schmitt}, {Pavlov},
  and {Misanovic}}}]{2012ApJ...745...99K}
\bibinfo{author}{\bibfnamefont{O.}~\bibnamefont{{Kargaltsev}}},
  \bibinfo{author}{\bibfnamefont{B.~M.} \bibnamefont{{Schmitt}}},
  \bibinfo{author}{\bibfnamefont{G.~G.} \bibnamefont{{Pavlov}}},
  \bibnamefont{and}
  \bibinfo{author}{\bibfnamefont{Z.}~\bibnamefont{{Misanovic}}},
  \bibinfo{journal}{\apj} \textbf{\bibinfo{volume}{745}}, \bibinfo{eid}{99}
  (\bibinfo{year}{2012}{\natexlab{b}}), \eprint{1110.5131}.

\bibitem[{\citenamefont{Collaboration}(2011)}]{HESSColl2011}
\bibinfo{author}{\bibfnamefont{H.}~\bibnamefont{Collaboration}}, in
  \emph{\bibinfo{booktitle}{A H.E.S.S. VHE gamma-ray source near the supernova
  remnant Kes 78}} (\bibinfo{year}{2011}),
  \urlprefix\url{http://www.mpi-hd.mpg.de/hfm/HESS/pages/home/som/2011/02/}.

\bibitem[{\citenamefont{{Abramowski} et~al.}(2012)\citenamefont{{Abramowski},
  {Acero}, {Aharonian}, {Akhperjanian}, {Anton}, {Balzer}, {Barnacka}, {Barres
  de Almeida}, {Becherini}, {Becker} et~al.}}]{2012AnA...537A.114A}
\bibinfo{author}{\bibfnamefont{A.}~\bibnamefont{{Abramowski}}},
  \bibinfo{author}{\bibfnamefont{F.}~\bibnamefont{{Acero}}},
  \bibinfo{author}{\bibfnamefont{F.}~\bibnamefont{{Aharonian}}},
  \bibinfo{author}{\bibfnamefont{A.~G.} \bibnamefont{{Akhperjanian}}},
  \bibinfo{author}{\bibfnamefont{G.}~\bibnamefont{{Anton}}},
  \bibinfo{author}{\bibfnamefont{A.}~\bibnamefont{{Balzer}}},
  \bibinfo{author}{\bibfnamefont{A.}~\bibnamefont{{Barnacka}}},
  \bibinfo{author}{\bibfnamefont{U.}~\bibnamefont{{Barres de Almeida}}},
  \bibinfo{author}{\bibfnamefont{Y.}~\bibnamefont{{Becherini}}},
  \bibinfo{author}{\bibfnamefont{J.}~\bibnamefont{{Becker}}},
  \bibnamefont{et~al.}, \bibinfo{journal}{\aap} \textbf{\bibinfo{volume}{537}},
  \bibinfo{eid}{A114} (\bibinfo{year}{2012}), \eprint{1111.2043}.

\bibitem[{\citenamefont{{Amanda Weinstein for the VERITAS
  Collaboration}}(2011)}]{2011arXiv1111.1034A}
\bibinfo{author}{\bibnamefont{{Amanda Weinstein for the VERITAS
  Collaboration}}}, \bibinfo{journal}{ArXiv e-prints}  (\bibinfo{year}{2011}),
  \eprint{1111.1034}.

\bibitem[{\citenamefont{{Aharonian}
  et~al.}(2008{\natexlab{c}})\citenamefont{{Aharonian}, {Akhperjanian},
  {Bazer-Bachi}, {Behera}, {Beilicke}, {Benbow}, {Berge}, {Bernl{\"o}hr},
  {Boisson}, {Bolz} et~al.}}]{2008AnA...481..401A}
\bibinfo{author}{\bibfnamefont{F.}~\bibnamefont{{Aharonian}}},
  \bibinfo{author}{\bibfnamefont{A.~G.} \bibnamefont{{Akhperjanian}}},
  \bibinfo{author}{\bibfnamefont{A.~R.} \bibnamefont{{Bazer-Bachi}}},
  \bibinfo{author}{\bibfnamefont{B.}~\bibnamefont{{Behera}}},
  \bibinfo{author}{\bibfnamefont{M.}~\bibnamefont{{Beilicke}}},
  \bibinfo{author}{\bibfnamefont{W.}~\bibnamefont{{Benbow}}},
  \bibinfo{author}{\bibfnamefont{D.}~\bibnamefont{{Berge}}},
  \bibinfo{author}{\bibfnamefont{K.}~\bibnamefont{{Bernl{\"o}hr}}},
  \bibinfo{author}{\bibfnamefont{C.}~\bibnamefont{{Boisson}}},
  \bibinfo{author}{\bibfnamefont{O.}~\bibnamefont{{Bolz}}},
  \bibnamefont{et~al.}, \bibinfo{journal}{\aap} \textbf{\bibinfo{volume}{481}},
  \bibinfo{pages}{401} (\bibinfo{year}{2008}{\natexlab{c}}),
  \eprint{0801.3555}.

\bibitem[{\citenamefont{Chaves}(2011)}]{chav}
\bibinfo{author}{\bibfnamefont{R.}~\bibnamefont{Chaves}}, in
  \emph{\bibinfo{booktitle}{Ph.D. Thesis, University of Heidelberg, Germany}}
  (\bibinfo{year}{2011}).

\bibitem[{\citenamefont{{Hui} and {Becker}}(2006)}]{2006AnA...448L..13H}
\bibinfo{author}{\bibfnamefont{C.~Y.} \bibnamefont{{Hui}}} \bibnamefont{and}
  \bibinfo{author}{\bibfnamefont{W.}~\bibnamefont{{Becker}}},
  \bibinfo{journal}{\aap} \textbf{\bibinfo{volume}{448}}, \bibinfo{pages}{L13}
  (\bibinfo{year}{2006}), \eprint{arXiv:astro-ph/0601189}.

\bibitem[{\citenamefont{{Hofverberg}}(2011)}]{2011ICRC....7..247H}
\bibinfo{author}{\bibfnamefont{P.}~\bibnamefont{{Hofverberg}}}, in
  \emph{\bibinfo{booktitle}{International Cosmic Ray Conference}}
  (\bibinfo{year}{2011}), vol.~\bibinfo{volume}{7} of
  \emph{\bibinfo{series}{International Cosmic Ray Conference}}, p.
  \bibinfo{pages}{247}, \eprint{1112.2901}.

\bibitem[{\citenamefont{{Helfand}
  et~al.}(2003{\natexlab{b}})\citenamefont{{Helfand}, {Ag{\"u}eros}, and
  {Gotthelf}}}]{2003ApJ...592..941H}
\bibinfo{author}{\bibfnamefont{D.~J.} \bibnamefont{{Helfand}}},
  \bibinfo{author}{\bibfnamefont{M.~A.} \bibnamefont{{Ag{\"u}eros}}},
  \bibnamefont{and} \bibinfo{author}{\bibfnamefont{E.~V.}
  \bibnamefont{{Gotthelf}}}, \bibinfo{journal}{\apj}
  \textbf{\bibinfo{volume}{592}}, \bibinfo{pages}{941}
  (\bibinfo{year}{2003}{\natexlab{b}}), \eprint{arXiv:astro-ph/0304302}.

\bibitem[{\citenamefont{Terrier}(2011)}]{Terrier2011}
\bibinfo{author}{\bibfnamefont{e.~a.} \bibnamefont{Terrier},
  \bibfnamefont{R.}}, in \emph{\bibinfo{booktitle}{1--5 August 2011, Stockholm,
  Sweden}} (\bibinfo{year}{2011}).

\bibitem[{\citenamefont{{Olbert}
  et~al.}(2003{\natexlab{a}})\citenamefont{{Olbert}, {Keohane}, {Arnaud},
  {Dyer}, {Reynolds}, and {Safi-Harb}}}]{2003ApJ...592L..45O}
\bibinfo{author}{\bibfnamefont{C.~M.} \bibnamefont{{Olbert}}},
  \bibinfo{author}{\bibfnamefont{J.~W.} \bibnamefont{{Keohane}}},
  \bibinfo{author}{\bibfnamefont{K.~A.} \bibnamefont{{Arnaud}}},
  \bibinfo{author}{\bibfnamefont{K.~K.} \bibnamefont{{Dyer}}},
  \bibinfo{author}{\bibfnamefont{S.~P.} \bibnamefont{{Reynolds}}},
  \bibnamefont{and}
  \bibinfo{author}{\bibfnamefont{S.}~\bibnamefont{{Safi-Harb}}},
  \bibinfo{journal}{\apjl} \textbf{\bibinfo{volume}{592}}, \bibinfo{pages}{L45}
  (\bibinfo{year}{2003}{\natexlab{a}}).

\bibitem[{\citenamefont{{Arzoumanian} et~al.}(2008)\citenamefont{{Arzoumanian},
  {Safi-Harb}, {Landecker}, {Kothes}, and {Camilo}}}]{2008ApJ...687..505A}
\bibinfo{author}{\bibfnamefont{Z.}~\bibnamefont{{Arzoumanian}}},
  \bibinfo{author}{\bibfnamefont{S.}~\bibnamefont{{Safi-Harb}}},
  \bibinfo{author}{\bibfnamefont{T.~L.} \bibnamefont{{Landecker}}},
  \bibinfo{author}{\bibfnamefont{R.}~\bibnamefont{{Kothes}}}, \bibnamefont{and}
  \bibinfo{author}{\bibfnamefont{F.}~\bibnamefont{{Camilo}}},
  \bibinfo{journal}{\apj} \textbf{\bibinfo{volume}{687}}, \bibinfo{pages}{505}
  (\bibinfo{year}{2008}), \eprint{0806.3766}.

\bibitem[{\citenamefont{{Gaensler} et~al.}(2006)\citenamefont{{Gaensler},
  {Chatterjee}, {Slane}, {van der Swaluw}, {Camilo}, and
  {Hughes}}}]{2006ApJ...648.1037G}
\bibinfo{author}{\bibfnamefont{B.~M.} \bibnamefont{{Gaensler}}},
  \bibinfo{author}{\bibfnamefont{S.}~\bibnamefont{{Chatterjee}}},
  \bibinfo{author}{\bibfnamefont{P.~O.} \bibnamefont{{Slane}}},
  \bibinfo{author}{\bibfnamefont{E.}~\bibnamefont{{van der Swaluw}}},
  \bibinfo{author}{\bibfnamefont{F.}~\bibnamefont{{Camilo}}}, \bibnamefont{and}
  \bibinfo{author}{\bibfnamefont{J.~P.} \bibnamefont{{Hughes}}},
  \bibinfo{journal}{\apj} \textbf{\bibinfo{volume}{648}}, \bibinfo{pages}{1037}
  (\bibinfo{year}{2006}), \eprint{arXiv:astro-ph/0601304}.

\bibitem[{\citenamefont{{Acciari} et~al.}(2009)\citenamefont{{Acciari}, {Aliu},
  {Arlen}, {Aune}, {Bautista}, {Beilicke}, {Benbow}, {Bradbury}, {Buckley},
  {Bugaev} et~al.}}]{2009ApJ...698L.133A}
\bibinfo{author}{\bibfnamefont{V.~A.} \bibnamefont{{Acciari}}},
  \bibinfo{author}{\bibfnamefont{E.}~\bibnamefont{{Aliu}}},
  \bibinfo{author}{\bibfnamefont{T.}~\bibnamefont{{Arlen}}},
  \bibinfo{author}{\bibfnamefont{T.}~\bibnamefont{{Aune}}},
  \bibinfo{author}{\bibfnamefont{M.}~\bibnamefont{{Bautista}}},
  \bibinfo{author}{\bibfnamefont{M.}~\bibnamefont{{Beilicke}}},
  \bibinfo{author}{\bibfnamefont{W.}~\bibnamefont{{Benbow}}},
  \bibinfo{author}{\bibfnamefont{S.~M.} \bibnamefont{{Bradbury}}},
  \bibinfo{author}{\bibfnamefont{J.~H.} \bibnamefont{{Buckley}}},
  \bibinfo{author}{\bibfnamefont{V.}~\bibnamefont{{Bugaev}}},
  \bibnamefont{et~al.}, \bibinfo{journal}{\apjl}
  \textbf{\bibinfo{volume}{698}}, \bibinfo{pages}{L133} (\bibinfo{year}{2009}),
  \eprint{0905.3291}.

\bibitem[{\citenamefont{{Tomsick} et~al.}(2012)\citenamefont{{Tomsick},
  {Bodaghee}, {Rodriguez}, {Chaty}, {Camilo}, {Fornasini}, and
  {Rahoui}}}]{2012ApJ...750L..39T}
\bibinfo{author}{\bibfnamefont{J.~A.} \bibnamefont{{Tomsick}}},
  \bibinfo{author}{\bibfnamefont{A.}~\bibnamefont{{Bodaghee}}},
  \bibinfo{author}{\bibfnamefont{J.}~\bibnamefont{{Rodriguez}}},
  \bibinfo{author}{\bibfnamefont{S.}~\bibnamefont{{Chaty}}},
  \bibinfo{author}{\bibfnamefont{F.}~\bibnamefont{{Camilo}}},
  \bibinfo{author}{\bibfnamefont{F.}~\bibnamefont{{Fornasini}}},
  \bibnamefont{and} \bibinfo{author}{\bibfnamefont{F.}~\bibnamefont{{Rahoui}}},
  \bibinfo{journal}{\apjl} \textbf{\bibinfo{volume}{750}}, \bibinfo{eid}{L39}
  (\bibinfo{year}{2012}), \eprint{1204.2836}.

\bibitem[{\citenamefont{{Harrus} et~al.}(2004)\citenamefont{{Harrus},
  {Bernstein}, {Slane}, {Gaensler}, {Hughes}, {Moffett}, and
  {Dodson}}}]{2004IAUS..218..203H}
\bibinfo{author}{\bibfnamefont{I.}~\bibnamefont{{Harrus}}},
  \bibinfo{author}{\bibfnamefont{J.~P.} \bibnamefont{{Bernstein}}},
  \bibinfo{author}{\bibfnamefont{P.~O.} \bibnamefont{{Slane}}},
  \bibinfo{author}{\bibfnamefont{B.}~\bibnamefont{{Gaensler}}},
  \bibinfo{author}{\bibfnamefont{J.~P.} \bibnamefont{{Hughes}}},
  \bibinfo{author}{\bibfnamefont{D.}~\bibnamefont{{Moffett}}},
  \bibnamefont{and} \bibinfo{author}{\bibfnamefont{R.}~\bibnamefont{{Dodson}}},
  in \emph{\bibinfo{booktitle}{Young Neutron Stars and Their Environments}},
  edited by \bibinfo{editor}{\bibfnamefont{F.}~\bibnamefont{{Camilo}}}
  \bibnamefont{and} \bibinfo{editor}{\bibfnamefont{B.~M.}
  \bibnamefont{{Gaensler}}} (\bibinfo{year}{2004}), vol. \bibinfo{volume}{218}
  of \emph{\bibinfo{series}{IAU Symposium}}, p. \bibinfo{pages}{203}.

\bibitem[{\citenamefont{{Slane} et~al.}(2012)\citenamefont{{Slane}, {Hughes},
  {Temim}, {Rousseau}, {Castro}, {Foight}, {Gaensler}, {Funk},
  {Lemoine-Goumard}, {Gelfand} et~al.}}]{2012ApJ...749..131S}
\bibinfo{author}{\bibfnamefont{P.}~\bibnamefont{{Slane}}},
  \bibinfo{author}{\bibfnamefont{J.~P.} \bibnamefont{{Hughes}}},
  \bibinfo{author}{\bibfnamefont{T.}~\bibnamefont{{Temim}}},
  \bibinfo{author}{\bibfnamefont{R.}~\bibnamefont{{Rousseau}}},
  \bibinfo{author}{\bibfnamefont{D.}~\bibnamefont{{Castro}}},
  \bibinfo{author}{\bibfnamefont{D.}~\bibnamefont{{Foight}}},
  \bibinfo{author}{\bibfnamefont{B.~M.} \bibnamefont{{Gaensler}}},
  \bibinfo{author}{\bibfnamefont{S.}~\bibnamefont{{Funk}}},
  \bibinfo{author}{\bibfnamefont{M.}~\bibnamefont{{Lemoine-Goumard}}},
  \bibinfo{author}{\bibfnamefont{J.~D.} \bibnamefont{{Gelfand}}},
  \bibnamefont{et~al.}, \bibinfo{journal}{\apj} \textbf{\bibinfo{volume}{749}},
  \bibinfo{eid}{131} (\bibinfo{year}{2012}), \eprint{1202.3371}.

\bibitem[{\citenamefont{{Olbert}
  et~al.}(2003{\natexlab{b}})\citenamefont{{Olbert}, {Keohane}, and
  {Gotthelf}}}]{2003AAS...203.3907O}
\bibinfo{author}{\bibfnamefont{C.~M.} \bibnamefont{{Olbert}}},
  \bibinfo{author}{\bibfnamefont{J.~W.} \bibnamefont{{Keohane}}},
  \bibnamefont{and} \bibinfo{author}{\bibfnamefont{E.~V.}
  \bibnamefont{{Gotthelf}}}, in \emph{\bibinfo{booktitle}{American Astronomical
  Society Meeting Abstracts}} (\bibinfo{year}{2003}{\natexlab{b}}),
  vol.~\bibinfo{volume}{35} of \emph{\bibinfo{series}{Bulletin of the American
  Astronomical Society}}, p. \bibinfo{pages}{1265}.

\bibitem[{\citenamefont{{Plucinsky} et~al.}(2002)\citenamefont{{Plucinsky},
  {Dickel}, {Slane}, {Edgar}, {Gaetz}, and {Smith}}}]{2002APS..APRN17037P}
\bibinfo{author}{\bibfnamefont{P.~P.} \bibnamefont{{Plucinsky}}},
  \bibinfo{author}{\bibfnamefont{J.~R.} \bibnamefont{{Dickel}}},
  \bibinfo{author}{\bibfnamefont{P.~O.} \bibnamefont{{Slane}}},
  \bibinfo{author}{\bibfnamefont{R.~J.} \bibnamefont{{Edgar}}},
  \bibinfo{author}{\bibfnamefont{T.~J.} \bibnamefont{{Gaetz}}},
  \bibnamefont{and} \bibinfo{author}{\bibfnamefont{R.~K.}
  \bibnamefont{{Smith}}}, in \emph{\bibinfo{booktitle}{APS Meeting Abstracts}}
  (\bibinfo{year}{2002}), p. \bibinfo{pages}{17037}.

\bibitem[{\citenamefont{{Slane}
  et~al.}(2004{\natexlab{b}})\citenamefont{{Slane}, {Gaensler}, {van der
  Swaluw}, {Hughes}, and {Jenkins}}}]{2004AAS...205.8405S}
\bibinfo{author}{\bibfnamefont{P.}~\bibnamefont{{Slane}}},
  \bibinfo{author}{\bibfnamefont{B.~M.} \bibnamefont{{Gaensler}}},
  \bibinfo{author}{\bibfnamefont{E.}~\bibnamefont{{van der Swaluw}}},
  \bibinfo{author}{\bibfnamefont{J.~P.} \bibnamefont{{Hughes}}},
  \bibnamefont{and} \bibinfo{author}{\bibfnamefont{J.~A.}
  \bibnamefont{{Jenkins}}}, in \emph{\bibinfo{booktitle}{American Astronomical
  Society Meeting Abstracts}} (\bibinfo{year}{2004}{\natexlab{b}}),
  vol.~\bibinfo{volume}{36} of \emph{\bibinfo{series}{Bulletin of the American
  Astronomical Society}}, p. \bibinfo{pages}{1481}.

\bibitem[{\citenamefont{{Acero} et~al.}(2012)\citenamefont{{Acero},
  {Djannati-Ata{\"i}}, {F{\"o}rster}, {Gallant}, {Renaud}, and {for the
  H.~E.~S.~S.~collaboration}}}]{2012arXiv1201.0481A}
\bibinfo{author}{\bibfnamefont{F.}~\bibnamefont{{Acero}}},
  \bibinfo{author}{\bibfnamefont{A.}~\bibnamefont{{Djannati-Ata{\"i}}}},
  \bibinfo{author}{\bibfnamefont{A.}~\bibnamefont{{F{\"o}rster}}},
  \bibinfo{author}{\bibfnamefont{Y.}~\bibnamefont{{Gallant}}},
  \bibinfo{author}{\bibfnamefont{M.}~\bibnamefont{{Renaud}}}, \bibnamefont{and}
  \bibinfo{author}{\bibnamefont{{for the H.~E.~S.~S.~collaboration}}},
  \bibinfo{journal}{ArXiv e-prints}  (\bibinfo{year}{2012}),
  \eprint{1201.0481}.

\bibitem[{\citenamefont{{Lemiere} et~al.}(2009)\citenamefont{{Lemiere},
  {Slane}, {Gaensler}, and {Murray}}}]{2009ApJ...706.1269L}
\bibinfo{author}{\bibfnamefont{A.}~\bibnamefont{{Lemiere}}},
  \bibinfo{author}{\bibfnamefont{P.}~\bibnamefont{{Slane}}},
  \bibinfo{author}{\bibfnamefont{B.~M.} \bibnamefont{{Gaensler}}},
  \bibnamefont{and} \bibinfo{author}{\bibfnamefont{S.}~\bibnamefont{{Murray}}},
  \bibinfo{journal}{\apj} \textbf{\bibinfo{volume}{706}}, \bibinfo{pages}{1269}
  (\bibinfo{year}{2009}), \eprint{0910.2652}.

\bibitem[{\citenamefont{{Mandelartz} and {Becker
  Tjus}}(2013)}]{2013arXiv1301.2437M}
\bibinfo{author}{\bibfnamefont{M.}~\bibnamefont{{Mandelartz}}}
  \bibnamefont{and} \bibinfo{author}{\bibfnamefont{J.}~\bibnamefont{{Becker
  Tjus}}}, \bibinfo{journal}{ArXiv e-prints}  (\bibinfo{year}{2013}),
  \eprint{1301.2437}.

\bibitem[{\citenamefont{{Abdo} et~al.}(2010)\citenamefont{{Abdo}, {Ackermann},
  {Ajello}, {Allafort}, {Baldini}, {Ballet}, {Barbiellini}, {Bastieri},
  {Bechtol}, {Bellazzini} et~al.}}]{2010ApJ...718..348A}
\bibinfo{author}{\bibfnamefont{A.~A.} \bibnamefont{{Abdo}}},
  \bibinfo{author}{\bibfnamefont{M.}~\bibnamefont{{Ackermann}}},
  \bibinfo{author}{\bibfnamefont{M.}~\bibnamefont{{Ajello}}},
  \bibinfo{author}{\bibfnamefont{A.}~\bibnamefont{{Allafort}}},
  \bibinfo{author}{\bibfnamefont{L.}~\bibnamefont{{Baldini}}},
  \bibinfo{author}{\bibfnamefont{J.}~\bibnamefont{{Ballet}}},
  \bibinfo{author}{\bibfnamefont{G.}~\bibnamefont{{Barbiellini}}},
  \bibinfo{author}{\bibfnamefont{D.}~\bibnamefont{{Bastieri}}},
  \bibinfo{author}{\bibfnamefont{K.}~\bibnamefont{{Bechtol}}},
  \bibinfo{author}{\bibfnamefont{R.}~\bibnamefont{{Bellazzini}}},
  \bibnamefont{et~al.}, \bibinfo{journal}{\apj} \textbf{\bibinfo{volume}{718}},
  \bibinfo{pages}{348} (\bibinfo{year}{2010}).

\bibitem[{\citenamefont{{Brun} et~al.}(2011)\citenamefont{{Brun}, {de Naurois},
  {Hofmann}, {Carrigan}, {Djannati-Ata{\"i}}, {Ohm}, and {for the
  H.~E.~S.~S.~Collaboration}}}]{2011arXiv1104.5003B}
\bibinfo{author}{\bibfnamefont{F.}~\bibnamefont{{Brun}}},
  \bibinfo{author}{\bibfnamefont{M.}~\bibnamefont{{de Naurois}}},
  \bibinfo{author}{\bibfnamefont{W.}~\bibnamefont{{Hofmann}}},
  \bibinfo{author}{\bibfnamefont{S.}~\bibnamefont{{Carrigan}}},
  \bibinfo{author}{\bibfnamefont{A.}~\bibnamefont{{Djannati-Ata{\"i}}}},
  \bibinfo{author}{\bibfnamefont{S.}~\bibnamefont{{Ohm}}}, \bibnamefont{and}
  \bibinfo{author}{\bibnamefont{{for the H.~E.~S.~S.~Collaboration}}},
  \bibinfo{journal}{ArXiv e-prints}  (\bibinfo{year}{2011}),
  \eprint{1104.5003}.

\bibitem[{\citenamefont{{H.E.S.S.~Collaboration}
  et~al.}(2011{\natexlab{b}})\citenamefont{{H.E.S.S.~Collaboration},
  {Abramowski}, {Acero}, {Aharonian}, {Akhperjanian}, {Anton}, {Balzer},
  {Barnacka}, {Barres de Almeida}, {Becherini} et~al.}}]{2011AnA...531A..81H}
\bibinfo{author}{\bibnamefont{{H.E.S.S.~Collaboration}}},
  \bibinfo{author}{\bibfnamefont{A.}~\bibnamefont{{Abramowski}}},
  \bibinfo{author}{\bibfnamefont{F.}~\bibnamefont{{Acero}}},
  \bibinfo{author}{\bibfnamefont{F.}~\bibnamefont{{Aharonian}}},
  \bibinfo{author}{\bibfnamefont{A.~G.} \bibnamefont{{Akhperjanian}}},
  \bibinfo{author}{\bibfnamefont{G.}~\bibnamefont{{Anton}}},
  \bibinfo{author}{\bibfnamefont{A.}~\bibnamefont{{Balzer}}},
  \bibinfo{author}{\bibfnamefont{A.}~\bibnamefont{{Barnacka}}},
  \bibinfo{author}{\bibfnamefont{U.}~\bibnamefont{{Barres de Almeida}}},
  \bibinfo{author}{\bibfnamefont{Y.}~\bibnamefont{{Becherini}}},
  \bibnamefont{et~al.}, \bibinfo{journal}{\aap} \textbf{\bibinfo{volume}{531}},
  \bibinfo{eid}{A81} (\bibinfo{year}{2011}{\natexlab{b}}), \eprint{1105.3206}.

\bibitem[{\citenamefont{{Rowell} et~al.}(2012)\citenamefont{{Rowell},
  {Naurois}, {Ata{\"i}}, {Gallant}, and
  {H.~E.~S.~S.~Collaboration}}}]{2012AIPC.1505..273R}
\bibinfo{author}{\bibfnamefont{G.}~\bibnamefont{{Rowell}}},
  \bibinfo{author}{\bibfnamefont{M.~D.} \bibnamefont{{Naurois}}},
  \bibinfo{author}{\bibfnamefont{A.~D.} \bibnamefont{{Ata{\"i}}}},
  \bibinfo{author}{\bibfnamefont{Y.}~\bibnamefont{{Gallant}}},
  \bibnamefont{and}
  \bibinfo{author}{\bibnamefont{{H.~E.~S.~S.~Collaboration}}}, in
  \emph{\bibinfo{booktitle}{American Institute of Physics Conference Series}},
  edited by \bibinfo{editor}{\bibfnamefont{F.~A.} \bibnamefont{{Aharonian}}},
  \bibinfo{editor}{\bibfnamefont{W.}~\bibnamefont{{Hofmann}}},
  \bibnamefont{and} \bibinfo{editor}{\bibfnamefont{F.~M.}
  \bibnamefont{{Rieger}}} (\bibinfo{year}{2012}), vol. \bibinfo{volume}{1505}
  of \emph{\bibinfo{series}{American Institute of Physics Conference Series}},
  pp. \bibinfo{pages}{273--276}.

\bibitem[{\citenamefont{{Sguera} et~al.}(2009)\citenamefont{{Sguera}, {Romero},
  {Bazzano}, {Masetti}, {Bird}, and {Bassani}}}]{2009ApJ...697.1194S}
\bibinfo{author}{\bibfnamefont{V.}~\bibnamefont{{Sguera}}},
  \bibinfo{author}{\bibfnamefont{G.~E.} \bibnamefont{{Romero}}},
  \bibinfo{author}{\bibfnamefont{A.}~\bibnamefont{{Bazzano}}},
  \bibinfo{author}{\bibfnamefont{N.}~\bibnamefont{{Masetti}}},
  \bibinfo{author}{\bibfnamefont{A.~J.} \bibnamefont{{Bird}}},
  \bibnamefont{and}
  \bibinfo{author}{\bibfnamefont{L.}~\bibnamefont{{Bassani}}},
  \bibinfo{journal}{\apj} \textbf{\bibinfo{volume}{697}}, \bibinfo{pages}{1194}
  (\bibinfo{year}{2009}), \eprint{0903.1763}.

\bibitem[{\citenamefont{{Wilbert}}(2012)}]{2012xrb..confE..64W}
\bibinfo{author}{\bibfnamefont{S.}~\bibnamefont{{Wilbert}}}, in
  \emph{\bibinfo{booktitle}{X-ray Binaries. Celebrating 50 Years Since the
  Discovery of Sco X-1}} (\bibinfo{year}{2012}).

\bibitem[{\citenamefont{{Halpern} and {Gotthelf}}(2010)}]{2010ApJ...725.1384H}
\bibinfo{author}{\bibfnamefont{J.~P.} \bibnamefont{{Halpern}}}
  \bibnamefont{and} \bibinfo{author}{\bibfnamefont{E.~V.}
  \bibnamefont{{Gotthelf}}}, \bibinfo{journal}{\apj}
  \textbf{\bibinfo{volume}{725}}, \bibinfo{pages}{1384} (\bibinfo{year}{2010}),
  \eprint{1008.2558}.

\bibitem[{\citenamefont{{Albert} et~al.}(2006)\citenamefont{{Albert}, {Aliu},
  {Anderhub}, {Antoranz}, {Armada}, {Asensio}, {Baixeras}, {Barrio}, {Bartelt},
  {Bartko} et~al.}}]{2006ApJ...643L..53A}
\bibinfo{author}{\bibfnamefont{J.}~\bibnamefont{{Albert}}},
  \bibinfo{author}{\bibfnamefont{E.}~\bibnamefont{{Aliu}}},
  \bibinfo{author}{\bibfnamefont{H.}~\bibnamefont{{Anderhub}}},
  \bibinfo{author}{\bibfnamefont{P.}~\bibnamefont{{Antoranz}}},
  \bibinfo{author}{\bibfnamefont{A.}~\bibnamefont{{Armada}}},
  \bibinfo{author}{\bibfnamefont{M.}~\bibnamefont{{Asensio}}},
  \bibinfo{author}{\bibfnamefont{C.}~\bibnamefont{{Baixeras}}},
  \bibinfo{author}{\bibfnamefont{J.~A.} \bibnamefont{{Barrio}}},
  \bibinfo{author}{\bibfnamefont{M.}~\bibnamefont{{Bartelt}}},
  \bibinfo{author}{\bibfnamefont{H.}~\bibnamefont{{Bartko}}},
  \bibnamefont{et~al.}, \bibinfo{journal}{\apjl}
  \textbf{\bibinfo{volume}{643}}, \bibinfo{pages}{L53} (\bibinfo{year}{2006}),
  \eprint{arXiv:astro-ph/0604197}.

\bibitem[{\citenamefont{{Misanovic} et~al.}(2011)\citenamefont{{Misanovic},
  {Kargaltsev}, and {Pavlov}}}]{2011ApJ...735...33M}
\bibinfo{author}{\bibfnamefont{Z.}~\bibnamefont{{Misanovic}}},
  \bibinfo{author}{\bibfnamefont{O.}~\bibnamefont{{Kargaltsev}}},
  \bibnamefont{and} \bibinfo{author}{\bibfnamefont{G.~G.}
  \bibnamefont{{Pavlov}}}, \bibinfo{journal}{\apj}
  \textbf{\bibinfo{volume}{735}}, \bibinfo{eid}{33} (\bibinfo{year}{2011}),
  \eprint{1101.1342}.

\bibitem[{\citenamefont{{Fujinaga} et~al.}(2013)\citenamefont{{Fujinaga},
  {Mori}, {Bamba}, {Kimura}, {Dotani}, {Ozaki}, {Matsuta}, {Puehlhofer},
  {Uchiyama}, {Hiraga} et~al.}}]{2013arXiv1301.5274F}
\bibinfo{author}{\bibfnamefont{T.}~\bibnamefont{{Fujinaga}}},
  \bibinfo{author}{\bibfnamefont{K.}~\bibnamefont{{Mori}}},
  \bibinfo{author}{\bibfnamefont{A.}~\bibnamefont{{Bamba}}},
  \bibinfo{author}{\bibfnamefont{S.}~\bibnamefont{{Kimura}}},
  \bibinfo{author}{\bibfnamefont{T.}~\bibnamefont{{Dotani}}},
  \bibinfo{author}{\bibfnamefont{M.}~\bibnamefont{{Ozaki}}},
  \bibinfo{author}{\bibfnamefont{K.}~\bibnamefont{{Matsuta}}},
  \bibinfo{author}{\bibfnamefont{G.}~\bibnamefont{{Puehlhofer}}},
  \bibinfo{author}{\bibfnamefont{H.}~\bibnamefont{{Uchiyama}}},
  \bibinfo{author}{\bibfnamefont{J.~S.} \bibnamefont{{Hiraga}}},
  \bibnamefont{et~al.}, \bibinfo{journal}{ArXiv e-prints}
  (\bibinfo{year}{2013}), \eprint{1301.5274}.

\bibitem[{\citenamefont{{H.E.S.S.~Collaboration}
  et~al.}(2011{\natexlab{c}})\citenamefont{{H.E.S.S.~Collaboration},
  {Abramowski}, {Acero}, {Aharonian}, {Akhperjanian}, {Anton}, {Balzer},
  {Barnacka}, {Barres de Almeida}, {Becherini} et~al.}}]{2011AnA...531L..18H}
\bibinfo{author}{\bibnamefont{{H.E.S.S.~Collaboration}}},
  \bibinfo{author}{\bibfnamefont{A.}~\bibnamefont{{Abramowski}}},
  \bibinfo{author}{\bibfnamefont{F.}~\bibnamefont{{Acero}}},
  \bibinfo{author}{\bibfnamefont{F.}~\bibnamefont{{Aharonian}}},
  \bibinfo{author}{\bibfnamefont{A.~G.} \bibnamefont{{Akhperjanian}}},
  \bibinfo{author}{\bibfnamefont{G.}~\bibnamefont{{Anton}}},
  \bibinfo{author}{\bibfnamefont{A.}~\bibnamefont{{Balzer}}},
  \bibinfo{author}{\bibfnamefont{A.}~\bibnamefont{{Barnacka}}},
  \bibinfo{author}{\bibfnamefont{U.}~\bibnamefont{{Barres de Almeida}}},
  \bibinfo{author}{\bibfnamefont{Y.}~\bibnamefont{{Becherini}}},
  \bibnamefont{et~al.}, \bibinfo{journal}{\aap} \textbf{\bibinfo{volume}{531}},
  \bibinfo{eid}{L18} (\bibinfo{year}{2011}{\natexlab{c}}), \eprint{1106.4069}.

\bibitem[{\citenamefont{{Chaves} et~al.}(2008)\citenamefont{{Chaves}, {Renaud},
  {Lemoine-Goumard}, and {Goret}}}]{2008AIPC.1085..372C}
\bibinfo{author}{\bibfnamefont{R.~C.~G.} \bibnamefont{{Chaves}}},
  \bibinfo{author}{\bibfnamefont{M.}~\bibnamefont{{Renaud}}},
  \bibinfo{author}{\bibfnamefont{M.}~\bibnamefont{{Lemoine-Goumard}}},
  \bibnamefont{and} \bibinfo{author}{\bibfnamefont{P.}~\bibnamefont{{Goret}}},
  in \emph{\bibinfo{booktitle}{American Institute of Physics Conference
  Series}}, edited by \bibinfo{editor}{\bibfnamefont{F.~A.}
  \bibnamefont{{Aharonian}}},
  \bibinfo{editor}{\bibfnamefont{W.}~\bibnamefont{{Hofmann}}},
  \bibnamefont{and} \bibinfo{editor}{\bibfnamefont{F.}~\bibnamefont{{Rieger}}}
  (\bibinfo{year}{2008}), vol. \bibinfo{volume}{1085} of
  \emph{\bibinfo{series}{American Institute of Physics Conference Series}}, pp.
  \bibinfo{pages}{372--375}.

\bibitem[{\citenamefont{{Hofverberg} et~al.}(2010)\citenamefont{{Hofverberg},
  {Chaves}, {Fiasson}, {Kosack}, {M{\'e}hault}, {de On{\~a} Wilhelmi}, and
  {H.E.S.S.~Collaboration}}}]{2010tsra.confE.196H}
\bibinfo{author}{\bibfnamefont{P.}~\bibnamefont{{Hofverberg}}},
  \bibinfo{author}{\bibfnamefont{R.~C.~G.} \bibnamefont{{Chaves}}},
  \bibinfo{author}{\bibfnamefont{A.}~\bibnamefont{{Fiasson}}},
  \bibinfo{author}{\bibfnamefont{K.}~\bibnamefont{{Kosack}}},
  \bibinfo{author}{\bibfnamefont{J.}~\bibnamefont{{M{\'e}hault}}},
  \bibinfo{author}{\bibfnamefont{E.}~\bibnamefont{{de On{\~a} Wilhelmi}}},
  \bibnamefont{and} \bibinfo{author}{\bibnamefont{{H.E.S.S.~Collaboration}}},
  in \emph{\bibinfo{booktitle}{25th Texas Symposium on Relativistic
  Astrophysics}} (\bibinfo{year}{2010}), \eprint{1104.5119}.

\bibitem[{\citenamefont{{Hoppe}}(2008)}]{2008ICRC....2..579H}
\bibinfo{author}{\bibfnamefont{S.}~\bibnamefont{{Hoppe}}}, in
  \emph{\bibinfo{booktitle}{International Cosmic Ray Conference}}
  (\bibinfo{year}{2008}), vol.~\bibinfo{volume}{2} of
  \emph{\bibinfo{series}{International Cosmic Ray Conference}}, pp.
  \bibinfo{pages}{579--582}, \eprint{0710.3528}.

\bibitem[{\citenamefont{{H.E.S.S.~Collaboration}
  et~al.}(2011{\natexlab{d}})\citenamefont{{H.E.S.S.~Collaboration}, {Acero},
  {Aharonian}, {Akhperjanian}, {Anton}, {Barres de Almeida}, {Bazer-Bachi},
  {Becherini}, {Behera}, {Bernl{\"o}hr} et~al.}}]{2011AnA...525A..45H}
\bibinfo{author}{\bibnamefont{{H.E.S.S.~Collaboration}}},
  \bibinfo{author}{\bibfnamefont{F.}~\bibnamefont{{Acero}}},
  \bibinfo{author}{\bibfnamefont{F.}~\bibnamefont{{Aharonian}}},
  \bibinfo{author}{\bibfnamefont{A.~G.} \bibnamefont{{Akhperjanian}}},
  \bibinfo{author}{\bibfnamefont{G.}~\bibnamefont{{Anton}}},
  \bibinfo{author}{\bibfnamefont{U.}~\bibnamefont{{Barres de Almeida}}},
  \bibinfo{author}{\bibfnamefont{A.~R.} \bibnamefont{{Bazer-Bachi}}},
  \bibinfo{author}{\bibfnamefont{Y.}~\bibnamefont{{Becherini}}},
  \bibinfo{author}{\bibfnamefont{B.}~\bibnamefont{{Behera}}},
  \bibinfo{author}{\bibfnamefont{K.}~\bibnamefont{{Bernl{\"o}hr}}},
  \bibnamefont{et~al.}, \bibinfo{journal}{\aap} \textbf{\bibinfo{volume}{525}},
  \bibinfo{eid}{A45} (\bibinfo{year}{2011}{\natexlab{d}}), \eprint{1010.4907}.

\bibitem[{\citenamefont{{Domainko} and {Ohm}}(2012)}]{2012AnA...545A..94D}
\bibinfo{author}{\bibfnamefont{W.}~\bibnamefont{{Domainko}}} \bibnamefont{and}
  \bibinfo{author}{\bibfnamefont{S.}~\bibnamefont{{Ohm}}},
  \bibinfo{journal}{\aap} \textbf{\bibinfo{volume}{545}}, \bibinfo{eid}{A94}
  (\bibinfo{year}{2012}), \eprint{1203.6545}.

\bibitem[{\citenamefont{{Van Etten} et~al.}(2009)\citenamefont{{Van Etten},
  {Funk}, and {Hinton}}}]{2009ApJ...707.1717V}
\bibinfo{author}{\bibfnamefont{A.}~\bibnamefont{{Van Etten}}},
  \bibinfo{author}{\bibfnamefont{S.}~\bibnamefont{{Funk}}}, \bibnamefont{and}
  \bibinfo{author}{\bibfnamefont{J.}~\bibnamefont{{Hinton}}},
  \bibinfo{journal}{\apj} \textbf{\bibinfo{volume}{707}}, \bibinfo{pages}{1717}
  (\bibinfo{year}{2009}), \eprint{0911.0700}.

\bibitem[{\citenamefont{{Oya} et~al.}(2013)\citenamefont{{Oya}, {Dalton},
  {Behera}, {Bordas}, {Djannati-Ata{\"i}}, {Hahn}, {Marandon}, {Schwanke},
  {Spengler}, and {H.~E.~S.~S.~Collaboration}}}]{2013arXiv1303.0979O}
\bibinfo{author}{\bibfnamefont{I.}~\bibnamefont{{Oya}}},
  \bibinfo{author}{\bibfnamefont{M.}~\bibnamefont{{Dalton}}},
  \bibinfo{author}{\bibfnamefont{B.}~\bibnamefont{{Behera}}},
  \bibinfo{author}{\bibfnamefont{P.}~\bibnamefont{{Bordas}}},
  \bibinfo{author}{\bibfnamefont{A.}~\bibnamefont{{Djannati-Ata{\"i}}}},
  \bibinfo{author}{\bibfnamefont{J.}~\bibnamefont{{Hahn}}},
  \bibinfo{author}{\bibfnamefont{V.}~\bibnamefont{{Marandon}}},
  \bibinfo{author}{\bibfnamefont{U.}~\bibnamefont{{Schwanke}}},
  \bibinfo{author}{\bibfnamefont{G.}~\bibnamefont{{Spengler}}},
  \bibnamefont{and}
  \bibinfo{author}{\bibnamefont{{H.~E.~S.~S.~Collaboration}}},
  \bibinfo{journal}{ArXiv e-prints}  (\bibinfo{year}{2013}),
  \eprint{1303.0979}.

\end{thebibliography}

\end{document}